\RequirePackage[T1]{fontenc}
\documentclass[12pt]{article}
\pdfoutput=1

\usepackage[height=8.85in,width=6.45in]{geometry}

\usepackage{times}

\usepackage{adjustbox}

\usepackage[utf8]{inputenc}
\usepackage{amsmath}
\usepackage{amssymb}
\usepackage{mathtools}
\numberwithin{equation}{section}
\usepackage{slashed}
\usepackage{braket}
\usepackage[svgnames]{xcolor}
\usepackage[colorlinks,citecolor=DarkGreen,linkcolor=FireBrick,linktocpage]{hyperref}
\usepackage{cite}
\usepackage{graphicx}
\usepackage{tikz}
\usepackage{tikz-cd}
\usepackage{spectralsequences}

\usepackage{courier}
\usepackage{bm}
\usepackage{dashbox}
\usepackage{subcaption}

\usepackage{mdframed}

\usepackage{blkarray}
\usepackage{arydshln}
\usepackage{dsfont}
\usetikzlibrary{patterns}

\newcommand{\ex}[1]{e^{#1}}
\newcommand{\fr}{\frac}
\newcommand{\pa}[1]{\left(#1 \right)}

\newcommand{\bb}[1]{\mathbb{#1}}

\newcommand{\kett}[1]{ \ket{#1}\!\rangle }

\def\tr{{\mathrm{tr}}}

\renewcommand{\title}[1]{\vbox{\center\LARGE{#1}}\vspace{5mm}}
\renewcommand{\author}[1]{\vbox{\center#1}\vspace{5mm}}
\newcommand{\address}[1]{\vbox{\center\em#1}}

\makeatletter
\newsavebox{\@brx}
\newcommand{\llangle}[1][]{\savebox{\@brx}{\(\m@th{#1\langle}\)}%
  \mathopen{\copy\@brx\kern-0.5\wd\@brx\usebox{\@brx}}}
\newcommand{\rrangle}[1][]{\savebox{\@brx}{\(\m@th{#1\rangle}\)}%
  \mathclose{\copy\@brx\kern-0.5\wd\@brx\usebox{\@brx}}}
\makeatother

\begin{document}

\begin{titlepage}
   \hfill     KYUSHU-HET-327, RIKEN-iTHEMS-Report-25\\
  
\title{Higher Structures on Boundary Conformal Manifolds:\\
Higher Berry Phase and Boundary Conformal Field Theory}

\author{Yichul Choi$^1$,
Hyunsoo Ha$^2$,
Dongyeob Kim$^2$,
Yuya Kusuki$^{3,4,5}$,
Shuhei Ohyama$^6$,
Shinsei Ryu$^2$}

\address{${ }^1${\small School of Natural Sciences, Institute for Advanced Study, Princeton, NJ 08540, USA}}
\address{${ }^2${\small Department of Physics, Princeton University, Princeton, 
NJ 08544, USA}}
\address{${ }^3${\small Institute for Advanced Study, 
Kyushu University, Fukuoka 819-0395, Japan}}
\address{${ }^4${\small Department of Physics, 
Kyushu University, Fukuoka 819-0395, Japan}}
\address{${ }^5${\small RIKEN Interdisciplinary Theoretical and Mathematical Sciences (iTHEMS), Wako, Saitama 351-0198, Japan}}
\address{${ }^6${\small University of Vienna, Faculty of Physics, Boltzmanngasse 5, A-1090 Vienna, Austria}}

\abstract

We introduce the notion of higher Berry connection and curvature in the space of conformal boundary conditions in (1+1)d conformal field theories (CFT), 
related to each other 
by exactly marginal boundary deformations,
 forming a ``boundary conformal manifold.''
Our definition builds upon previous works on tensor networks, such as matrix product states (MPS), where the triple inner product or multi-wavefunction overlap plays the key geometric role.
On the one hand, our boundary conformal field theory (BCFT) formulation of higher Berry phase provides a new analytic tool to study families of invertible phases in condensed matter systems.
On the other hand, it uncovers a new geometric structure on the moduli space of conformal boundary conditions, beyond the usual Riemannian structure defined through the Zamolodchikov metric.
When the boundary conformal manifold has an interpretation as the position moduli space of a D-brane, our higher Berry connection coincides with the NS-NS $B$-field in string theory.
The general definition does not require such an interpretation and is formulated purely field-theoretically, in terms of correlation functions of boundary-condition-changing (bcc) operators.
We also explore a connection between higher Berry connections and functional Berry connections in the loop spaces of boundary conformal manifolds.

\end{titlepage}

\eject

\tableofcontents

\section{Introduction}

Spaces of exactly marginal deformations
in conformal field theory (CFT) hold rich geometric structures and physical significance.
They are often referred to as ``conformal manifolds''.\footnote{
Although this terminology is standard, for the uninitiated readers we note that ``conformal manifolds'' in this context are not to be confused with ``manifolds with conformal structure'' in conformal geometry.}
The 2-point correlation functions of exactly marginal operators define the so-called Zamolodchikov metric on a conformal manifold \cite{Zamolodchikov:1986gt,Seiberg:1988pf}, and the corresponding connection as well as the Riemann tensor can be expressed in terms of integrated correlation functions of exactly marginal operators \cite{Kutasov:1988xb,Ranganathan:1992nb,Ranganathan:1993vj,Friedan:2012hi}.
This gives conformal manifolds the structure of a Riemannian manifold locally around a generic point in the space of CFTs which are related by exactly marginal deformations.\footnote{Conformal manifolds are also equipped with the structure of a vector bundle for more general operators \cite{deBoer:2008ss,Balthazar:2022hzb}.}
More globally, conformal manifolds may have different types of singularities (see, for instance, \cite{Seiberg:1988pf,Ginsparg:1987eb,Candelas:1990rm}).
The existence of a conformal manifold puts stringent constraints on the CFT data \cite{Behan:2017mwi,Hollands:2017chb,Bashmakov:2017rko,Green:2010da,Gomis:2015yaa,Gomis:2016sab,Seiberg:2018ntt}, and the geometry of conformal manifolds plays important roles in string theory and holography \cite{Ooguri:2006in,Perlmutter:2020buo,Ooguri:2024ofs}.

Similarly, one may also consider families of conformal boundary conditions 
which are related by exactly marginal boundary deformations.
Such boundary
marginal deformations were extensively studied in \cite{Recknagel:1998ih} (see \cite{Callan:1994ub} for an earlier work), and more recently, in \cite{Gaberdiel:2008fn,Karch:2018uft,Herzog:2023dop,Bartlett-Tisdall:2023ghh}.
As in the bulk case, the space of boundary exactly marginal deformations, called a ``boundary conformal manifold'' or the moduli space of boundary conditions, is locally equipped with the structure of a Riemannian manifold given by the Zamolodchikov metric \cite{PhysRevLett.129.201603,Herzog:2023dop}.
Unlike the bulk conformal manifolds which are rare
in non-supersymmetric theories, boundary (and more general defect) conformal manifolds are common,
as they can arise from continuous global symmetries which are explicitly broken by the boundary (or defect).\footnote{When the broken symmetry is an ordinary Lie group, Ref. \cite{Herzog:2023dop} showed that the boundary conformal manifold is a homogeneous coset space.
More generally, when a continuous non-invertible global symmetry is explicitly broken by the boundary, the boundary conformal manifold can be an orbifold \cite{Antinucci:2025uvj} (c.f. \cite{Damia:2023gtc}).
One interesting such example appears in \cite{Gaberdiel:2001zq}, where the space of conformal boundary conditions for a $c=1$ compact free boson at a fractional radius was shown to be an orbifold of a three-sphere $S^3$. See Section \ref{sec:non-chiral}.}

In a related but distinct context, recent studies in topological phases of matter have focused
on the geometric properties of the moduli space of short-range entangled states. Intriguingly, in extended quantum systems, such as lattice quantum many-body systems and quantum field
theories, the structure of the Berry phases (a.k.a geometric phase) becomes significantly richer.
Specifically, in (1+1)-dimensional gapped short-range entangled systems, the conventional Berry
connection (locally a 1-form in parameter space) generalizes to a 2-form, leading to the notion of a higher Berry phase – See, e.g., \cite{
Kitaev_Freed60,
  KS20-1,KS20-2,
Hsin_2020,
Cordova_2020a, Cordova_2020b,
Kapustin:2022apy,
Choi:2022odr,
Wen_2023,
OTS23,
beaudry2023homotopical,
qi2023charting,
spiegel2023calgebraicapproachparametrizedquantum,
shiozaki2023higher,
sommer2024higher_a,
sommer2024higher_b,
2024CMaPh.405..191A,
Geiko:2024cra,
Manjunath_2025}.
Ref.\ \cite{ohyama2023higher} introduced a formulation of the higher
Berry phase in (1+1)d using matrix product states (MPS), 
based on a notion of triple inner product,
or multi-wavefunction overlap.
This is 
a generalization of the usual inner product
to more than two quantum states,
and 
reveals the
higher-geometric structure (mathematically, so-called a gerbe structure) over the parameter space.
This structure can, for example, distinguish and classify parameterized families of many-body quantum systems, describe many-body analogs of the Thouless pump (“higher Thouless pump”),
characterize 
diabolic points in parameter space, and define topological invariants associated with
symmetry-protected topological (SPT) phases.

Although CFT is a primary tool for analyzing gapless critical systems, conformal boundary states \cite{Cardy:2004hm} can be thought of as gapped short-range entangled states in (1+1)d 
\cite{qi2012general, miyaji2014boundarystatesholographicduals, Cho_2017, Cardy_2017}.
Therefore, it is natural to expect that the higher Berry phase can be defined over boundary conformal manifolds, 
although previous studies did not address 
such higher geometric structures.
One of the main obstructions is the lack of a definition of the triple inner product in quantum field theory.

In this work, we formulate and analyze the higher Berry connection, curvature, and associated topological invariants for such boundary conformal manifolds in (1+1)d CFTs.
Specifically, we define the triple inner product of conformal boundary states from correlation functions of boundary-condition-changing (bcc) operators.
This is to be contrasted with the previous works where the geometric structure of a conformal manifold is studied using correlation functions of exactly marginal operators.
One of the concrete motivations for our definition comes from the similarity between the MPS formulation of higher Berry phase and the cubic interaction term in Witten's open string field theory \cite{Witten:1985cc}, which was observed in \cite{ohyama2023higher}.

In the context of condensed matter physics, our result provides new insight into the geometric and topological structure of higher Berry phases in quantum many-body systems, with potential applications especially in analytical settings.
In string theory, our result has the following interpretation.
When the (1+1)d CFT is a part of the worldsheet CFT, its boundary conformal manifold defines the moduli space of D-branes \cite{Schomerus:2002dc}.
A subset of the moduli space comes from the position moduli of the D-brane, which correspond to directions in spacetime that are perpendicular to the D-brane.
In examples we consider, when the boundary conformal manifold can be regarded as a position moduli space of a D-brane, we find that our higher Berry connection coincides with the components of the NS-NS $B$-field along the normal directions to the D-brane. 

The rest of the paper is organized as follows:
In Section \ref{sec: Basic idea}, we explain the main ideas and provide the definition of higher Berry connection as well as curvature on boundary conformal manifolds.
We review the definition of the triple inner product previously introduced in the context of tensor network, and the MPS formulation of the higher Berry phase in Section \ref{sec:triple_MPS}.
In Section \ref{sec:boundary_conformal}, we review basic aspects of boundary conformal field theory (BCFT), boundary conformal manifolds, and bcc operators, while drawing analogies between various tensor network and BCFT concepts.
In Section \ref{sec:berry_BCFT}, we define the higher Berry connection and curvature on boundary conformal manifolds.
Section \ref{sec:examples} provides several concrete examples.
In Section \ref{sec:bosons}, we consider multiple copies of free compact boson fields with $B$-field couplings, i.e., the Narain CFTs.
In Section \ref{sec:WZW}, we study $\hat{\mathfrak{g}}_k$ Wess-Zumino-Witten (WZW) models.
Section \ref{sec:fermions} discusses the $\mathrm{SO}(N)$ family of conformal boundary conditions for $N$-copies of non-chiral Majorana fermions.
In Section \ref{sec:non-chiral}, 
we discuss the so-called non-chiral deformation,
set apart from the other examples
since  
the boundary conformal manifold takes the form of an orbifold.
In Section \ref{sec:SPT}, we explain the parallel between (1+1)d SPT phases and symmetric conformal boundary conditions.
In Section \ref{Loop space connections on boundary conformal manifolds},
we connect our formulation of the higher Berry connection to the connection on loop space.
In particular, we use the so-called transgression/regression relation to relate them.
In Section \ref{sec:summary}, we summarize our results and conclude with a few future problems.
Finally, Appendix \ref{app:thouless} discusses an ordinary Thouless charge pump in $c=1$ free compact boson BCFT,
and Appendix \ref{app:flow} discusses a general method to compute the ground state energy flow under chiral deformations using the bootstrap approach.

\section{Basic ideas}
\label{sec: Basic idea}

In this section, we first recall the necessary ingredients of the MPS formulation of the higher Berry phase following Ref.\ \cite{ohyama2023higher}.\footnote{The short description given here
skips some details of the MPS formalism 
such as injectivity, the thermodynamic limit in MPS, 
MPS gauge, and canonical forms.
More details can be found in \cite{ohyama2023higher}.}
We then review basic aspects of BCFT and boundary conformal manifolds, making analogies between BCFT and tensor network concepts when possible.
Based on the analogy and the MPS formulation of the higher Berry phase, we define the higher Berry connection and curvature on boundary conformal manifolds in terms of correlation functions of bcc operators.
We also comment on the modulated boundary condition, which is a key technical element in our calculations in Section \ref{sec:examples}, and is also essential in the discussion of ``functional Berry phase'' in the loop space in Section \ref{Loop space connections on boundary conformal manifolds}.

\subsection{Triple inner product in MPS} 
\label{sec:triple_MPS}

\begin{figure}[t]
           \centering
           \begin{align*}
              \Psi_{\alpha}&=\, \adjincludegraphics[scale=0.8,trim={10pt 10pt 10pt 10pt},valign = c]{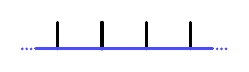}\\
              \Psi_{\alpha} * \Psi_{\beta}&=\,\adjincludegraphics[scale=0.8,trim={10pt 10pt 10pt -5pt},valign = c]{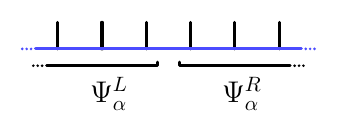}
              \;*\;
              \adjincludegraphics[scale=0.8,trim={10pt 10pt 10pt -5pt},valign = c]{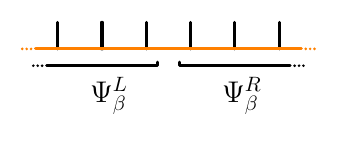}\\
              &=\,
              \adjincludegraphics[scale=0.8,trim={10pt 10pt 10pt -5pt},valign = c]{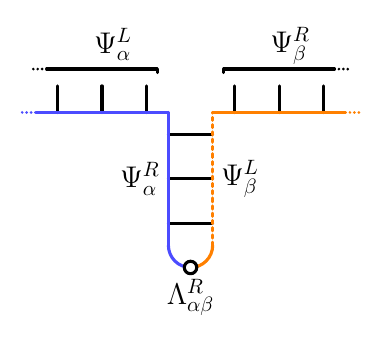}
              =\quad 
              \adjincludegraphics[scale=0.8,trim={10pt 10pt 10pt -5pt},valign = c]{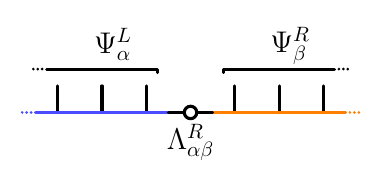}\\
             T_{\alpha\beta}&=\adjincludegraphics[scale=1.0,trim={10pt 10pt 10pt 10pt},valign = c]{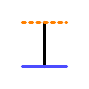},
             \quad
              \adjincludegraphics[scale=1.0,trim={10pt 10pt 10pt 10pt},valign = c]{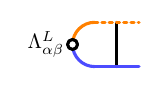}
              =\adjincludegraphics[scale=1.0,trim={10pt 10pt 10pt 10pt},valign = c]{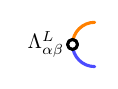},
              \quad
              \adjincludegraphics[scale=1.0,trim={10pt 10pt 10pt 10pt},valign = c]{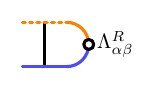}
              =\adjincludegraphics[scale=1.0,trim={10pt 10pt 10pt 10pt},valign = c]{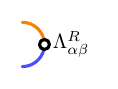} 
           \end{align*}
       \caption{Matrix product states,
    star product ($*$),
    the (mixed) transfer matrix, and
    the left- and right-fixed points of the transfer matrix.
    Along the dotted lines, the relevant MPS tensors are conjugated.
    }
    \label{fig: str}
\end{figure}

\begin{figure}[t]
           \centering
           \begin{align*}
              \int \Psi_{\alpha} 
               &= \int \adjincludegraphics[scale=0.8,trim={10pt 10pt 10pt 10pt},valign = c]{MPS1.pdf}
               = \adjincludegraphics[scale=0.8,trim={10pt 10pt 10pt 10pt},valign = c]{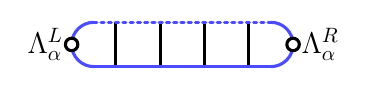}
              \\
              \int\Psi_{\alpha}*\Psi_{\beta}*\Psi_{\gamma}
              &=\int\adjincludegraphics[scale=0.7,trim={10pt 10pt 10pt 10pt},valign = t]{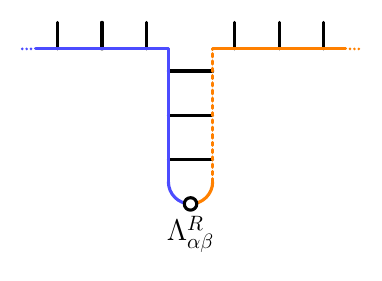} 
              \;*\; 
              \adjincludegraphics[scale=0.7,trim={10pt 10pt 10pt 10pt},valign = c]{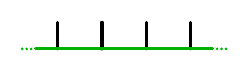}\\
              &=\int\adjincludegraphics[scale=0.7,trim={10pt 10pt 10pt 10pt},valign = t]{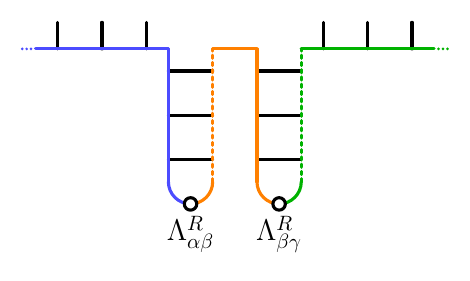}
              =
              \adjincludegraphics[scale=0.7,trim={10pt 10pt 10pt 10pt},valign = t]{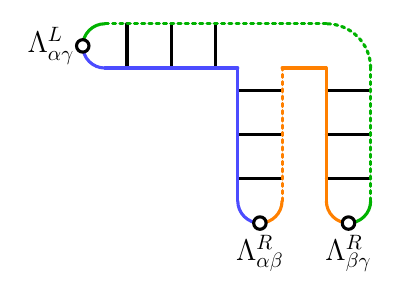}
           \end{align*}
       \caption{The integration ($\int$), the star product of three MPS
  $\Psi_{\alpha}$, $\Psi_{\beta}$, $\Psi_{\gamma}$,
    and the triple inner product.
       }
       \label{fig: trns}
\end{figure}

Similar to the regular Berry phase (and the Berry connection and curvature) 
that can be formulated in terms of the inner product of two quantum mechanical states, 
Ref.\ \cite{ohyama2023higher} introduced the triple inner product (or multi-wavefunction overlap) as a key ingredient in defining the higher Berry phase.
To define the triple inner product, 
below we first introduce the star product $*$, and integration $\int$, for infinite MPS.
These concepts are summarized in Figures \ref{fig: str} and \ref{fig: trns}.

To set the stage, we consider a family of (1+1)d invertible, translationally invariant 
states that admit MPS representations,
parameterized over the parameter space $M$.
Importantly, the MPS representation of a given quantum many-body state is not unique:
in addition to the overall phase ambiguity inherent to quantum wavefunctions, there is an intrinsic gauge ambiguity of MPS representations (MPS gauge ambiguity for short)\cite{Perez-Garcia:2006nqo,P_rez_Garc_a_2008}.
While this MPS gauge choice is arbitrary (unless constrained by, e.g., a global symmetry),  
for a parameterized family of MPS, there may be an obstruction to using a single gauge over the parameter space
that preserves a constant bond dimension and smooth tensor components.
This situation is analogous to the case of a parameterized family of quantum mechanical states
with non-zero Chern number,
where no smooth, global gauge choice exists.
We therefore anticipate the necessity to use multiple patches to describe a parameterized family of MPS,
$M=\cup_{\alpha} U_{\alpha}$.

The patches generically intersect, and we focus on double and triple intersections,
$U_{\alpha}\cap U_{\beta} \neq \emptyset$,
$U_{\alpha}\cap U_{\beta} \cap U_{\gamma}\neq \emptyset$.
At a point in such an intersection, 
there are two or three MPS representations representing the same physical state.
We denote an MPS constructed on patch $U_{\alpha}$ by $\Psi_{\alpha}$.
More concretely, $\Psi_{\alpha}$ is represented by a set of MPS tensors
$\{A^i_{\alpha}\}$.
For two MPS $\Psi_{{\alpha}}$ and $\Psi_{{\beta}}$ from different patches
$U_{\alpha}$ and $U_{\beta}$, 
the star product  $\Psi_{{\alpha}} *  \Psi_{{\beta}}$ 
is defined by first splitting $\Psi_{{\alpha}}$ and 
$\Psi_{{\beta}}$ into their left and right pieces, denoted by
$\Psi^L_{\alpha}, \Psi^R_{\alpha}$ and $\Psi^L_{\beta}$, $\Psi^R_{\beta}$, respectively.
In the product $\Psi_{{\alpha}} * \Psi_{{\beta}}$, 
$\Psi^R_{\alpha}$ and $\Psi^L_{\beta}$ are ``glued,'' i.e., contracted.
In this process, the MPS matrices 
$\{A^i_{\beta}\}$ on the left part of $\Psi_{\beta}$ are first converted 
to their conjugates $\{A^{i*}_{\beta}\}$ (``bras'')
and then contracted with the right part of $\Psi_{\alpha}$
(Fig.\ \ref{fig: str}).
Intuitively, we regard physical indices in $\Psi^L_{\alpha}$ 
and $\Psi^R_{\alpha}$ as row (input) and column (output) 
indices of an infinite matrix,
or a semi-infinite matrix product operator.
Accordingly, the star product $*$ can be interpreted as 
matrix multiplication
of two infinite-dimensional matrices.
The star product is associative,
$(\Psi_{{\alpha}}* \Psi_{{\beta}}) * 
\Psi_{{\gamma}}
=
\Psi_{{\alpha}}* (\Psi_{{\beta}} *  \Psi_{{\gamma}})$,
but not commutative.

Using the transfer matrix of MPS, the glued part can be reduced to a single tensor,
$\Lambda^R_{\alpha\beta}$,
which is the (unique) right fixed point of the (mixed) transfer matrix.
Here, we denote a transfer matrix 
made of two MPS $\Psi_{\alpha}$
and $\Psi_{\beta}$ as
$T_{\alpha\beta}$,
and the left- and right-fixed points as 
$\Lambda^{L}_{\beta\alpha}$
and $\Lambda^R_{\alpha\beta}$
(Fig.\ \ref{fig: str}).
The resulting state 
$\Psi_{\alpha\beta} :=\Psi_{{\alpha}}* \Psi_{{\beta}}$ 
can be thought of as an MPS in the so-called mixed gauge 
(also known as mixed canonical form),
where different MPS canonical gauges are used for the left and right halves.
The two gauges are ``connected'' or ``interfaced'' by $\Lambda^R_{\alpha\beta}$.

Another important ingredient is integration $\int$.
To define the integration of $\Psi_{\alpha}$, $\int \Psi_{\alpha}$, 
we first note that only the connectivity of the tensor network matters, allowing us to freely deform it geometrically.
We thus fold 
the MPS $\Psi_{{\alpha}}$ at the center and contract
the left and right parts, denoted by
$\Psi^L_{\alpha}$ and $\Psi^R_{\alpha}$, respectively
(Fig.\ \ref{fig: trns}).
The integration $\int \Psi_{\alpha}$
can be represented in terms of left- and right-fixed points of the transfer matrix as 
$\int \Psi_{\alpha} =\mathrm{tr}\, (\Lambda^L_{\alpha} \Lambda^R_{\alpha})$,
where $\Lambda^{L/R}_{\alpha}\equiv \Lambda^{L/R}_{\alpha\alpha}$
and the trace is taken over the bond Hilbert space of MPS.
Similarly, the inner product of two MPS
is given by
$\int \Psi_{\alpha}*\Psi_{\beta}
=
\mathrm{tr}\,(
\Lambda^L_{\beta\alpha}\Lambda^R_{\alpha\beta})
$.

By combining these ingredients, 
we can now introduce the integral of the triple product
$\Psi_{{\alpha}}* \Psi_{{\beta}} * \Psi_{{\gamma}}$,
which we call the triple inner product,  
\begin{align}
    \int
    \Psi_{\alpha} * \Psi_{\beta} * \Psi_{\gamma}
    =
    \mathrm{tr}\,(\Lambda^L_{\alpha\gamma} \Lambda^R_{\alpha\beta}\Lambda^R_{\beta\gamma}) 
    \equiv
    c_{\alpha\beta\gamma}\, .
\label{eq: triple inner product}
\end{align}
One can check the cyclicity
$
\int \Psi_{\alpha} * \Psi_{\beta} * \Psi_{\gamma}
=
\int \Psi_{\beta} * \Psi_{\gamma} * \Psi_{\alpha}
$.
From a geometrical point of view, this is nothing but a gerbe structure (e.g.\ \cite{Brylinski}).
The collection of $U(1)$ phases $c_{\alpha\beta\gamma}$ on triple intersections
defines an element in $H^3(M; \mathbb{Z})$,
which is a topological invariant of a gerbe -- the Dixmier-Douady class.\footnote{See \cite{ohyama2023higher,beaudry2023homotopical,beaudry2025classifyingspacephasesmatrix} for an MPS-based understanding of $K(\mathbb{Z};3)$.}
This is analogous to the Chern class associated to $H^2(M;\mathbb{Z})$
which is a topological invariant of a parameterized family of quantum mechanical systems.
For the later comparison with BCFT,
it is also useful to note that the product of
$\Psi_{{\alpha\beta}}$ and $\Psi_{{\beta\gamma}}$ is given by
\begin{align}
\Psi_{{\alpha\beta}}* \Psi_{{\beta\gamma}}=
\Psi_{{\alpha}}
* 
\Psi_{{\beta}} * 
\Psi_{{\beta}}
*
\Psi_{{\gamma}}
=
c_{\alpha\beta\gamma}\Psi_{{\alpha\gamma}} \, .
\end{align}
Hence, the triple inner product of three MPS $\Psi_{\alpha},\Psi_{\beta}, \Psi_{\gamma}$
can also be viewed as the regular inner product  of two non-uniform states, 
$\Psi_{\alpha\beta}$ and $\Psi_{\beta\gamma}$.

In the above setup, we considered the triple inner product of
physically equivalent states that are in different MPS gauges.
This is similar to Wu-Yang's description of a magnetic monopole
\cite{1975PhRvD..12.3845W}.
On the other hand, the triple inner product can also 
be applied in a different manner \cite{shiozaki2023higher}:
Here, we consider 
three physically distinct states that are close in the parameter space,
$\Psi(x_0), \Psi(x_1), \Psi(x_2)$ ($x_{0}, x_1, x_2\in M$).
Since each pair of states is sufficiently close, the mixed transfer matrix between them is expected to have a unique eigenvector with an eigenvalue close to $1$.
Therefore, naively, one may expect that the higher holonomy along the small triangle spanned by $x_0,x_1$ and $x_2$ can be computed by evaluating the trace of the product of the left/right fixed points $\Lambda_{x_0 x_1}^{L/R},\Lambda_{x_1 x_2}^{L/R},\Lambda_{x_2 x_0}^{L/R}$ of the transfer matrices between each pair of states:
\begin{align}
\mathrm{Arg}\operatorname{tr}\left[\Lambda_{x_0 x_1}^{L} \Lambda_{x_1 x_2}^{R} \Lambda_{x_2 x_0}^{R}\right]\, .
\end{align}
Due to an ambiguity in the discretization procedure, the above method is, in its current form, not valid. However, it has been confirmed that a slightly modified expression
\begin{align} \label{eq: triple inner product modified}
\mathrm{Arg}\operatorname{tr}\left[
\left(\Lambda_{x_0}^{L}\right)^{1/3} 
\Lambda_{x_0 x_1}^R 
\left(\Lambda_{x_1}^{L}\right)^{1/3} 
\Lambda^R_{x_1 x_2} 
\left(\Lambda_{x_2}^{L}\right)^{1/3} 
\Lambda^R_{x_2 x_0}\right]
\end{align}
allows us to compute the higher Berry phase numerically
\cite{shiozaki2023higher}.\footnote{Indeed, one can check that \eqref{eq: triple inner product modified} is equivalent to the argument of \eqref{eq: triple inner product} when $x_0=x_1=x_2$.}
It is worth mentioning that this computation is a direct generalization of the Fukui–Hatsugai–Suzuki's method\cite{Fukui_2005}
for the regular Berry phase (connection) on 
discretized parameter space, and that the quantization of the integrated value of the higher Berry curvature computed from the above quantity is manifestly guaranteed.

We also recall that
$-\log\, (\Lambda_x^{L})$, (the minus of) the logarithm of the left fixed point of the (single) transfer matrix at $x$, is nothing but the entanglement Hamiltonian when the MPS is bipartitioned into two parts.
In the current context, it plays the role of a cutoff or regularization factor
in \eqref{eq: triple inner product modified}.
While \eqref{eq: triple inner product} does not depend on 
the entanglement spectrum due to the normalization $\mathrm{tr}\, (\Lambda^L_{\alpha})=1$, \eqref{eq: triple inner product modified}
does.
In general, the spectrum of $\Lambda^L_x$
can be rather complicated and sensitive to 
microscopic details, even though the topological invariant extracted from the triple inner product is expected to be
robust. 
However, 
for integrable lattice quantum models {\it away from criticality}, 
it is known that the entanglement spectrum for a half-space (i.e., the spectrum of Baxter's corner transfer matrix) exhibits 
a regular structure and, in fact, coincides with the spectrum of the corresponding BCFT
\cite{Cardy:1989vyr,
Cardy:1989da,
PhysRevB.35.2105,
Saleur:1988zx,
Calabrese_2010}.
In the next section, when we formulate the higher Berry phase for boundary conformal manifolds, we will see that the counterpart of $\Lambda^L_x$ is given by the Hamiltonian of BCFT defined on an interval.

\subsection{Boundary conformal manifold and bcc operators} \label{sec:boundary_conformal}

The intimate connections between boundary states and gapped states \cite{qi2012general, miyaji2014boundarystatesholographicduals, Cho_2017, Cardy_2017} suggest that there should be a notion of ``higher Berry phase'' in the space of conformal boundary conditions.
A concrete realization of this idea is hinted at by the formal resemblance of the triple inner product of MPS reviewed above, with the cubic interaction vertex in Witten's open string field theory \cite{Witten:1985cc}.
This observation was made in \cite{ohyama2023higher}, and in fact, the star product and integration were originally defined in \cite{Witten:1985cc} for open string fields in a curiously similar way the corresponding operations are defined for MPS.

The cubic interaction vertex in open string field theory is in the end determined by disk 3-point correlation functions of boundary operators of the worldsheet BCFT (see, for instance, \cite{Erler:2019vhl}).
This gives us one of the direct motivations for our definition of the higher Berry connection on boundary conformal manifolds.

To set the notations, we first review basic aspects of BCFT and boundary conformal manifolds.
We then describe the general definition of the higher Berry connection and curvature for the boundary conformal manifolds in the next subsection, which are determined by the 3-point correlation functions of (lightest) bcc operators connecting nearby boundary conditions on the boundary conformal manifold.

\paragraph{Conformal boundary conditions}

A conformal boundary state $\ket{B_\alpha}$ satisfies
\begin{equation} \label{eq:conformal_bc}
    \left( L_n - \bar{L}_{-n} \right) \ket{B_\alpha} = 0 \quad \text{for all $n \in \mathbb{Z}$} \,,
\end{equation}
where $L_n$ and $\bar{L}_n$ are left- and right-moving Virasoro generators, respectively.
The boundary condition is conformal, since it preserves the (single) Virasoro symmetry \cite{Cardy:2004hm}.  
Here, we do not assume that the boundary condition $B_\alpha$ preserves any additional extended chiral algebra.\footnote{In particular, in rational CFTs, the number of conformal boundary conditions preserving the extended chiral algebra is finite (with a fixed gluing condition, see \eqref{eq:twisted_gluing}) \cite{Fuchs_2002}, and such boundary conditions cannot form a nontrivial conformal manifold.}
With a slight abuse of terminology, we use boundary \emph{conditions} and boundary \emph{states} interchangeably.

Given two conformal boundary conditions $B_\alpha$ and $B_\beta$, we have the interval Hilbert space of the CFT with the left boundary condition $B_\alpha$ and the right boundary condition $B_\beta$, which we denote as $\mathcal{H}_{\alpha\beta} $.
The boundary states mutually satisfy the Cardy condition \cite{Cardy:1989ir,Cardy:2004hm},
\begin{equation}
    \langle B_\alpha | \tilde{q}^{\frac{1}{2} (L_0 + \bar{L}_0 - \frac{c}{12})} | B_\beta \rangle = \mathrm{Tr}_{\mathcal{H}_{\alpha \beta}} \left( q^{L_0 - \frac{c}{24}} \right),
\end{equation}
where the right-hand side has an expansion in terms of the Virasoro characters with non-negative integer coefficients.
Here, $c = c_L = c_R$ is the central charge of the CFT,\footnote{To admit a conformal boundary condition, the bulk CFT must have a vanishing chiral central charge, $c_L - c_R = 0$. This follows from the condition \eqref{eq:conformal_bc}.} and $q=e^{-\pi \delta}$, $\tilde{q} = e^{-4\pi/\delta}$ with $\delta = \beta/L$ being the ratio between the length $L$ of the interval and the circumference $\beta$ of the thermal circle.\footnote{
This is the regular interpretation of the compact direction in the context of boundary criticality in (1+1)d quantum many-body systems. 
For our application to gapped ground states, the compact direction plays the role of spatial direction.}
The boundary conditions furthermore satisfy the Cardy-Lewellen sewing constraints \cite{Lewellen:1991tb}, which however will not play much role below.

A conformal boundary condition $B_\alpha$ is called \emph{simple} if the interval Hilbert space $\mathcal{H}_{\alpha \alpha}$ has a unique ground state.\footnote{A simple boundary condition has been also called \emph{fundamental} or \emph{elementary} in the literature.}
Upon the state-operator correspondence, this is equivalent to saying that the only topological (i.e., zero scaling dimension) local operator on the boundary is the identity operator.
A simple boundary condition cannot be written as a direct sum of other boundary conditions (see \cite{Choi:2023xjw} and references therein).

\paragraph{Exactly marginal boundary deformations}
We consider deforming (or perturbing) a conformal boundary condition by adding a boundary term $\delta S$ to the action $S_0$,
\begin{equation}
S=S_0 + \delta S = S_0 + \sum_a \xi^a \int_{\Sigma} \frac{dx}{2\pi}\, \psi_a(x)\,,
\end{equation}
where $\Sigma$ is the boundary of the spacetime manifold, $\psi_a$ are boundary operators, and $\xi^a$ are the couplings.
Here, a boundary operator refers to an operator that lives only on the boundary,  
and it is obtained from a state in the interval Hilbert space with identical boundary conditions on the two ends through the state-operator correspondence.
It was shown in \cite{Recknagel:1998ih} that, if the boundary operators $\psi_a$ all have conformal dimensions $\Delta_a=1$ (i.e., they are marginal), and are mutually local among themselves (referred to as being ``self-local''), then the deformation generated by the boundary operators becomes an exactly marginal deformation, to all orders in conformal perturbation theory.
In other words, for boundary perturbations, the “self-locality” ensures that the beta functions of coupling constants $\xi^a$ vanish to all orders. 
The theory obtained by an exactly marginal perturbation remains conformally invariant, and exactly marginal deformations generate a ``boundary conformal manifold,'' where the coupling constants $\xi^a$ play the role of coordinates on the conformal manifold.

An exactly marginal deformation, by definition, preserves the conformal symmetry~(\ref{eq:conformal_bc}),  
but it generally breaks other symmetries.  
Consider a CFT with a chiral algebra $\mathcal{A} \otimes \mathcal{A}$ generated by currents $W^a$ and $\overline{W}^a$.
In order to preserve the symmetry, it is necessary to impose the following gluing condition on the currents at the boundary,
\begin{equation}\label{eq:bdyW}
W^a(z) - \overline{W}^a(\bar{z})   = 0\, .
\end{equation}
Let us see how an exactly marginal deformation affects the gluing condition.
An exactly marginal deformation modifies the correlation functions as follows,
\begin{equation}
\begin{aligned}
\braket{X} &\to  \braket{P \ex{\delta S}  X}\\
&=\sum_n \xi^n \int \cdots \int_{x_i < x_{i+1}} \fr{dx_1}{2\pi} \cdots \fr{dx_n}{2\pi} \braket{\psi(x_1) \cdots \psi(x_n) X},
\end{aligned}
\end{equation}
where, for simplicity, we assume that the conformal manifold is one-dimensional.  
$X$ represents a composite operator consisting of local operators,  
including both bulk operators and boundary operators.
We define a path-ordered exponential $Pe$.
Since boundary operators live on a one-dimensional space,  
they are generally non-commutative; therefore, it is necessary to introduce path-ordering.
The integrals associated with the deformation generally diverge,  
and one has to regularize the integrals.  
The regularization procedure is essentially the same as that for bulk deformations,  
namely, we remove small neighborhoods around the insertion points of local operators from the integration domain.  
We refer the readers to \cite{Recknagel:1998ih} for further details.  
To investigate how an exactly marginal deformation affects the gluing condition of chiral algebra currents,
we express (\ref{eq:bdyW}) as follows,
\begin{equation}
\lim_{\delta \to 0} \pa{W^a(z+2i\delta) - \overline{W}^a(\bar{z}-2i\delta)  } = 0\, ,
\end{equation}
where the boundary is located at $z = \bar{z}$.
In this setup, the exactly marginal deformation is implemented by inserting the path ordered exponential  
through the gap between $W^a$ and $\overline{W}^a$,
\begin{equation}
\lim_{\delta \to 0} P \ex{\delta S} \pa{W^a(z+2i\delta) - \overline{W}^a(\bar{z}-2i\delta)  } = 0\, .
\end{equation}
One simple example of exactly marginal deformations can be obtained by a current deformation (also called chiral deformation) 
$\int_{\Sigma} \fr{dx}{2\pi} \xi \cdot J(x)$,
which leads to the following simple gluing condition,
\begin{equation}\label{eq:BC}
\pa{\ex{2\pi i \xi\cdot J_0} W_n^a \ex{-2 \pi i \xi\cdot J_0} - (-1)^{s_a} \overline{W}^a_{-n} }\ket{B_\xi}=0\, ,
\end{equation}
where $s_a$ is a spin of the current $W^a$, $W^a_n$ are the modes of the current $W^a(z)$, and $U_\xi \equiv \ex{2 \pi i \xi\cdot J_0}$.\footnote{
When defining chiral deformations (or current deformations),  
we specifically use current operators with spin $s_a = 1$ to deform the theory,  
however in general, currents are not restricted to $s_a = 1$.  
}
It implies that the chiral deformation amounts to a rotation of the gluing condition on the current by an inner automorphism generated by conjugation with $U_\xi$.
Consequently, if we denote the reference boundary state by $\ket{B_0}$,  
the deformed boundary state is simply given by
\begin{equation} \label{eq:chiral_boundary_state}
\ket{B_\xi} = U_\xi \ket{B_0}.
\end{equation}
This fact shows that the chiral deformation does not change the spectrum of boundary operators.
In this paper, we will mostly focus on this chiral deformation in CFTs with extra currents.

There exist exactly marginal boundary operators that do not correspond to currents,  
and the exactly marginal deformations induced by such operators  
are called non-chiral deformations.
Unlike chiral deformations, non-chiral deformations significantly change the spectrum of the boundary theory.
Non-chiral deformations, for example, often enlarge the boundary moduli space of the $c=1$ free boson CFT to an orbifold of $\mathrm{SU}(2)$~\cite{Callan:1994ub} (see also~\cite{Recknagel:1998ih, Gaberdiel:2001zq, Janik:2001hb}). 
We will explain such deformations in detail later in Section \ref{sec:non-chiral} through a concrete example.

\paragraph{Boundary conformal manifold}

We consider a family of conformal boundary conditions, denoted by $M$, of a fixed (1+1)d CFT,\footnote{For the most part of the paper, we focus on unitary, compact, bosonic CFTs with a unique ground state in the circle Hilbert space, and compact boundary conditions (see \cite{Janik:2001hb} for examples of non-compact boundary conditions in compact CFTs). In Section \ref{sec:fermions}, we discuss free fermion CFTs.
}
which are related to each other by boundary exactly marginal deformations as above.
We call $M$ a boundary conformal manifold or the moduli space of boundary conditions.
Locally, near a generic point, $M$ has the structure of a Riemannian manifold determined by the Zamolodchikov metric \cite{Zamolodchikov:1986gt,Seiberg:1988pf,Kutasov:1988xb,PhysRevLett.129.201603,Herzog:2023dop}.
Globally, $M$ is not necessarily a smooth manifold, but may have singularities. (We discuss one such example in Section \ref{sec:non-chiral}.)
Given a point $\alpha \in M$, we denote the corresponding conformally invariant boundary state as
\begin{equation} \label{eq:boundary_states}
    \ket{B_\alpha} \,, \quad \alpha \in M \,.
\end{equation}
The $g$-function (or the boundary entropy) \cite{Affleck:1991tk} of the boundary condition $B_\alpha$ is constant over the boundary conformal manifold,
\begin{equation} \label{eq:g-function}
    g_M \equiv \langle 0 | B_\alpha \rangle ~~\text{is constant over $M$} \,.
\end{equation}
This is a consequence of the $g$-theorem \cite{Friedan:2003yc,Casini:2016fgb,Cuomo:2021rkm}.
In most cases of interest, boundary condition $B_\alpha$ is simple at generic points $\alpha \in M$ on the boundary conformal manifold.
However, it may become non-simple at special points, especially at singular points \cite{Gaberdiel:2001xm}.

In the analogy with tensor networks, the family of boundary states
$|B_{\alpha}\rangle$ corresponds to 
the family of MPS $\Psi_\alpha$.
Here, the simplicity of the boundary condition corresponds to the injectivity of MPS.
While in the MPS context, 
the subscript $\alpha$ is associated 
to the MPS gauge redundancy, 
in the following, we use the label $\alpha$
just to distinguish different boundary states (i.e., different points on the boundary conformal manifold).
As explained in the previous subsection,
the triple inner product in the MPS context can be applied to 
two different situations; 
(i) three physically identical states in different MPS gauges, and (ii) three physically distinct states that are close in the parameter space.
Our formulation of higher Berry phase in BCFT, to be spelled out momentarily, is close in spirit to the case (ii). 
We note that in the BCFT context, without additional symmetries, it is unclear how physically identical boundary states can be in different gauges.
We discuss the BCFT analogue of the case (i) in Section \ref{sec:SPT} in the presence of a global symmetry.

\paragraph{Boundary-condition-changing operators}

In BCFT, there are boundary-condition-changing (bcc) operators that play important roles  
\cite{Cardy:1986gw, Cardy:1989ir}.\footnote{See also \cite{Brehm_2022,brehm2024latticemodelscftsurfaces,cheng2025precisionreconstructionrationalcft} for recent interesting applications of bcc operators to tensor networks and lattice systems.}
We have the state-operator correspondence between the states in $\mathcal{H}_{\alpha \beta}$ and 
bcc operators between $B_\alpha$ and $B_\beta$,
\begin{equation}
    \mathcal{H}_{\alpha \beta} \ni \ket{\mathcal{O}_{\alpha \beta}} \leftrightarrow ~~\text{bcc operator $\mathcal{O}_{\alpha \beta}(x)$ between $B_\alpha$ and $B_\beta$}\,.  
\end{equation}
On the right-hand side, the CFT is defined on the upper half-plane, the boundary is along the real line, and $x \in \mathbb{R}$ is the location of the bcc operator $\mathcal{O}_{\alpha\beta}(x)$.
Along the real line, the boundary condition $B_\alpha$ is imposed to the left of $x$, and $B_\beta$ is imposed to the right of $x$.

When a pair of simple boundary conditions $B_\alpha$ and $B_\beta$ are sufficiently close to each other in the moduli space $M$, we expect there to be a unique ground state in the interval Hilbert space $\mathcal{H}_{\alpha \beta}$, obtained from the ``spectral flow'' of the identity operator.
That is, we may start from the Hilbert space $\mathcal{H}_{\alpha \alpha}$, which has a unique ground state if $B_\alpha$ is simple, and then adiabatically deform the boundary condition on the right end of the interval to obtain the Hilbert space $\mathcal{H}_{\alpha \beta}$.
Assuming that $B_\alpha$ and $B_\beta$ are sufficiently close to each other in $M$ so that there is no level crossing during the adiabatic deformation, we find that $\mathcal{H}_{\alpha\beta}$ has a unique ground state.
We denote this ground state and the corresponding bcc operator as
\begin{equation} \label{eq:lightest_BCC}
     \mathcal{H}_{\alpha \beta} \ni \ket{\psi_{\alpha  \beta}} \leftrightarrow \psi_{\alpha \beta} (x) \,.
\end{equation}
The scaling dimension of the operator $\psi_{\alpha \beta} (x)$ is denoted as $\Delta_{\alpha \beta}$, and we have
$\Delta_{\alpha  \beta} = \Delta_{\beta  \alpha}$.
In particular, the operator $\psi_{\alpha \alpha}$ is the identity operator on the boundary $B_\alpha$, with scaling dimension $\Delta_{\alpha \alpha} = 0$.
On the other hand, when $B_\alpha$ and $B_\beta$ are different, $\Delta_{\alpha \beta}$ must be strictly larger than zero. Otherwise, one could freely move the bcc operator, which would contradict $B_\alpha \neq B_\beta$.

\begin{figure}[t]
       \begin{minipage}{0.4\textwidth}
           \centering
           \begin{equation*}
              \adjincludegraphics[scale=1.0,trim={20pt 10pt 10pt 10pt},valign = c]{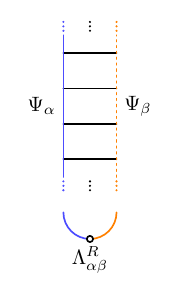}
              \longleftrightarrow\;
              \adjincludegraphics[scale=1.0,trim={15pt 10pt 10pt 10pt},valign = c]{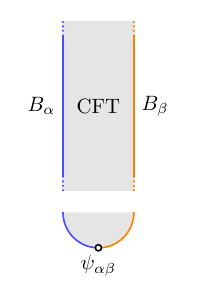}
           \end{equation*}
           \caption*{(a)}    
       \end{minipage}
       \hspace{30pt}
       \begin{minipage}{0.45\textwidth}
           \centering
           \begin{equation*}
              \adjincludegraphics[scale=1.1,trim={20pt 10pt 10pt 10pt},valign = c]{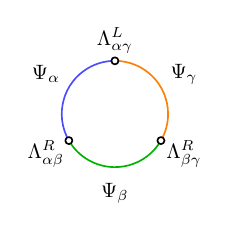}
              \longleftrightarrow\;\;
              \adjincludegraphics[scale=1.1,trim={20pt 10pt 10pt 10pt},valign = c]{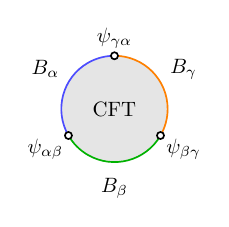}
           \end{equation*}
           \caption*{(b)}
       \end{minipage}
       \caption{
       \label{fig: concept}
       Conceptual correspondence between tensor networks and BCFT. (a) The left-hand side represents the mixed transfer matrix between $\Psi_\alpha$ and $\Psi_\beta$, and the right-hand side represents the time evolution by the CFT Hamiltonian on an interval with boundary conditions $\ket{B_{\alpha}}$ and $\ket{B_{\beta}}$ imposed. Corresponding to $\Lambda_{\alpha\beta}^{R}$ (or $\Lambda^L_{\alpha\beta}$) being realized as the fixed point of the mixed transfer matrix, the bcc operator $\psi_{\alpha\beta}$ is realized as the ground state of the Hilbert space on the interval. 
      (b) The triple inner product is given by the product of the three fixed point tensors. Therefore, in BCFT, it is expected that the higher Berry phase is computed from the 3-point correlation function of the bcc operators.}
       \label{fig: TN vs BCFT}
\end{figure}

Given the identification 
$\Psi_{\alpha}\leftrightarrow |B_{\alpha}\rangle$,
it is natural to identify 
the mixed gauge MPS $\Psi_{\alpha\beta}$
or its central tensor $\Lambda^{R}_{\alpha\beta}$
as the bcc operator $\psi_{\alpha\beta}$.
We also recall that the fixed point tensor 
$\Lambda^{R,L}_{\alpha\beta}$ is the fixed point 
of the transfer matrix $T_{\alpha\beta}$.
This also nicely echoes with the state interpretation of 
$\psi_{\alpha\beta}$ using the CFT state-operator correspondence; 
in this interpretation, the bcc operator $\psi_{\alpha\beta}$ is the ground state  
of the transfer matrix 
defined for an infinite strip region 
sandwiched by the boundary conditions $\alpha$ and $\beta$ (Fig.\ \ref{fig: TN vs BCFT}(a)).
The dependence of \eqref{eq: triple inner product modified} on the entanglement spectrum parallels the fact that the conformal family of the bcc operator $\psi_{\alpha\beta}$ represents the spectrum of the transfer matrix. 
Moreover, as noted at the end of the previous subsection, the entanglement spectrum for a half-space in certain integrable models coincides with the spectrum of the corresponding BCFT.
This lends further support to 
our BCFT formulation of the higher Berry phase.

The star product 
$\Psi_{\alpha\beta}* \Psi_{\beta\gamma} = c_{\alpha\beta\gamma}\Psi_{\alpha\gamma}$
represents the operator product expansion (OPE) of two bcc operators 
$\psi_{\alpha\beta}$ and $\psi_{\beta\gamma}$. 
With proper regularization (Euclidean evolution),
the triple inner product corresponds to the partition function on a disk with 
boundary conditions specified by 
$\alpha, \beta$ and $\gamma$, with insertions of three bcc operators.
See Figure \ref{fig: concept}(b).

From this observation, 
we claim that the higher Berry phase on a boundary conformal manifold can be extracted from the OPE coefficient of bcc operators.
Alternatively, we can calculate the higher Berry phase by considering 
the disk partition function with the insertion of three bcc operators.
We make this more precise in the next subsection.

\subsection{Higher Berry connection and curvature in BCFT}
\label{sec:berry_BCFT}

We now provide the definition of higher Berry connection and curvature on boundary conformal manifolds, based on the above considerations and the analogy with tensor networks.
We mostly work within a small local neighborhood inside the conformal manifold $M$.
Specifically, we take a local coordinate patch $U$ inside the boundary conformal manifold $M$, in such a way that every boundary condition $B_\alpha$ with $\alpha \in U$ is simple, and furthermore that the interval Hilbert spaces $\mathcal{H}_{\alpha \beta}$ have unique ground states with the corresponding lightest bcc operators $\psi_{\alpha \beta} (x)$ in \eqref{eq:lightest_BCC} well-defined for all $\alpha, \beta \in U$, without any level-crossing as we continuously vary $\alpha$ and $\beta$ in $U$.

\paragraph{Correlation functions of bcc operators}
We normalize the 2-point correlation functions of lightest bcc operators $\psi_{\alpha\beta}$ on the upper half-plane such that 
\begin{equation} \label{eq:BCC_normalization}
    \langle \psi_{\alpha \beta} (x_1) \psi_{\beta \alpha} (x_2) \rangle_{{\it UHP}} = \frac{g_M}{(x_2 - x_1)^{2\Delta_{\alpha \beta}}} \,,
\end{equation}
where $g_M$ is the $g$-function \eqref{eq:g-function}, and $x_1 < x_2$.
By conformally mapping the upper half-plane to a unit disk $D^2$,\footnote{The explicit conformal transformation from the upper half-plane, $z = x+ iy$, $y \geq 0$, to the unit disk $\zeta = re^{i\theta}$, $|\zeta| \leq 1$, is given by \begin{equation*}
    \zeta = \frac{1+iz}{1-iz} \,.
\end{equation*}
The real line is mapped to the boundary of the disk by $x = \mathrm{tan}\left(\frac{\theta}{2}\right)$.} the 2-point function becomes
\begin{equation} \label{eq:2pt_normalization}
    \langle \psi_{\alpha \beta} (\theta_1) \psi_{\beta \alpha} (\theta_2) \rangle_{D^2} = g_M \left[ 2\mathrm{sin}\left(\frac{\theta_2 - \theta_1}{2}\right) \right]^{-2\Delta_{\alpha \beta}} \,,
\end{equation}
where $\left|2\mathrm{sin}\left(\frac{\theta_2 - \theta_1}{2}\right)\right|$ is the chord distance between the two bcc operators on the boundary of the disk.
When $\alpha = \beta$, $\psi_{\alpha \alpha} = 1$ is simply the identity operator, and \eqref{eq:2pt_normalization} reduces to $\langle 1 \rangle_{D^2} = g_M$.
That is, we normalize the disk partition function with no other operator insertions to be equal to the $g$-function of the boundary condition, unless otherwise stated.\footnote{The overall normalization of the disk partition function depends on the choice of the Euler counterterm, and can be freely tuned.}

The normalization convention \eqref{eq:BCC_normalization} does not fix the phase ambiguity
\begin{equation} \label{eq:phase_ambiguity}
    \psi_{\alpha \beta} (x) \rightarrow  e^{i\Lambda (\alpha ,  \beta)} \psi_{\alpha \beta} (x) \,,
\end{equation}
with arbitrary $e^{i\Lambda (\alpha ,  \beta)} = e^{-i\Lambda (\beta ,  \alpha)} \in \mathrm{U}(1)$.
This will be related to the gauge ambiguity of the higher Berry connection below.

To define the higher Berry connection, we take the 3-point correlation function of bcc operators on a unit disk ($0 \leq \theta_1 <\theta_2 <\theta_3 < 2\pi$),
\begin{align}
\begin{split}
    &\langle \psi_{\alpha  \beta} (\theta_1) \psi_{ \beta  \gamma} (\theta_2) \psi_{\gamma  \alpha} (\theta_3) \rangle_{D^2} \\
    &~~~ = c( \alpha , \beta , \gamma ) g_M  \left[ 2\mathrm{sin}\left(\frac{\theta_2 - \theta_1}{2}\right) \right]^{-\Delta_{\alpha \beta \gamma}} \left[ 2\mathrm{sin}\left(\frac{\theta_3 - \theta_2}{2}\right) \right]^{-\Delta_{\beta  \gamma \alpha}} \left[ 2\mathrm{sin}\left(\frac{\theta_1 - \theta_3}{2}\right) \right]^{-\Delta_{\gamma \alpha \beta }} \,.
\end{split}
\end{align}
Here, $\Delta_{\alpha \beta \gamma} = \Delta_{\alpha \beta} + \Delta_{\beta \gamma} - \Delta_{\gamma \alpha}$ and so on, and $c(\alpha , \beta, \gamma)$ is the OPE coefficient involving the three bcc operators.
The OPE coefficient is in general a complex number, and we denote its phase part as $\phi (\alpha ,\beta ,\gamma)$.
That is,
\begin{equation} \label{eq:OPE_phase}
    c(\alpha , \beta , \gamma) = |c(\alpha , \beta , \gamma)| e^{ i \phi( \alpha , \beta , \gamma) } \,, \quad \phi( \alpha , \beta , \gamma)  \in \mathbb{R}/2\pi\mathbb{Z} \,.
\end{equation}
The OPE coefficient satisfies the cyclicity property,
\begin{equation}
    c(\alpha , \beta , \gamma) = c( \beta , \gamma, \alpha ) = c(\gamma ,\alpha , \beta ) \,,
\end{equation}
due to the rotation symmetry of the disk, and furthermore,
\begin{equation}
    c(\alpha , \alpha , \alpha) = 1 \,.
\end{equation}
We will assume that the OPE coefficient is a smooth function of its arguments (or at least differentiable sufficiently many times) inside the local patch $U$.

\paragraph{Higher Berry connection and curvature}
We now state the definition of higher Berry connection and curvature.
From the OPE coefficient of bcc operators, we define the higher Berry connection on the boundary conformal manifold as
\begin{align} \label{eq:def_berry_connection}
\begin{split}
    \mathcal{B} &=\frac{1}{2!} \mathcal{B}_{ij} (\alpha) d\alpha^i \wedge d\alpha^j \\
    &\equiv -\frac{i}{2!} \left[ 
        \frac{\partial^2 \phi(\alpha , \beta , \gamma)}{\partial \beta^i \partial \gamma^j} - (i \leftrightarrow j)
    \right]_{\beta = \gamma = \alpha} d\alpha^i \wedge d\alpha^j \,,
\end{split}
\end{align}
where $\phi (\alpha, \beta ,\gamma)$ is the phase of the OPE coefficient, given in \eqref{eq:OPE_phase}.
Here and in the following,
$\alpha^i, \beta^i, \gamma^i$ represent the coordinates on $M$ at points $\alpha,\beta,\gamma$, respectively.
The 3-form higher Berry curvature is given by
\begin{equation} \label{eq:def_berry_curvature}
    \mathcal{H} = d\mathcal{B} = \frac{1}{3!} \mathcal{H}_{ijk} (\alpha) d\alpha^i \wedge d\alpha^j \wedge d \alpha^k \,,
\end{equation}
where $d$ is the exterior derivative on $M$.
As we explained, the fact that we use the 3-point correlation function of bcc operators to define the higher Berry connection on the boundary conformal manifold is motivated by various intuitions from the tensor network.
More concretely, when $\alpha, \beta, \gamma \in M$ are infinitesimally close to each other inside the boundary conformal manifold, we interpret the phase $e^{i\phi(\alpha,\beta,\gamma)}$ of the OPE coefficient as the holonomy of the putative higher Berry connection along a small triangle formed by the three points $\alpha, \beta, \gamma$, in analogy with the tensor network case \cite{shiozaki2023higher}, which was briefly reviewed around \eqref{eq: triple inner product modified}.
This interpretation leads us to the definition \eqref{eq:def_berry_connection} of the higher Berry connection.
In Section \ref{sec:examples}, we explicitly compute \eqref{eq:def_berry_connection} and \eqref{eq:def_berry_curvature} in several examples, and confirm that we obtain sensible answers.

We may also write the definition of the higher Berry connection \eqref{eq:def_berry_connection} in a more natural form.
From the analogy with tensor networks, we define the \emph{triple inner product of boundary conditions} as
\begin{equation}
    \langle B_\alpha , B_\beta , B_\gamma \rangle \equiv c(\alpha, \beta, \gamma) \,.
\end{equation}
When it is possible to normalize the OPE coefficient $c (\alpha ,\beta , \gamma)$ to be a pure phase inside a local patch $\alpha , \beta , \gamma \in U$, we can write the definition of the higher Berry connection \eqref{eq:def_berry_connection} equivalently using the triple inner product as
\begin{equation} \label{eq:connection_triple}
    \mathcal{B} = -i \langle B_\alpha , d B_\alpha , \wedge d B_\alpha \rangle \,,
\end{equation}
where $\wedge$ denotes the wedge product on the boundary conformal manifold.
Equation \eqref{eq:connection_triple} resembles more the familiar definition of an ordinary Berry connection in quantum mechanics, $\mathcal{A} = -i \langle \psi_\alpha| d \psi_{\alpha} \rangle$, defined for a family of quantum states $\ket{\psi_\alpha}$. 
This suggests that \eqref{eq:def_berry_connection} is a natural ``higher'' generalization of the ordinary Berry connection in quantum mechanics.

\paragraph{Gauge ambiguity of higher Berry connection}
Let us comment on the gauge ambiguity of the higher Berry connection.
Recall that we have the freedom to redefine the overall phase factors of the bcc operators, as in \eqref{eq:phase_ambiguity}.
We would like to keep the fact that $\psi_{\alpha\beta}$ for $\alpha = \beta$ is an identity operator, and also the normalization convention for the 2-point function in \eqref{eq:2pt_normalization}.
This requires $\Lambda (\alpha , \alpha) = 0$ and $\Lambda (\alpha , \beta) = - \Lambda (\beta ,\alpha)$ mod $2\pi$ for all $\alpha , \beta \in M$.
We will also require that the phase redefinition $\Lambda (\alpha,\beta)$ is a smooth function of its arguments inside the local patch $\alpha, \beta \in U$.

Under the redefinition \eqref{eq:phase_ambiguity} of bcc operators, the phase of the OPE coefficient changes as
\begin{equation} \label{eq:phi_gauge}
    \phi (\alpha ,\beta ,\gamma) \rightarrow \phi (\alpha ,\beta ,\gamma) + \Lambda (\alpha ,\beta) + \Lambda (\beta ,\gamma) + \Lambda (\gamma ,\alpha) \,.
\end{equation}
This translates into the shift of the higher Berry connection \eqref{eq:def_berry_connection},
\begin{equation} \label{eq:gauge_1}
    \mathcal{B} \rightarrow \mathcal{B} +\frac{1}{2} \left[ 
        \frac{\partial^2 \Lambda (\beta , \gamma)} {\partial \beta^i \partial \gamma^j} - (i\leftrightarrow j)
    \right]_{\beta = \gamma  = \alpha} d\alpha^i \wedge d\alpha^j \,.
\end{equation}
To put \eqref{eq:gauge_1} into a more natural form, recall that physically, we interpret the phase $e^{i\phi(\alpha,\beta,\gamma)}$ as the holonomy of the higher Berry connection along a small triangle formed by $\alpha, \beta, \gamma \in M$.
Then, from \eqref{eq:phi_gauge}, it is natural to interpret the phase $e^{i\Lambda (\alpha,\beta)}$ as the \emph{holonomy of a 1-form gauge parameter}, which we denote as $\lambda = \lambda_i (\alpha) d \alpha^i$, along a path from $\alpha$ to $\beta$ in $M$.
In other words, we write $\Lambda (\alpha , \beta) = \int_{\alpha}^\beta \lambda_i (\gamma) d\gamma^i$, where
the precise integration path is not important when the two points $\alpha$ and $\beta$ are infinitesimally close to each other.
Note that this naturally satisfies the requirements $\Lambda (\alpha , \alpha) =0$ and $\Lambda (\alpha , \beta) = -\Lambda (\beta, \alpha)$.
In terms of the 1-form connection $\lambda$, the gauge transformation \eqref{eq:gauge_1} of the higher Berry connection becomes
\begin{equation}
    \mathcal{B}_{ij} \rightarrow \mathcal{B}_{ij} + \frac{\partial \lambda_j}{\partial \alpha^i } - \frac{\partial \lambda_i}{\partial \alpha^j } \,,
\end{equation}
or equivalently,
\begin{equation} \label{eq:gauge_2}
    \mathcal{B} \rightarrow \mathcal{B} + d\lambda \,.
\end{equation}
This is the familiar gauge transformation of a 2-form connection by a 1-form gauge parameter.\footnote{The gauge parameter $\lambda$ itself, being a 1-form connection, has its own gauge transformations. 
This comes from the ``pointwise'' phase redefinitions of the bcc operators of the form $\psi_{\alpha \beta} (x) \rightarrow e^{-i\eta (\alpha)} e^{i\eta (\beta)} \psi_{\alpha \beta} (x)$, which corresponds to  $\lambda \rightarrow \lambda + d \eta$.
Such pointwise phase redefinitions do not generate a nontrivial gauge transformation of $\mathcal{B}$, as can be seen from \eqref{eq:gauge_1}, and hence define redundancies in the gauge parameter $\lambda$.}
The higher Berry curvature \eqref{eq:def_berry_curvature} is invariant under the gauge transformation \eqref{eq:gauge_2}.

\paragraph{Modulated boundary condition}

As explained around \eqref{eq:chiral_boundary_state},
many of the families of boundary states we study in this work can be written as
$| B_{\xi}\rangle = U_{\xi} |B_0\rangle$
where $\xi=(\xi^1,\xi^2,\cdots) \in M$
is a set of parameters.  
Here, $|B_0\rangle$ is some reference boundary state, 
and $U_{\xi}$ is a unitary line operator,
which implements exactly marginal deformations.
(This should be contrasted with boundary interactions
and loop operators considered, e.g., in 
\cite{Bachas_2004},
which undergo RG flow.)
Schematically, 
$U_{\xi}$ is given by $U_{\xi}
\sim \exp [i \oint d\theta \xi^a \psi_a(\theta)]$,
where $\psi_a$ 
are exactly marginal boundary operators and 
$\xi^a$ are exactly marginal boundary coupling constants.
The disk partition function is given by 
$\langle 0| B_{\xi}\rangle$.

As we will discuss later through examples, (lightest) bcc operators can be understood as a ``termination'' of the line operator.
They can also be understood as arising from 
modulated or spatially inhomogeneous boundary states.
Here, we make the boundary couplings space-dependent, 
$\xi^a\to \xi^a(\theta)$, 
$U_{\xi} \to U(\xi(\theta))$ 
where
\begin{align}
U(\xi(\theta)) = \exp \Big[i \oint d\theta  \xi^a(\theta) \psi_a(\theta)\Big]\, . 
\end{align}
Correspondingly, 
we define a modulated boundary 
state by
$|B(\xi(\theta))\rangle \equiv 
U(\xi(\theta)) |B_0\rangle$.
(We will often write $|B(\xi)\rangle$ for notational simplicity.)
Bcc operators can be interpreted as creating  
a ``sharp'' domain wall between 
boundaries with $\xi$ and $\xi'$,
and hence correspond to 
a step-function-like configuration of boundary couplings.
Using general modulated boundary states, we will also discuss a connection between the higher Berry connection and the loop space connection
in Section \ref{Loop space connections on boundary conformal manifolds}.

\section{Examples} \label{sec:examples}

In this section, we discuss several examples.
In various (1+1)d CFTs, we explicitly compute the higher Berry connection and curvature, defined in \eqref{eq:def_berry_connection} and \eqref{eq:def_berry_curvature}, from the correlation functions of bcc operators.
As we mentioned, this is conceptually analogous to computing the triple inner product of three physically distinct tensor network states, which are nearby in the parameter space.
Examples we discuss include compact boson (Narain) CFTs, Wess-Zumino-Witten (WZW) models, and free fermions.
The Narain CFT and WZW model results suggest that the higher Berry connection is given by the NS-NS $B$-field in appropriate situations, as mentioned in the introduction. 

We also discuss $G$-symmetric conformal boundary conditions and the correspondence to (1+1)d $G$-symmetric SPT phases. 
Here, $G$ is a non-anomalous internal symmetry in the case of BCFT, or an on-site unitary symmetry in the case of tensor network.
While in these cases we do not have continuously deformable boundary moduli parameters,
we can relate
the triple inner product 
to topological invariants (``SPT invariants'').
Specifically, a disk amplitude with a $G$-symmetric conformal boundary condition decorated with symmetry line defects computes the corresponding SPT invariant valued in group cohomology.
This is a BCFT analogue of the triple inner product of a $G$-symmetric (injective) MPS in three different MPS gauges.

\subsection{Compact bosons with $B$-field} \label{sec:bosons}

We first consider the Narain CFT with an arbitrary number of compact boson fields $X^a$, $a=1, 2, \cdots , N$. 
The Euclidean action is given by
\begin{align} \label{eq:boson_B}
\begin{split}
    S &= \frac{1}{4\pi} \int d^2 \sigma \left[ 
        G_{ab} \partial_\mu X^a \partial^\mu X^b - i \epsilon^{\mu \nu}  B_{ab} \partial_\mu X^a \partial_\nu X^b   
    \right] \\
    &= \frac{1}{4\pi} \int \left[ 
        G_{ab} d X^a \wedge \star d X^b - i B_{ab} d X^a \wedge d X^b   
    \right] \,,
\end{split}
\end{align}
where $G_{ab}$ is the target space metric, and $B_{ab}$ is the Kalb-Ramond $B$-field.
The compact boson fields are normalized so that they are $2\pi$-periodic, $X^a \sim X^a + 2\pi$.
In this convention, the $B$-field is a periodic variable with $B_{ab} \sim B_{ab} + 1$.

The $B$-field term is locally a total derivative, but it has nontrivial effects when the boson fields have nonzero winding numbers or if we consider boundary conditions of the theory.
In particular, we will see that it appears as the higher Berry connection in the space of Dirichlet boundary conditions.
The family of Dirichlet boundary conditions is given by setting the boson fields to be constant along the boundary,
\begin{equation} \label{eq:Dirichlet_B}
    X^a |= \xi^a \,,
\end{equation}
for some $\xi^a \sim \xi^a + 2\pi$.
Here, $X^a|$ 
represents the field configuration on the boundary.
The moduli space of boundary conditions given by the parameters $\xi^a$ is an $N$-dimensional torus $T^N$.
Since these are Dirichlet boundary conditions, we may identify the boundary moduli space with the target space of the sigma model \eqref{eq:boson_B}.

Let $\psi_{\xi \xi'}$ denote the (lightest) bcc operator between two Dirichlet boundary conditions $X^a| = \xi^a$ and $X^a| = \xi'^a$.
Our goal is to derive the OPE coefficient between three such bcc operators, and then to obtain the higher Berry connection defined in \eqref{eq:def_berry_connection} from the OPE coefficient.
Below, we compute the OPE coefficient, first by an explicit path integral, and then from canonical quantization by identifying the bcc operator as an appropriate vertex operator \cite{1994JPhA...27.5375A,Oshikawa_1997}.

\subsubsection{Path integral}

We first compute the disk 3-point correlation function of the bcc operators by explicitly evaluating the path integral.
To do so, it is convenient to consider a more general \emph{modulated} (or inhomogeneous) boundary condition.
Specifically, let $D^2= \{ \zeta = re^{i\theta} : |\zeta| \leq 1 \} $ be a unit disk.
On the boundary of the disk, we impose a modulated Dirichlet boundary condition,
\begin{equation} \label{eq:boson_modulated}
    X^a|_{\partial D^2} = \xi^a (\theta) \,.
\end{equation}
We will see that correlation functions of bcc operators can be computed by letting the modulated parameter $\xi^a (\theta)$ have step-function-like discontinuities at the points where the bcc operators are inserted.\footnote{
In Appendix of \cite{Oshikawa_1997}, discontinuous boundary conditions were considered to compute the 2-point function of bcc operators in free boson theory.}
For simplicity, let us set the target space metric to be $G_{ab} = R^2 \delta_{ab}$, with $R$ being the radius of the target space circles.
We now compute the disk partition function of the Narain CFT with the general modulated boundary condition \eqref{eq:boson_modulated}.

To begin with, we note that the $B$-field term in the action \eqref{eq:boson_B} localizes to the boundary of the disk.
Even though the $B$-field term is locally a total derivative, on a general Riemann surface with boundaries, this is not necessarily the case due to global issues.
However, on a disk, there are no nontrivial cycles, and the compact boson fields do not admit winding configurations (unless we insert vertex operators with nontrivial winding numbers).
Therefore, the $B$-field dependence of the disk path integral with the modulated boundary condition \eqref{eq:boson_modulated} is simply
\begin{equation}
    \exp \left[
        \frac{i B_{ab}}{4\pi} \int_0^{2\pi} d\theta \xi^a \partial_\theta \xi^b
    \right] \,.
\end{equation}

Next, let us compute the $B$-field independent part of the disk partition function. 
Write the free boson fields as
\begin{equation} \label{eq:bessel_fourier}
    X^a (r, \theta) = X_{cl}^a (r,\theta) + X_q^a (r,\theta)
    \,.
\end{equation}
Here, $X_{cl}^a (r,\theta)$ is the solution to the classical equation of motion satisfying the modulated boundary condition,
\begin{equation} \label{eq:X_cl}
    \nabla^2 X_{cl}^a = 0 \,, \quad X_{cl}^a |_{\partial D^2} = \xi^a (\theta) \,,
\end{equation}
and $X^a_q$ are fluctuations, which vanish at the boundary,
\begin{equation}
    X^a_q |_{\partial D^2} = 0 \,.
\end{equation}
The classical solution is given by
\begin{equation} \label{eq:X_cl_2}
    X_{cl}^a (r, \theta) = \sum_{m \in \mathbb{Z}} \xi^a_m r^{|m|} e^{-im \theta} \,,
\end{equation}
where $\xi^a_m$ are Fourier modes of the modulated parameter,
\begin{equation} \label{eq:fourier_xi}
    \xi^a_m = \frac{1}{2\pi} \int_{0}^{2\pi} d\theta \xi^a (\theta) e^{im\theta} \,.
\end{equation}

The disk partition function with the modulated boundary condition \eqref{eq:boson_modulated} is given by
\begin{equation} \label{eq:boson_modulated_Z}
    Z[D^2; \xi^a (\theta)] = \exp \left[
        \frac{i B_{ab}}{4\pi} \int_0^{2\pi} d\theta \xi^a \partial_\theta \xi^b
    \right]  \times \exp \left( -S_{cl} [\xi^a (\theta)] \right) \times  \left( \mathrm{Det}' \left( -R^2 \nabla^2 /4\pi  \right) \right)^{-N/2} \,,
\end{equation}
where $S_{cl}[\xi^a (\theta)]$ is the saddle point action evaluated at the classical solution \eqref{eq:X_cl_2} (while setting $B_{ab}=0$), which is a functional of the modulation parameter $\xi^a (\theta)$, and the path integral over the fluctuations $X_q^a$ gives the determinant of the Laplacian on the disk (the prime in $\mathrm{Det}'$ means that the zero mode is excluded).

The saddle point action can be computed explicitly from the classical solution \eqref{eq:X_cl_2},
and we obtain
\begin{equation} \label{eq:X_saddle_action}
    \exp \left( -S_{cl} [\xi^a (\theta)] \right) = \exp \left[ -R^2
        \sum_{a=1}^N  \sum_{m=1}^{\infty} m \xi_m^a \xi_{-m}^a
    \right] \,.
\end{equation}
We now proceed to derive correlation functions of bcc operators.

Let us first demonstrate the idea by computing the 2-point function of the bcc operators from \eqref{eq:boson_modulated_Z}, and then matching with the known scaling dimension.
To compute the 2-point function,  we let the modulated parameter $\xi^a (\theta)$ be
\begin{equation} \label{eq:xi_2pt}
    \xi^a (\theta) = \begin{cases}
        \xi_{(1)}^a &\quad \text{if $0 \leq \theta \leq \theta_1$ or $\theta_2 < \theta \leq 2\pi$} \,,\\
        \xi_{(2)}^a &\quad \text{if $\theta_1 < \theta \leq \theta_2$} \,,
    \end{cases}
\end{equation}
where $\xi_{(i)}^a$'s are constants, and $0 \leq \theta_1 < \theta_2 < 2\pi$.
We assume $|\xi_{(2)}^a - \xi_{(1)}^a| < 2\pi$ for all $a = 1, 2, \cdots , N$.\footnote{If $|\xi_{(2)}^a - \xi_{(1)}^a| > 2\pi$ for some $a$, it corresponds to inserting bcc operators with higher scaling dimensions
\cite{Oshikawa_1997}.}
The disk partition function with the discontinuous modulation parameter \eqref{eq:xi_2pt} computes the 2-point function
\begin{equation}
    \langle \psi_{\xi_{(1)} \xi_{(2)}} (\theta_1) \psi_{\xi_{(2)} \xi_{(1)}} (\theta_2 ) \rangle_{D^2}
\end{equation}
after a suitable wavefunction renormalization of bcc operators, as we explain below.

First, the $B$-field dependence of the disk partition function is
\begin{align} \label{eq:X_2pt_B}
\begin{split}
    \frac{i B_{ab}}{4\pi} \int_0^{2\pi} d\theta \xi^a \partial_\theta \xi^b 
        &= \frac{i B_{ab}}{4\pi} (\xi_{(2)}^a - \xi_{(1)}^a) (\xi_{(2)}^b - \xi_{(1)}^b)
        = 0 \,.
\end{split}
\end{align}
In evaluating \eqref{eq:X_2pt_B}, we had to insert the exact values of the modulated parameter $\xi^a (\theta)$ at the discontinuities.
In \eqref{eq:xi_2pt}, we made a particular choice for these values.
In general, different choices will change the disk partition function by overall phase factors localized at the discontinuities.
This corresponds to the freedom to redefine the overall phase of the bcc operators, see \eqref{eq:phase_ambiguity}.
For now, we proceed with the particular choice \eqref{eq:xi_2pt}.

Next, we compute the $B$-field independent part of the classical saddle point action.
Fourier modes of the modulated parameter $\xi^a (\theta)$ in \eqref{eq:xi_2pt} are given by
\begin{align}
\begin{split}
    \xi^a_m 
    &= \frac{1}{2\pi i m} \left( 
        e^{im\theta_2} - e^{im\theta_1}
    \right) \left(
        \xi^a_{(2)} - \xi^a_{(1)}
    \right) \,.
\end{split}
\end{align}
From \eqref{eq:X_saddle_action}, we obtain
\begin{align}
\begin{split}
    S_{cl} [\xi^a (\theta)] 
    &=  2R^2 \left( \frac{\xi_{(2)}^a- \xi_{(1)}^a}{2\pi} \right)^2 \sum_{m=1}^{\infty} \frac{1}{m} \left( 1 - \mathrm{cos} (m(\theta_2 - \theta_1 )) \right) \,.
\end{split}
\end{align}
We regularize the divergent sum $\sum_{m=1}^\infty 1/m$ by introducing a short-distance regulator $\epsilon > 0$,
\begin{equation}
    \sum_{m=1}^\infty \frac{(1-\epsilon)^m}{m} = -\mathrm{log}\epsilon \,.
\end{equation}
The remaining infinite sum is convergent,
\begin{equation}
    \sum_{m=1}^{\infty} \frac{1}{m} \mathrm{cos} (m(\theta_2 - \theta_1 )) = - \mathrm{log} \left[ 2 \mathrm{sin}\left( \frac{\theta_2 - \theta_1 }{2} \right) \right] \,.
\end{equation}

We interpret the disk partition function with the modulated boundary condition given by \eqref{eq:xi_2pt}, divided by the appropriate power of the determinant of the Laplacian operator on the disk,\footnote{With this normalization convention, for constant $\xi^a (\theta)$, the disk partition function is different from the $g$-function of the corresponding boundary state.} as the 2-point correlation function of ``bare'' bcc operators,
\begin{align}
\begin{split}
    \langle \psi^{bare}_{\xi_{(1)} \xi_{(2)}} (\theta_1) \psi^{bare}_{\xi_{(2)} \xi_{(1)}} (\theta_2 ) \rangle_{D^2} &= \frac{Z[D^2; \xi^a (\theta)]}{ \left( \mathrm{Det}' \left( -R^2 \nabla^2 /4\pi  \right) \right)^{-N/2}} \\
    &= \left[ 
        \frac{2}{\epsilon} \mathrm{sin} \left( \frac{\theta_2 - \theta_1 }{2} \right)
    \right]^{-2R^2 \left( \frac{\xi_{(2)}^a- \xi_{(1)}^a}{2\pi} \right)^2} \,.
\end{split}
\end{align}
The dependence on the regulator $\epsilon$ can be removed by the wavefunction renormalization of the bcc operators \cite{Oshikawa_1997}.
We define the renormalized bcc operator as
\begin{equation}
    \psi_{\xi_{(1)} \xi_{(2)}} = \epsilon^{-R^2 \left( \frac{\xi_{(2)}^a- \xi_{(1)}^a}{2\pi} \right)^2} \psi^{bare}_{\xi_{(1)} \xi_{(2)}} \,.
\end{equation}
The 2-point function of renormalized bcc operators does not depend on $\epsilon$, and we may take $\epsilon \rightarrow 0$,
\begin{equation}
    \langle \psi_{\xi_{(1)} \xi_{(2)}} (\theta_1) \psi_{\xi_{(2)} \xi_{(1)}} (\theta_2 ) \rangle_{D^2} = \left[ 
        2 \mathrm{sin} \left( \frac{\theta_2 - \theta_1 }{2} \right)
    \right]^{-2R^2 \left( \frac{\xi_{(2)}^a- \xi_{(1)}^a}{2\pi} \right)^2} \,.
\end{equation}

The 2-point correlation function has the expected conformal structure for boundary primary operators, and we read off the scaling dimension of the bcc operator,
\begin{equation}
    \Delta_{\xi_{(1)} \xi_{(2)}} = \frac{R^2}{4\pi^2} \left( \xi^a_{{(2)}} - \xi^a_{{(1)}} \right)^2 \,.
\end{equation}
The scaling dimension does not depend on the $B$-field.
Furthermore, it correctly reproduces the known scaling dimension (see, for instance, \cite{Affleck_1994,Oshikawa_1997,Gaberdiel:2001zq}).

We now apply the same method to compute the 3-point correlation function of bcc operators.
Similar to the 2-point function case, we impose a discontinuous modulated boundary condition, given by
\begin{equation}
    \xi^a (\theta) = \begin{cases}
        \xi_{(1)}^a &\quad \text{if $0\leq \theta \leq \theta_1$ or $\theta_3 < \theta \leq 2\pi$} \,,\\
        \xi_{(2)}^a &\quad \text{if $\theta_1 < \theta \leq \theta_2$} \,, \\
        \xi_{(3)}^a &\quad \text{if $\theta_2 < \theta \leq \theta_3$} \,,
    \end{cases}
\end{equation}
where $0 \leq \theta_1 < \theta_2 < \theta_3 < 2\pi$.
The precise values of $\xi^a (\theta)$ at the discontinuities are again related to the overall phase of the bcc operators.
Using the general formula for the disk partition function \eqref{eq:boson_modulated_Z}, we obtain the 3-point function of (renormalized) bcc operators,
\begin{align}
\begin{split}
    &\langle \psi_{\xi_{(1)} \xi_{(2)}} (\theta_1) \psi_{\xi_{(2)} \xi_{(3)}} (\theta_2 ) \psi_{\xi_{(3)} \xi_{(1)}} (\theta_3 ) \rangle_{D^2} \\
    &= 
    \exp \left[ 
        \frac{iB_{ab}}{4\pi} \left(
            \xi_{(1)}^a \xi_{(2)}^b + \xi_{(2)}^a \xi_{(3)}^b + \xi_{(3)}^a \xi_{(1)}^b
        \right)
    \right]
    \times 
    \left[ 
        2 \mathrm{sin} \left( \frac{\theta_2 - \theta_1 }{2} \right)
    \right]^{-\Delta_{\xi_{(1)} \xi_{(2)} \xi_{(3)}}} \\
    &\quad \times \left[ 
        2 \mathrm{sin} \left( \frac{\theta_3 - \theta_2 }{2} \right)
    \right]^{-\Delta_{\xi_{(2)} \xi_{(3)} \xi_{(1)}}}  \left[ 
        2 \mathrm{sin} \left( \frac{\theta_3 - \theta_1 }{2} \right)
    \right]^{-\Delta_{\xi_{(3)} \xi_{(1)} \xi_{(2)}}}  \,,
\end{split}
\end{align}
where $\Delta_{\xi_{(i)} \xi_{(j)} \xi_{(k)}} = \Delta_{\xi_{(i)} \xi_{(j)} } + \Delta_{\xi_{(j)} \xi_{(k)} } - \Delta_{\xi_{(i)} \xi_{(k)} }$.

From the definition of the higher Berry connection \eqref{eq:def_berry_connection}, we conclude that
\begin{equation} \label{eq:boson_B1}
    \mathcal{B} = \frac{1}{2} \frac{B_{ab}}{2\pi} d\xi^a \wedge d\xi^b \,.
\end{equation}
Namely, the higher Berry connection coincides with the Kalb-Ramond $B$-field in this case.
The connection is flat, and the curvature vanishes.
Locally in the parameter space, this means that we may set $\mathcal{B} = 0$ by an appropriate gauge transformation.
In particular, if we perform a phase redefinition of the bcc operators,
\begin{equation}
    \psi_{\xi_{(1)} \xi_{(2)}} \rightarrow \exp \left[-\frac{iB_{ab}}{4\pi} \xi_{(1)}^a \xi_{(2)}^{b} \right] \psi_{\xi_{(1)} \xi_{(2)}} \,,
\end{equation}
then the OPE coefficient becomes real, and we get $\mathcal{B} = 0$.
However, the higher Berry connection \eqref{eq:boson_B1} is globally still nontrivial, since it has nonzero holonomies along nontrivial 2-cycles inside the parameter space $T^N$.

\subsubsection{Canonical quantization approach}

This model can also be analyzed using canonical quantization
\cite{Giveon_1994}.
In particular, the bcc operators can be explicitly constructed 
\cite{Affleck_1994, Oshikawa_1997, affleck2009quantumimpurityproblemscondensed}.
The starting Lagrangian is  
\begin{align}
\mathcal{L} &=
\frac{1}{4\pi}
\delta_{ab}
\left[
\partial_{t} {X}^{a} \partial_{t}{X}^{b}
-
\partial_{\theta} {X}^{a} \partial_{\theta}{X}^{b}
\right]
+ \frac{2}{4\pi}
B_{ab}
\partial_{t}{X}^{a}
\partial_{\theta}{X}^{b}\, , 
\end{align}
where $\theta \in [0,2\pi]$ parameterizes 
a spatial circle of length $2\pi$ and $X^{a}$ are compact boson fields satisfying 
$
X^{a} \sim 
X^{a} + 2 \pi N^{\alpha} e^a_{\alpha},
$
$ 
N^{\alpha} \in \mathbb{Z}\, .
$
(Here $e^a_{\alpha}$ specifies the compactification lattice, which will not play a central role in what follows.)
The boson fields $X^a$ and their conjugate momenta 
\begin{align}
P_{a} 
= \frac{\partial {\cal L}}{\partial \dot{{X}}^{a}}
=
\frac{1}{2\pi}
\left[
\delta_{ab}
\dot{{X}}^{b}
+
B_{ab}
\partial_{\theta}{X}^{b}
\right]\, ,
\end{align}
satisfy the canonical equal time commutation relations,
\begin{align}
&\left[
{X}^{a}(t,\theta),
X^{b}(t,\theta^{\prime})
\right]
=
0\, ,
\quad
\left[
{X}^{a}(t,\theta),
P_{b}(t,\theta^{\prime})
\right]
=
i \delta^{a}_{b} 
\delta(\theta-\theta^{\prime})\, ,
\nonumber \\
&\left[
{P}_{a}(t,\theta), P_{b}(t,\theta^{\prime})
\right]
=
\frac{ i B_{ab}}{\pi}
\partial_{\theta}
\delta(\theta-\theta')\, , 
\end{align}
where 
$\delta(\theta)=
(1/2\pi)\sum_{n\in \mathbb{Z}}e^{ i n \theta}
$
is the periodic delta function on a circle.

Since the fields $X^a$ commute with each other, they can be simultaneously localized; that is, they can have definite c-number expectation values. A boundary state corresponding to a Dirichlet boundary condition is defined by
\begin{align}
\left[ X^a(\theta) - \xi^a \right]
|B_\xi \rangle = 0\, . 
\end{align}
The solution to this equation can be written 
in a coherent state form
\cite{Callan:1988wz}.
The boundary state $|B_{\xi}\rangle$
can be written, using a reference state $|B_0\rangle$
and a unitary operator $U_{\xi}$,
as
$|B_{\xi}\rangle = U_{\xi}|B_0\rangle$,
where the unitary operator $U_{\xi}$ is given by 
\begin{align}
U_{\xi} = 
\exp \Big[-i \xi^{a} \oint d\theta\, P_{a}(\theta)\big]\, .
\end{align}

The bcc operator can then be obtained
as the end point of the line operator $U_{\xi}$.
By introducing the dual field 
$\widehat{X}$
by 
$P_a(\theta) \equiv 
(1/2\pi)
\partial_{\theta} \widehat{X}_a(\theta)$,
we find
\begin{align}
\exp\Big[
-i  \xi^a\int_{\theta_2}^{\theta_1} d\theta\, P_{a}(\theta)\Big]
&\sim 
e^{
-\frac{i \xi^a}{2\pi} \widehat{X}_a(\theta_1)
}
\,
e^{
+\frac{i  \xi^a}{2\pi} \widehat{X}_a(\theta_2)
}\, .
\end{align}
We thus identify the bcc operator from $|B_0\rangle$ to $|B_{\xi}\rangle$ as (normal-ordering is implicit)
\begin{align}
\psi_{\xi}(\theta)
\equiv
e^{
-\frac{i  \xi^a}{2\pi} \widehat{X}_a(\theta)
}\, .
\end{align}
Note that the overall phase of the operator is intrinsically ambiguous. 
To calculate the higher Berry phase, we examine the boundary OPE of the two bcc operators:
\begin{align}
\lim_{\theta'\to \theta}
\psi_{ \xi}(\theta)\,
\psi_{ \xi'}(\theta')
=
e^{i\phi(\xi,\xi')}
\psi_{\xi+ \xi'}(\theta)\, .
\end{align}
The phase factor $\phi$ can be calculated from the commutator of the dual field as  
\begin{align}
i\phi(\xi, \xi')
&=
\frac{ \xi^{a}
 \xi^{\prime b} }{(2\pi)^2}
\big[
\widehat{X}_a(\theta),
\widehat{X}_{b}(\theta')
\big]
=
-i
\frac{  \xi^{a} B_{ab}
 \xi^{\prime b} }{2\pi}
 \mathrm{sgn}\,(\theta-\theta')\, , 
\end{align}
which agrees with \eqref{eq:boson_B1}.

Finally, we can also study the mixed Dirichlet-Neumann boundary condition:
\begin{align}
(\partial_{\theta} X^a(\theta) + B_{ab} \partial_t X^b(\theta))|N_{\xi}\rangle
=0\, .
\end{align}
Although this boundary condition does not make it explicit, the boundary state $|N_\xi\rangle$ can depend on parameters $\xi$. 
Following an analysis similar to the one above, we find that the bcc operators commute in this case. As a result, there is no nontrivial higher Berry phase.

\subsection{WZW models} \label{sec:WZW}

Here, we discuss families of conformal boundary conditions in WZW models parametrized by the group manifold, and show that there is a nontrivial higher Berry curvature in the space of boundary conditions, as defined in \eqref{eq:def_berry_curvature}, given by the Wess-Zumino (WZ) term.\footnote{This result is also compatible with the anomaly inflow argument \cite{Debray:2023ior}.}

Specifically, we take the bulk theory to be the $\hat{\mathfrak{g}}_k$ WZW model based on the diagonal modular invariant torus partition function.
The Lie algebra $\mathfrak{g}$ is assumed to be simple and compact. 
The affine Kac-Moody algebra is given by
\begin{equation} \label{eq:KM}
    [J_m^a , J_n^b] = km \delta^{ab} \delta_{m+n,0} + i f^{abc} J_{m+n}^c \,,
\end{equation}
where $k \in \mathbb{Z}_{\geq 1}$ is the level, and $f^{abc}$ is the structure constant of the Lie algebra.
Repeated Lie algebra indices are always summed over.

Let
\begin{equation} \label{eq:vacuum_Cardy}
    \ket{B_0} = \sum_i S_{0i}^{1/2} | i \rangle \!\rangle
\end{equation}
be the Cardy boundary state \cite{Cardy:1989ir,Cardy:2004hm} corresponding to the identity representation.
Here, $i$'s label chiral primaries of the current algebra $\hat{\mathfrak{g}}_k$, $S_{ij}$ is the modular S-matrix, and $| i \rangle \! \rangle$ is the Ishibashi state corresponding to the $i$'th primary, satisfying the (untwisted) gluing condition $(J^a_n + \bar{J}^a_{-n})| i \rangle \!\rangle = 0$ for all $n$.
By acting the left-moving $G$-symmetry on the identity Cardy state, we obtain a family of boundary states parametrized by the group manifold $G$,\footnote{On isolated points on the group manifold, the boundary state can again be a Cardy state (with untwisted gluing condition). For instance, in the case of the $\hat{\mathfrak{su}}(2)_k$ WZW models, $\ket{B_g}$ with $g = -\mathds{1}_{2\times 2}$ is a Cardy boundary state corresponding to the spin $\frac{k}{2}$ representation of the current algebra. See Section \ref{sec:su2_more}.}
\begin{equation} \label{eq:WZW_parameter_space}
    \ket{B_g} = U_g \ket{B_0} \,, \quad g \in G \,.
\end{equation}
Here, $U_g$ is the left-moving $G$-symmetry operator.
For a group element $g = \exp ( i\omega^a T^a)$,\footnote{In this section, instead of $\xi$, 
we use $g$ and $\omega$ to represent the parameterization of boundary states, respecting the group nature of the parameter space.} where $T^a$ are the Lie algebra generators,\footnote{\label{footnote killing}Our convention for the Lie algebra generators is always such that $K(T^a, T^b) = \delta^{ab}$ where $K(X,Y) =\frac{1}{2h^\vee} \mathrm{tr}\left( \mathrm{ad}(X) \mathrm{ad}(Y) \right)$ is the (renormalized) Killing form, with $h^\vee$ the dual coxeter number. Accordingly, we do not distinguish upper and lower Lie algebra indices.} the symmetry operator is given by
\begin{equation} \label{eq:U_g}
    U_g = \exp \left[ 
        i \omega^a \oint dz J^a (z) 
    \right] \,.
\end{equation}
This family of boundary conditions corresponds to the Dirichlet boundary condition for the $G$-valued sigma model field.
The $g$-function of the boundary conditions is given by
\begin{equation} \label{eq:g_WZW}
    \langle 0 | B_g \rangle = S_{00}^{1/2}
\end{equation}
for all $g \in G$.
For each $g \in G$, the boundary state $\ket{B_g}$ satisfies the twisted gluing condition for the left- and right-moving currents,
\begin{equation} \label{eq:twisted_gluing}
    \left(  \Omega (g) (J^a_n ) +\bar{J}^a_{-n} \right) \ket{B_g} = 0 \,,
\end{equation}
where $\Omega (g)$ is the gluing matrix \cite{Green:1995ga,Recknagel:1997sb,Recknagel_1999},
\begin{equation}
\Omega (g) : J^a_n \mapsto U_g J_n^a U_g^{-1}\,.
\end{equation}
The boundary state $\ket{B_g}$ preserves one copy of the current algebra for every $g\in G$, but the precise linear combination of left- and right-moving currents that is preserved depends on $g$.

\subsubsection{Modulated boundary condition}

To compute the 3-point function of bcc operators, we first consider the disk partition function of the WZW model with a general modulated boundary condition.
Let $0 \leq \theta \leq 2\pi$ be the coordinate along the boundary of a unit disk $D^2$.
Along the boundary of the disk, we impose a $\theta$-dependent boundary condition
\begin{equation} \label{eq:WZW_modulated_boundary}
    \ket{B(g(\theta))} \,, \quad g(\theta) = \exp \left[ 
        i (\omega^a  + \delta \omega^a (\theta) ) T^a 
    \right] \,.
\end{equation}
We treat the modulation $\delta \omega^a (\theta)$ as a small fluctuation around the reference point $\exp \left[ i \omega^a T^a  \right] \in G$ in the parameter space.
The disk partition function corresponds to the overlap of the modulated boundary state with the vacuum state,
\begin{equation} \label{eq:WZW_modulated_folded}
    \langle 0 | B(g(\theta)) \rangle = \langle 0 | \exp \left[
        i \int_{0}^{2\pi} d\theta (\omega^a + \delta \omega^a (\theta)) J^a (\theta)
    \right] | B_0 \rangle \,.
\end{equation}
For the purpose of obtaining the higher Berry connection \eqref{eq:def_berry_connection}, it is sufficient to compute \eqref{eq:WZW_modulated_folded} perturbatively up to second order in the modulation parameter $\delta \omega^a$, and we will do so below.

\begin{figure}[t!]
\centering
\includegraphics[width=.8\textwidth]{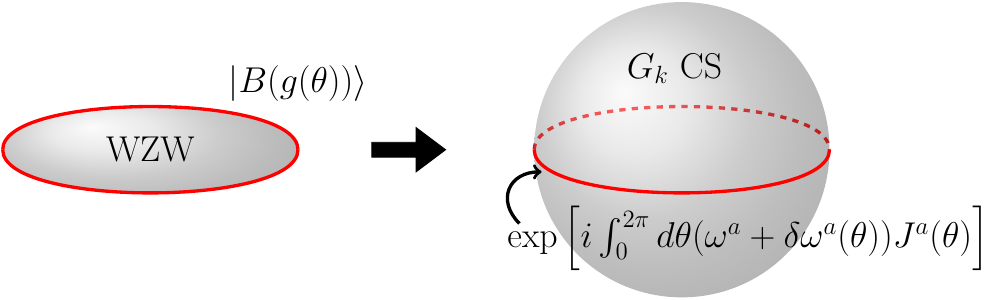}
\caption{The disk partition function of the WZW model with the modulated boundary condition $|B(g(\theta))\rangle$ (left) is the same as the partition function of the $G_k$ Chern-Simons theory on a three-dimensional ball (right), where along the equator of the boundary of the ball wraps around the modulated line operator $\exp \left[i \int_{0}^{2\pi} d\theta (\omega^a + \delta \omega^a (\theta)) J^a (\theta)\right]$.}\label{fig:doubling}
\end{figure}

The calculation simplifies if we apply the \emph{doubling trick}, following \cite{Kapustin:2010if}, or a slight generalization thereof, as shown in Figure \ref{fig:doubling}.
Specifically, we use the fact that the disk partition function of the WZW model with the modulated boundary condition \eqref{eq:WZW_modulated_boundary} is equal to the partition function of the $G_k$ Chern-Simons theory defined on a 3-dimensional ball $B^3$.
On the boundary of $B^3$, the northern and southern hemispheres host left- and right-moving degrees of freedom of the WZW model, respectively.
Along the equator of $\partial B^3 = S^2$, we insert the modulated line operator $\exp \left[i \int_{0}^{2\pi} d\theta (\omega^a + \delta \omega^a (\theta)) J^a (\theta)\right]$.
Squashing the ball $B^3$ to a disk $D^2$ by gluing the northern and southern hemispheres, we recover the WZW model on the disk with the modulated boundary condition.

On the three-dimensional Chern-Simons theory side, the path integral over the southern hemisphere prepares the chiral vacuum state $\ket{0}_L$ in the disk Hilbert space of the Chern-Simons theory, which furnishes the vacuum representation of the current algebra \cite{Elitzur:1989nr}.
Similarly, the path integral over the northern hemisphere prepares the bra state ${}_L\bra{0}$.
Finally, the modulated line operator along the equator acts on the disk Hilbert space of the Chern-Simons theory.
All in all, the $B^3$ partition function of the Chern-Simons theory, which is equal to the $D^2$ partition function of the WZW model, corresponds to the following expectation value of the modulated line operator,
\begin{equation} \label{eq:WZW_modulated_unfolded}
    {}_L \langle 0 | \exp \left[i \int_{0}^{2\pi} d\theta (\omega^a + \delta \omega^a (\theta)) J^a (\theta)\right] | 0 \rangle_L \,.
\end{equation}

We normalize the chiral vacuum state by
\begin{equation} \label{eq:chiral_vacuum}
    {}_L \langle 0 | 0 \rangle_L = S_{00}^{1/2} \,,
\end{equation}
such that for $\delta \omega^a = 0$, the $B^3$ partition function of the Chern-Simons theory and the $D^2$ partition function of the WZW model are equal to the $g$-function of the boundary state in \eqref{eq:g_WZW}.

Let us now compute \eqref{eq:WZW_modulated_unfolded} up to second order in the modulation parameter $\delta \omega^a$.
The mode expansion of the current operator is given by
\begin{equation}\label{eq: Fourier J}
    J^a (\theta) = \frac{1}{2\pi} \sum_{m \in \mathbb{Z}} J^a_m e^{im\theta} \,,
\end{equation}
and we denote the Fourier modes of the modulation parameter $\delta \omega^a ( \theta)$ as 
\begin{equation}
    \delta \omega^a_m = \frac{1}{2\pi}\int_0^{2\pi} d\theta \delta \omega^a (\theta) e^{im \theta} \,,
\end{equation}
similar to the free boson case \eqref{eq:fourier_xi}.
Inserting into \eqref{eq:WZW_modulated_unfolded}, we obtain
\begin{align}\label{eq: exp expansion}
\begin{split}
    &{}_L \langle 0 | \exp \left[i \int_{0}^{2\pi} d\theta (\omega^a + \delta \omega^a (\theta)) J^a (\theta)\right] | 0 \rangle_L \\
    &={}_L\langle 0 | \exp \left[ 
        i \left( 
            \omega^a J^a_0 + \sum_{m \in \mathbb{Z}} \delta \omega^a_m J_m^a
        \right)
    \right] | 0 \rangle_L \\
    &= \sum_{r=0}^{\infty} \frac{i^r}{r !} {}_L\langle 0 | \left( 
            \omega^{a_1} J^{a_1}_0 + \sum_{m_1 \in \mathbb{Z}} \delta \omega^{a_1}_{m_1} J_{m_1}^{a_1}
        \right) \times \cdots \times
        \left( 
            \omega^{a_r} J^{a_r}_0 + \sum_{m_r \in \mathbb{Z}} \delta \omega^{a_r}_{m_r} J_{m_r}^{a_r}
        \right) | 0 \rangle_L \,.
\end{split}
\end{align}
Using $J_m^a \ket{0}_L = 0$ for $m \geq 0$, $(J_m^a)^\dagger = J_{-m}^a$, and the Kac-Moody algebra \eqref{eq:KM}, this becomes
\begin{align} \label{eq:WZW_modulated_1}
\begin{split}
    &S_{00}^{-1/2} \times {}_L \langle 0 | \exp \left[i \int_{0}^{2\pi} d\theta (\omega^a + \delta \omega^a (\theta)) J^a (\theta)\right] | 0 \rangle_L \\
     &= 1 - \frac{1}{2} k\sum_{m=1}^{\infty} m \delta \omega_{m}^{a} \delta \omega_{-m}^{a}  \\
    &~~~~ +\frac{1}{6} k \left( \sum_{m=1}^{\infty} m \delta \omega_{m}^{a_1} \delta \omega_{-m}^{a_3} \right) \omega^{a_2} f^{a_1 a_2 a_3} \\
    &~~~~- k\sum_{r=4}^{\infty} \frac{(-1)^r}{r!} \left(
         \sum_{m=1}^{\infty} m \delta \omega_{m}^{a_1} \delta \omega_{-m}^{a_r}
    \right) \omega^{a_2} \cdots \omega^{a_{r-1}} f^{a_{r-1} a_r c_1}f^{a_{r-2} c_1 c_2} \cdots f^{a_{3} c_{r-4} c_{r-3}}f^{a_{2} c_{r-3} a_1} \\  &~~~~+ \mathcal{O} ( \delta \omega^3  )
    \,.
\end{split}
\end{align}
The factor of $S_{00}^{-1/2}$ is due to our normalization convention \eqref{eq:chiral_vacuum} for the chiral vacuum state.

The infinite sum expression \eqref{eq:WZW_modulated_1} can be simplified by defining a matrix
\begin{equation} \label{eq:X}
    X^{bc} = -\omega^a f^{abc} \,.
\end{equation}
We have
\begin{align} \label{eq:WZW_modulated_2}
\begin{split}
    &S_{00}^{-1/2} \times {}_L\langle 0 | \exp \left[ 
        i \int_{0}^{2\pi} d\theta (\omega^a + \delta\omega^a (\theta) ) J^a (\theta)
    \right] | 0 \rangle_L \\
    &~~~~= 1 - k \left(
         \sum_{m=1}^{\infty} m \delta \omega_{m}^{a} \delta \omega_{-m}^{b} \right) \left(\sum_{r=2}^{\infty} \frac{1}{r!} (X^{r-2})^{ba}
    \right) + \mathcal{O} ( \delta \omega^3  ) \\
    &~~~~=1 - k \left(
         \sum_{m=1}^{\infty} m \delta \omega_{m}^{a} \delta \omega_{-m}^{b} \right) \left(  X^{-2} (e^X - \mathds{1} - X ) 
    \right)^{ba} + \mathcal{O} ( \delta \omega^3  ) \,.
\end{split}
\end{align}
We now use this to compute correlation functions of bcc operators, by making the modulation parameter discontinuous. 

\subsubsection{2-point functions}

Let us briefly discuss 2-point functions of bcc operators, before we move on to 3-point functions and higher Berry connection.
The general formula for the modulated boundary condition \eqref{eq:WZW_modulated_2} allows us to compute the arbitrary $n$-point function of bcc operators.
In particular, from the 2-point function, we obtain the scaling dimension of bcc operators, perturbatively up to order $\delta \omega^2$.
We will compare this to the known scaling dimension of bcc operators.
This serves as a consistency check of the formula \eqref{eq:WZW_modulated_2}.

Let
\begin{equation} \label{eq:g1g2}
    g_1 = \exp \left[ i (\omega^a + \delta \omega_{(1)}^a )  T^a \right] \,, \quad g_2 = \exp \left[ i (\omega^a + \delta \omega_{(2)}^a )  T^a \right] 
\end{equation}
be two group elements, both close to $\exp \left[ i \omega^a T^a \right] \in G$ on the group manifold.
The 2-point function of bcc operators takes the form
\begin{equation} \label{eq:wzw_disk_2pt}
    \langle \psi_{g_1 g_2} ( \theta_1 ) \psi_{g_2 g_1} (\theta_2) \rangle_{D^2} = S_{00}^{1/2} 
    \left[ 
    \, 2\mathrm{sin}\left(\frac{\theta_2 - \theta_1}{2}\right) 
    \right]
    ^{-2\Delta_{g_1 g_2}} \,,
\end{equation}
where $\Delta_{g_1 g_2}$ is the scaling dimension.
To compute the 2-point function from the general formula \eqref{eq:WZW_modulated_2}, we let the modulation parameter be
\begin{equation}
    \delta \omega^a (\theta) = \begin{cases}
        \delta \omega_{(1)}^a ~~ \text{if $0 \leq \theta \leq \theta_1$ or $\theta_2 < \theta \leq 2\pi$} \,, \\
        \delta \omega_{(2)}^a ~~ \text{if $\theta_1 < \theta \leq \theta_2$} \,,
    \end{cases}
\end{equation}
where $0 \leq \theta_1 < \theta_2 < 2\pi$.
Fourier modes of the modulation parameter are given by
\begin{align}
\begin{split}
    \delta \omega^a_m 
    &= \frac{1}{2\pi im} \left( 
        e^{im \theta_2} - e^{im \theta_1} 
    \right)  \left( 
        \delta \omega_{(2)}^a - \delta \omega_{(1)}^a
    \right) \,,
\end{split}
\end{align}
for $m \neq 0$.
In \eqref{eq:WZW_modulated_2}, we need to compute
\begin{align}
\begin{split}
    &\sum_{m=1}^{\infty} m \delta \omega^{a}_m \delta \omega^{b}_{-m} \\
    &=\frac{1}{2\pi^2} (\delta \omega_{(2)}^{a} - \delta \omega_{(1)}^{a}) (\delta \omega_{(2)}^{b} - \delta \omega_{(1)}^{b}) \left( 
        \sum_{m=1}^{\infty} \frac{1}{m} - \sum_{m=1}^{\infty} \frac{1}{m} \mathrm{cos} [m (\theta_2 - \theta_1)]
    \right) \,.
\end{split}
\end{align}
The infinite sum $\sum_{m=1}^\infty 1/m$ diverges, and the divergence needs to be absorbed into the wavefunction renormalization of the bcc operator, similar to the compact boson case \cite{Oshikawa_1997}.
Since the procedure is exactly the same as in the compact boson example in Section \ref{sec:bosons}, we do not repeat it here, and the bcc operators are understood to be appropriately renormalized below.

The second infinite sum gives
\begin{equation}
    \sum_{m=1}^{\infty} \frac{1}{m} \mathrm{cos} [m(\theta_2 - \theta_1 )] = - \mathrm{log} \left[ 2 \mathrm{sin}\left( \frac{\theta_2 - \theta_1 }{2} \right) \right] \,,
\end{equation}
which also appeared in the compact boson example.
Therefore, the 2-point function of bcc operators, computed using the general formula \eqref{eq:WZW_modulated_2}, perturbatively up to order $\delta \omega^2$, is given by
\begin{align}
     &S_{00}^{-1/2} \times \langle \psi_{g_1 g_2} ( \theta_1 ) \psi_{g_2 g_1} (\theta_2) \rangle_{D^2}
     \nonumber \\
     &=  1 - \frac{k}{2\pi^2}  \mathrm{log}  \left[ 2 \mathrm{sin}\left( \frac{\theta_2 - \theta_1 }{2} \right) \right] (\delta \omega_{(2)}^{a} - \delta \omega_{(1)}^{a}) (\delta \omega_{(2)}^{b} - \delta \omega_{(1)}^{b}) \left( X^{-2} (e^X - \mathds{1} - X ) 
    \right)^{ba} 
    \nonumber \\
    &\quad 
    + \mathcal{O} ( \delta \omega^3  ) \,,
\end{align}
where $X^{bc} = -\omega^a f^{abc}$ was defined in \eqref{eq:X}.
Comparing this with the general form of the disk 2-point function \eqref{eq:wzw_disk_2pt}, we conclude that the scaling dimension of the bcc operator is given by
\begin{equation} \label{eq:WZW_dimension_pert}
    \Delta_{g_1 g_2} = \frac{k}{4\pi^2} (\delta \omega_{(2)}^{a} - \delta \omega_{(1)}^{a}) (\delta \omega_{(2)}^{b} - \delta \omega_{(1)}^{b}) \left( X^{-2} (e^X - \mathds{1} - X ) 
    \right)^{ba} + \mathcal{O} ( \delta \omega^3  ) \,.
\end{equation}

Let us now compare the perturbative result \eqref{eq:WZW_dimension_pert} with the known scaling dimension of the bcc operators.
To be concrete, we focus on the $\hat{\mathfrak{su}}(2)_1$ WZW model.
The case of arbitrary $\hat{\frak{g}}_k$ can be checked in a similar manner.
For $\hat{\mathfrak{su}}(2)_1$, the exact scaling dimension of the lightest bcc operator was computed in \cite{Gaberdiel:2001xm,Gaberdiel:2001zq}.
It is given by
\begin{equation} \label{eq:su2_dimension}
    \Delta_{g_1 g_2} =  \frac{1}{4\pi^2} \left[\mathrm{cos}^{-1} \left( \frac{1}{2} \mathrm{tr}(g_1^{-1} g_2) \right) \right]^2\,,
\end{equation}
for all $g_1, g_2 \in G$, and the trace is taken in the fundamental representation.
Inserting \eqref{eq:g1g2} for $g_1$ and $g_2$, and then expanding up to second order in $\delta \omega^a$, it becomes
\begin{align} \label{eq:su2_dimension_expanded}
    \Delta_{g_1 g_2} &= \frac{1}{4\pi^2} \left[
        \mathrm{sin}^2 \frac{(\omega^2)^{1/2}}{2} \frac{(\delta\omega_{(2)} - \delta\omega_{(1)})^2}{\omega^2} + \left(
        \frac{1}{4}\omega^2 - \mathrm{sin}^2 \frac{(\omega^2)^{1/2}}{2}
    \right) \frac{[\omega \cdot (\delta\omega_{(2)} - \delta\omega_{(1)})]^2}{(\omega^2)^2}
    \right] 
    \nonumber \\
    &\quad + \mathcal{O}(\delta \omega^3) \,.
\end{align}
For ease of notation, we have abbreviated $\omega^2 \equiv \omega^a \omega^a$, $(\delta\omega_{(2)} - \delta\omega_{(1)})^2 \equiv (\delta\omega_{(2)} - \delta\omega_{(1)})^a (\delta\omega_{(2)} - \delta\omega_{(1)})^a$, and $\omega \cdot (\delta\omega_{(2)} - \delta\omega_{(1)}) \equiv \omega^a (\delta\omega_{(2)} - \delta\omega_{(1)})^a$.

In the perturbative formula \eqref{eq:WZW_dimension_pert} for the scaling dimension, by inserting $k=1$ and $X^{bc} = \omega^a f^{abc} = \sqrt{2}\omega^a \epsilon^{abc}$,\footnote{Our convention for $\mathfrak{su}(2)$ Lie algebra generators is $T^a = \sigma^a /\sqrt{2}$, $a=1,2,3$, and the structure constants are given by $f^{abc} = \sqrt{2} \epsilon^{abc}$.} it is now straightforward to check that we correctly reproduce the known result \eqref{eq:su2_dimension_expanded}.
The conformal dimension of the bcc operator in a general WZW model is given in (\ref{eq:bcc_flow}) of Appendix \ref{app:flow}, and one can confirm that (\ref{eq:WZW_dimension_pert}) holds for general WZW models as well.
This demonstrates the validity of the general perturbative formula \eqref{eq:WZW_modulated_2} for the modulated boundary condition obtained using the doubling trick, and also that we can compute correlation functions of bcc operators by considering appropriately discontinuous modulations of the boundary condition, similar to the previous compact boson example.

\subsubsection{3-point functions}

3-point (and higher point) correlation functions of general bcc operators in WZW models are not known in the literature.
Here, we compute the 3-point function perturbatively using our general formula \eqref{eq:WZW_modulated_2}, and from that deduce the higher Berry connection, defined in \eqref{eq:def_berry_connection}.

Let
\begin{equation} \label{eq:g1g2g3}
    g_1 = \exp \left[ i (\omega^a + \delta \omega_{(1)}^a )  T^a \right] \,, ~~ g_2 = \exp \left[ i (\omega^a + \delta \omega_{(2)}^a )  T^a \right]  \,, ~~ g_3 = \exp \left[ i (\omega^a + \delta \omega_{(3)}^a )  T^a \right]
\end{equation}
be group elements, labeling three different conformal boundary conditions, which are all nearby in the parameter space.
To compute the disk 3-point function of bcc operators,
\begin{align}
\begin{split}
    &\langle \psi_{g_1 g_2} (\theta_1) \psi_{g_2 g_3} (\theta_2) \psi_{g_3 g_1} (\theta_3) \rangle \\
    &= S_{00}^{1/2}
    c(g_1 , g_2 , g_3) \left[ 2\mathrm{sin}\left(\frac{\theta_2 - \theta_1}{2}\right) 
    \right]^{-\Delta_{g_1 g_2 g_3}} \left[ 2\mathrm{sin}\left(\frac{\theta_3 - \theta_2}{2}\right) 
    \right]^{-\Delta_{g_2 g_3 g_1}} \left[ 2\mathrm{sin}\left(\frac{\theta_3 - \theta_1}{2}\right) \right]^{-\Delta_{g_3 g_1 g_2}} \,,
\end{split}
\end{align}
where $\Delta_{g_i g_j g_k} = \Delta_{g_i g_j} + \Delta_{g_j g_k} - \Delta_{g_k g_i}$, from the general formula \eqref{eq:WZW_modulated_2}, we take the modulation parameter to be
\begin{equation}
    \delta \omega^a (\theta) = \begin{cases}
        \delta \omega_{(1)}^a ~~ \text{if $0 \leq \theta \leq \theta_1$ or $\theta_3 < \theta \leq 2\pi$} \,, \\
        \delta \omega_{(2)}^a ~~ \text{if $\theta_1 < \theta \leq \theta_2$} \,, \\
        \delta \omega_{(3)}^a ~~ \text{if $\theta_2 < \theta \leq \theta_3$} \,,
    \end{cases}
\end{equation}
where $0 \leq \theta_1 < \theta_2 < \theta_3 < 2\pi$.
Fourier modes of the modulation parameter are
\begin{align}
\begin{split}
    \delta \omega^a_m 
    &= \frac{1}{2\pi im} \left( 
        \delta \omega_{(13)}^a e^{im\theta_1} + \delta \omega_{(21)}^a e^{im\theta_2} + \delta \omega_{(32)}^a e^{im\theta_3}
    \right) \,,
\end{split}
\end{align}
where $\delta \omega^a_{(ij)} \equiv \delta \omega^a_{(i)} - \delta \omega^a_{(j)}$, and $m\neq 0$.
We then compute
\begin{align} \label{eq:wzw_3pt_fourier_sum}
    &(2\pi)^2\sum_{m=1}^{\infty} m \delta \omega_m^{a} \delta \omega_{-m}^{b}
    \nonumber \\
    &=  (\delta \omega_{(31)}^{a} \delta \omega_{(31)}^{b} + \delta \omega_{(12)}^{a} \delta \omega_{(12)}^{b} + \delta \omega_{(23)}^{a} \delta \omega_{(23)}^{b}  ) \sum_{m=1}^{\infty} \frac{1}{m} - \frac{i\pi}{2}( \delta \omega_{(1)}^{[a} \delta \omega_{(2)}^{b]} + \delta \omega_{(2)}^{[a} \delta \omega_{(3)}^{b]} + \delta \omega_{(3)}^{[a} \delta \omega_{(1)}^{b]} )
    \nonumber \\
    &~~~~ + 2\delta \omega_{(1)}^{a} \delta \omega_{(1)}^{b} \mathrm{log}\left[ 2 \mathrm{sin}\left( \frac{\theta_2 -\theta_1}{2} \right) \right] + 2\delta \omega_{(2)}^{a} \delta \omega_{(2)}^{b} \mathrm{log}\left[ 2 \mathrm{sin}\left( \frac{\theta_3 -\theta_2}{2} \right) \right] 
    \nonumber \\
    &~~~~+ 2\delta \omega_{(3)}^{a} \delta \omega_{(3)}^{b} \mathrm{log}\left[ 2 \mathrm{sin}\left( \frac{\theta_3 -\theta_1}{2} \right) \right] \,.
\end{align}
Here, $\delta \omega^{[a}_{(i)} \delta \omega^{b]}_{(j)} \equiv \delta \omega^{a}_{(i)} \delta \omega^{b}_{(j)} - \delta \omega^{a}_{(j)} \delta \omega^{b}_{(i)}$.
The divergent sum $\sum_{m=1}^\infty 1/m$ is again absorbed into the renormalization of bcc operators.
The chord distance dependent terms give rise to the usual conformal structure of the 3-point correlation function of boundary primary operators.
On the other hand, the purely imaginary term comes from the expansion of the OPE coefficient, since it is independent of $\theta$'s.

Specifically, by inserting \eqref{eq:wzw_3pt_fourier_sum} into \eqref{eq:WZW_modulated_2}, we obtain the perturbative expansion of the OPE coefficient of bcc operators,
\begin{align} \label{eq:WZW_OPE}
     c(g_1 , g_2 , g_3) 
     &=  1 + \frac{ik}{8\pi} ( \delta \omega_{(1)}^{[a} \delta \omega_{(2)}^{b]} + \delta \omega_{(2)}^{[a} \delta \omega_{(3)}^{b]} + \delta \omega_{(3)}^{[a} \delta \omega_{(1)}^{b]} ) \left( X^{-2} (e^X - \mathds{1} - X ) 
    \right)^{ba} 
    \nonumber \\
    &
    \quad + \mathcal{O}(\delta \omega^3) \,.
\end{align}
This gives us enough information to (locally) determine the higher Berry connection in the space of conformal boundary conditions \eqref{eq:WZW_parameter_space}.

\subsubsection{Higher Berry connection and curvature}

The family of boundary conditions \eqref{eq:WZW_parameter_space}, as we briefly mentioned, corresponds to Dirichlet boundary conditions for the sigma model field valued in the group manifold $G$. 
The boundary parameter space in this case may be identified with the target space $G$ of the sigma model.
It is therefore natural to expect that the space of boundary conditions has a nontrivial gerbe structure \cite{Alvarez:1984es,gawkedzki1988topological,Kapustin:1999di,Freed:1999vc,Carey:2002xp,Carey:2004xt,Schreiber:2005mi,Waldorf:2006zz,runkel2008gerbeholonomysurfacesdefectnetworks},
with the curvature given by the 
WZ term,
\begin{equation} \label{eq:WZW_term}
    {\cal H}_{\mathrm{WZ}} = \frac{k}{12\pi} \mathrm{tr} \left(
        g^{-1} dg \wedge g^{-1} dg  \wedge g^{-1} dg 
    \right) \,.
\end{equation}
We now verify that this is indeed the case.

From the general definition given in \eqref{eq:def_berry_connection}, and the OPE coefficient \eqref{eq:WZW_OPE} of the bcc operators, we obtain the higher Berry connection
\begin{equation}
    \mathcal{B} = 
    \frac{1}{2} \mathcal{B}_{ab} (\omega) d\omega^a \wedge d\omega^b = \frac{k}{4\pi}\left( X^{-2} (e^X - \mathds{1} - X ) 
    \right)^{ba} d\omega^a \wedge d\omega^b \,.
\end{equation}
Note that from the definition $X^{bc} = - \omega^a f^{abc}$, it follows that even powers of $X$ are symmetric, and odd powers of $X$ are anti-symmetric.
Therefore, we can also write
\begin{align} \label{eq:B_wzw}
\begin{split}
    \mathcal{B} &
    = \frac{k}{4\pi} \sum_{r=2}^{\infty} \frac{(X^{r-2})^{ba}}{r!} d\omega^a \wedge d\omega^b 
    = \frac{k}{4\pi} \sum_{r=1}^{\infty} \frac{(X^{2r-1})^{ba}}{(2r+1)!} d\omega^a \wedge d\omega^b
    \\
    &= \frac{k}{4\pi} \left(X^{-2}(\mathrm{sin}X -  X ) \right)^{ba} d\omega^a \wedge d\omega^b \,.
\end{split}
\end{align}

Now, we claim that the corresponding higher Berry curvature 
$d{\cal B}$
is given by ${\cal H}_{\mathrm{WZ}}$ in
\eqref{eq:WZW_term}.
We will explicitly show this for the case of $\hat{\mathfrak{su}}(2)_k$.
By inserting $X^{bc} = - \sqrt{2} \omega^a \epsilon^{abc}$ in \eqref{eq:B_wzw}, we obtain
\begin{equation}
    \mathcal{B} = \frac{k}{8\pi (\omega \cdot \omega)^{3/2}} \left[ 
        \mathrm{sin}(\sqrt{2\omega \cdot \omega}) - \sqrt{2\omega \cdot \omega}
    \right] \omega^c \epsilon^{cba} d\omega^a \wedge d\omega^b \,,
\end{equation}
where $\omega \cdot \omega \equiv \omega^a \omega^a$.
Differentiating, we get
\begin{equation} \label{eq:WZW_dB}
    \mathcal{H} = d\mathcal{B} = \frac{k}{\sqrt{2}\pi \omega \cdot \omega} \mathrm{sin}^2 \left( \sqrt{\frac{\omega \cdot \omega}{2}} \right) d\omega^1 \wedge d\omega^2 \wedge d\omega^3 \,.
\end{equation}
Let us compare this with the WZ term
\eqref{eq:WZW_term}.
The Lie algebra generators in our convention are given by $T^a = \sigma^a /\sqrt{2}$, and we have
\begin{equation}
    g = \exp \left[ i \omega^a T^a \right] = \mathrm{cos} \left( \sqrt{\frac{\omega \cdot \omega}{2}} \right) + i\frac{\omega^a \sigma^a}{\sqrt{\omega \cdot \omega}} \mathrm{sin} \left( \sqrt{\frac{\omega \cdot \omega}{2}} \right)\, .
\end{equation}
Plugging this into \eqref{eq:WZW_term}, 
we confirm that ${\cal H}_{\mathrm{WZ}}$ agrees with
\eqref{eq:WZW_dB}.
Therefore, we see that the higher Berry curvature in the space of boundary conditions \eqref{eq:WZW_parameter_space} is given by the WZ term, as anticipated.
Here, the higher Berry connection again coincides with the Kalb-Ramond $B$-field.

\subsubsection{Other boundaries in $\hat{\mathfrak{su}}(2)_k$} \label{sec:su2_more}

So far, we considered the family of boundary states which arises from the vacuum Cardy state \eqref{eq:vacuum_Cardy} by acting with the left-moving $G$ global symmetry of the WZW model.
As we mentioned, these boundary conditions correspond to the Dirichlet boundary conditions of the $G$-valued sigma model field.
Let us briefly comment on other families of boundary conditions, where we start from a generic Cardy state and act by the left-moving $G$ symmetry. 
The analysis is essentially unchanged.
For concreteness, we focus on the (diagonal) $\hat{\mathfrak{su}}(2)_k$ WZW models.\footnote{Various other conformal boundary conditions of the $\hat{\mathfrak{su}}(2)_k$ WZW models, beyond the ones we consider in this work, are discussed in \cite{Maldacena:2001ky,Blakeley:2007gu} (see also \cite{Kudrna:2021rzd}).}

There are $k+1$ Cardy boundary states in the $\hat{\mathfrak{su}}(2)_k$ WZW model, labeled by the $\mathrm{SU}(2)$ spin $J = 0, \frac{1}{2}, \cdots , \frac{k}{2}$.
Let us denote them as $\ket{B_0 ; J}$, where $\ket{B_0 ; J=0} \equiv \ket{B_0}$ was given in \eqref{eq:vacuum_Cardy}.
Explicitly, we have
\begin{equation}
    \ket{B_0 ; J} = \sum_{K} \frac{S_{JK}}{\sqrt{S_{0K}}} | K \rangle\! \rangle \,,
\end{equation}
where
\begin{equation}
    S_{JK} = \sqrt{\frac{2}{k+2}} \mathrm{sin} \frac{(2J+1)(2K+1)\pi}{k+2}
\end{equation}
is the S-matrix, and $| K \rangle \!\rangle$ is the spin-$\frac{K}{2}$ Ishibashi state.
We then obtain families of boundary states by acting with the symmetry operator $U_g$ in \eqref{eq:U_g},
\begin{equation} \label{eq:B_g_su2}
    \ket{B_g ; J} = U_g \ket{B_0 ; J} \,, \quad g \in G \,.
\end{equation}
Not all pairs of $g$ and $J$ correspond to different boundary conditions.
Instead, there is an identification \cite{Kudrna:2021rzd},
\begin{equation} \label{eq:su2_identification}
    \ket{B_{-g} ; J} = \ket{B_g ; \frac{k}{2} - J} \,.
\end{equation}
This is because the element $-\mathds{1}_{2\times 2} \in \mathrm{SU}(2)$ in the center of the left-moving $\mathrm{SU}(2)$ symmetry corresponds to the $\mathbb{Z}_2$ Verlinde line of the $\hat{\mathfrak{su}}(2)_k$ WZW model, which preserves the current algebra and maps the spin-$J$ Cardy state to the spin-$(\frac{k}{2} - J)$ Cardy state.

For the case of $J = 0$, the family of boundary states \eqref{eq:B_g_su2} forms the $\mathrm{SU}(2) \cong S^3$ group manifold, as we discussed.
The spin-$\frac{k}{2}$ Cardy state is also included inside this boundary conformal manifold due to the identification \eqref{eq:su2_identification}.
More generally, starting from any $J \neq \frac{k}{4}$ Cardy state and acting with the left-moving $\mathrm{SU}(2)$ symmetry, we generate the $\mathrm{SU}(2)$ group manifold, which includes the spin-$(\frac{k}{2}-J)$ Cardy state.

On the other hand, when $k$ is even, the spin-$\frac{k}{4}$ Cardy state is invariant under the action of the $\mathbb{Z}_2$ Verlinde line, i.e. the center of the left-moving $\mathrm{SU}(2)$ symmetry.
It is not invariant under any larger subgroup of $\mathrm{SU}(2)$, since the boundary conditions $\ket{B_g , J}$ satisfy the twisted gluing condition \eqref{eq:twisted_gluing} which depends on the adjoint action of $g \in \mathrm{SU}(2)$ on the current algebra.
Therefore, the family \eqref{eq:B_g_su2} in the case of $J = \frac{k}{4}$ generates a boundary conformal manifold given by $\mathrm{SU}(2)/\mathbb{Z}_2 = \mathrm{SO}(3) \cong \mathbb{RP}^3$.

Summarizing, the moduli space of conformal boundary conditions in the $\hat{\mathfrak{su}}(2)_k$ WZW model includes
\begin{equation} \label{eq:su2_moduli_space}
\begin{cases}
    (S^3)^{\sqcup \frac{k+1}{2}} & \text{if $k$ is odd} \,,\\
    (S^3)^{\sqcup \frac{k}{2}} \sqcup \mathbb{RP}^3 & \text{if $k$ is even} \,.
\end{cases}
\end{equation}
We note that different connected components are not parts of a bigger, connected moduli space, and they are truly disconnected. 
This can be seen from the fact that the $g$-functions are different for different connected components.
Specifically, the $g$-function of the family of boundaries \eqref{eq:B_g_su2} is given by
\begin{equation}
    g_J = \left( \frac{2}{k+2} \right)^{1/4} \frac{\mathrm{sin} \frac{(2J+1)\pi}{k+2}}{\mathrm{sin}^{1/2} \frac{\pi}{k+2}} \,.
\end{equation}
Note that $g_J = g_{\frac{k}{2} -J}$, which is consistent with the fact that the spin-$J$ and spin-$(\frac{k}{2} - J)$ Cardy states belong to the same boundary conformal manifold.

The computation of the correlation functions of bcc operators for the families of boundary conditions \eqref{eq:B_g_su2}, using the modulated boundary condition and doubling trick, is almost identical to the previous case with $J=0$.
The only difference is that, in the doubled, Chern-Simons theory picture (see Figure \ref{fig:doubling}), we have an additional spin-$J$ Wilson line of the $\mathrm{SU}(2)_k$ Chern-Simons theory which wraps around the equator of the boundary $S^2$ \cite{Kapustin:2010if}.\footnote{The Wilson line operator can be either put slightly below or above the modulated line operator in Figure \ref{fig:doubling}. This is because, unlike the modulated line operator which is stuck on the boundary $\partial B^3 = S^2$, the Wilson line lives in the bulk and can be freely moved into the bulk. From the original (1+1)d CFT picture before the doubling, the Wilson line corresponds to a topological Verlinde line, which commutes with both the left- and right-moving current operators, and therefore also with the modulated line operator.}
This additional Wilson line can be shrunk to, say, the south pole of the $S^2$, which gives an additional factor of the quantum dimension of the Wilson line.
This accounts for the fact that the $g$-function of the spin-$J$ Cardy state is the $g$-function of the spin-0 Cardy state multiplied by the quantum dimension of the spin-$J$ Wilson line.
In our normalization convention, this does not change the OPE coefficient of the bcc operators.
Therefore, we conclude that each connected component of the boundary moduli space in \eqref{eq:su2_moduli_space} has a nontrivial higher Berry curvature, given by the $\mathrm{SU}(2)_k$ WZ term.

Let us briefly comment on the quantization of the higher Berry curvature.
Although in this work we do not provide a general proof that the flux of higher Berry curvature must be quantized, on general grounds we expect it to hold, at least when the boundary moduli space is globally a smooth manifold, which is the case here.\footnote{In Section \ref{sec:non-chiral}, we discuss an example of a singular boundary conformal manifold which hosts a fractional higher Berry curvature.}
In particular, for each disconnected component in \eqref{eq:su2_moduli_space}, we have
\begin{equation} \label{eq:flux_su2}
    \mathbb{Z} \ni \frac{1}{2\pi}\int_{M} \mathcal{H}_{\mathrm{WZ}} = \begin{cases}
        k & \text{for $M = S^3$ and $k\in\mathbb{Z}_{\geq 1}$} \,,\\
        \frac{k}{2} & \text{for $M = \mathbb{RP}^3$ and $k\in2\mathbb{Z}_{\geq 1}$} \,.
    \end{cases}
\end{equation}
Here, $\mathcal{H}_{\mathrm{WZ}}$ is the higher Berry curvature which is given by the $\mathrm{SU}(2)_k$ WZ term. 
We used the fact that the WZ term in this case is equal to $k$ times the volume form on $S^3$.
When we integrate over $\mathbb{RP}^3$, we get a factor of $\frac{1}{2}$ due to the $\mathbb{Z}_2$ identification.
In \eqref{eq:su2_moduli_space}, note that $\mathbb{RP}^3$ appears as a disconnected component of the boundary conformal manifold if and only if $k$ is even, whereas $S^3$ exists for all $k$.
Therefore, we see that the flux \eqref{eq:flux_su2} of the higher Berry curvature is always an integer.

\subsection{Free fermions} \label{sec:fermions}

As a special case of the general WZW model discussion in the previous subsection, we briefly mention the example with multiple flavors of free Majorana fermions.
Consider $N$ copies of left- and right-moving (1+1)d Majorana fermions,
\begin{equation} \label{eq:fermion_action}
    S= \int dx dt \sum_{\alpha=1}^N \left[ 
        \frac{i}{2} \chi_L^\alpha (\partial_{t} + \partial_x) \chi_L^\alpha + \frac{i}{2} \chi_R^\alpha (\partial_{t} - \partial_x) \chi_R^\alpha
    \right] \,.
\end{equation}
The theory admits an $\mathrm{O}(N)$ family of conformal boundary conditions, given by
\begin{equation} \label{eq:fermion_ON}
    \chi_L^{\alpha} | = M_{\alpha \beta} \chi_R^{\beta}| \,,
\end{equation}
for an arbitrary $\mathrm{O}(N)$ matrix $M$.
We focus only on the identity component $\mathrm{SO}(N) \subset \mathrm{O}(N)$.

The theory enjoys the $\mathrm{O}(N) \times \mathrm{O}(N)$ global symmetry, acting on left- and right-moving fermions.
The left-moving current is given by
\begin{equation}
    J_L^{\alpha \beta} = 2\pi \chi^\alpha_L \chi_L^\beta \,,
\end{equation}
and it generates the $\hat{\mathfrak{so}}(N)_1$ current algebra.
The family of boundary conditions \eqref{eq:fermion_ON} completely breaks the left-moving (and also the right-moving) $\mathrm{O}(N)$ global symmetry, and they get mapped into each other under the $\mathrm{O}(N)$ action, similar to \eqref{eq:WZW_parameter_space} in the general WZW model case.

Correlation functions of bcc operators can again be computed using the doubling trick.
The non-chiral theory \eqref{eq:fermion_action} on the upper half-plane is mapped to a chiral theory by the standard relation
\begin{equation}
    \chi^\alpha (z) = \begin{cases}
        \chi_L^\alpha (z) ~~~&\text{if $\mathrm{Im}z > 0$} \,, \\
        \chi_R^\alpha (\bar{z}) ~~~&\text{if $\mathrm{Im}z < 0$} \,. 
\end{cases}
\end{equation}
The disk partition function of the non-chiral theory, with a modulated boundary condition with a small modulation parameter $\delta \omega^{\alpha \beta}$, corresponds to the sphere partition function of the chiral theory with a modulated line operator inserted along the equator,
\begin{equation}
\label{eq: fermion}
    {}_{NS} \langle 0 | \exp \left[i \int_{0}^{2\pi} d\theta (\omega^{\alpha \beta} + \delta \omega^{\alpha \beta} (\theta)) (2\pi \chi^\alpha \chi^\beta) (\theta)\right] | 0 \rangle_{NS} \,.
\end{equation}
Here, $\ket{0}_{NS}$ is the NS-sector vacuum state of the $N$ chiral Majorana fermions.
Correlation functions of bcc operators are obtained by letting the modulation parameter $\delta \omega^{\alpha \beta}$ have step-function-like discontinuities as before.
The computation is exactly the same as the general WZW model case, and we omit the details.
We conclude that the space of free fermion boundary conditions \eqref{eq:fermion_ON} has a higher Berry curvature given by the $\mathrm{SO}(N)_1$ WZ term, which is nontrivial for $N\geq 3$.\footnote{Unlike in the bosonic case, the $\mathrm{SO}(N)_1$ WZ term for the fermions defines an element in the appropriate Anderson dual of the spin bordism group. 
See \cite{Lee:2020ojw} for more details.
} 

\subsection{Non-chiral deformation} \label{sec:non-chiral}

There are cases where self-local boundary operators of conformal dimension one exist  
that do not come from bulk (local) current operators,  
and the exactly marginal deformations induced by such operators  
were called non-chiral deformations in \cite{Recknagel:1998ih}.
Here, we consider such non-chiral deformations in $c=1$ free boson CFTs that were studied in \cite{Callan:1993mw,Callan:1994ub}, and the higher Berry phase of the corresponding boundary conformal manifolds.

\paragraph{Boundary conformal manifolds at $c=1$}

To begin with, let us briefly recall the overall structure of boundary conformal manifolds of the compact boson CFTs at different compactification radii.
First, the free boson CFT at the self-dual radius $R=1$ exhibits an enhanced current algebra $\hat{\mathfrak{su}}(2)_1$. 
This enhancement gives rise to new exactly marginal operators on, say, a Neumann boundary condition which come from the restriction of $\mathrm{SU(2)}$ current operators to the boundary.  
By rotating the Neumann boundary codition with such exactly marginal deformations,  
one can reach any of the Dirichlet boundary conditions.  
The boundary states of the compact free boson at the self-dual radius are classified in \cite{Gaberdiel:2001xm},
and the full moduli space is given by the group manifold $\mathrm{SU}(2)$.

More generally, when the compactification radius $R = p/q \in \mathbb{Q}$ is a rational number for positive coprime integers $p$ and $q$, it was shown in  
\cite{Gaberdiel:2001zq} that the moduli space of boundary conditions includes
\begin{equation} \label{eq:orbifold}
    M = \frac{\mathrm{SU}(2)}{\mathbb{Z}_p \times \mathbb{Z}_q} \,.
\end{equation}
The $\mathbb{Z}_p \times \mathbb{Z}_q$ quotient is according to the identifications
\begin{equation} \label{eq:SU2_quotient}
    g = \begin{pmatrix}
        a & b \\
        -b^* & a^*
    \end{pmatrix} \in \mathrm{SU}(2) \,, \quad a \sim a \exp \left(\frac{2\pi i}{q}\right) \,, \quad b \sim b \exp \left(\frac{2\pi i}{p}\right) \,,
\end{equation}
where $|a|^2 + |b|^2 = 1$.\footnote{The orifold family of boundary conditions \eqref{eq:orbifold} can also be understood as coming from the $\mathrm{SU}(2)$ family at the self-dual radius $R=1$ upon gauging the $\mathbb{Z}_p$ subgroup of the winding symmetry and the $\mathbb{Z}_q$ subgroup of the momentum symmetry \cite{Collier:2021ngi}, which maps the bulk theory to be at radius $R= p/q$.}
Importantly, the action of $\mathbb{Z}_p \times \mathbb{Z}_q$ on $\mathrm{SU}(2)$ has fixed points, given by either $a=0$ or $b=0$, and the orbifold \eqref{eq:orbifold} possesses conical singularities.
For instance, along the one-dimensional locus $b=0$ inside the moduli space $M$, we have a conical singularity with deficit angle $2\pi ( 1 - 1/p)$, and the corresponding boundary conditions are superpositions of $p$ Dirichlet boundary conditions.
Similarly, the locus $a=0$ defines another conical singularity, and they correspond to superpositions of $q$ Neumann boundary conditions.
In particular, we observe that along the singular locus in $M$, the boundary conditions are non-simple, whereas away from singularities the boundary conditions are always simple.
In \cite{Gaberdiel:2001zq}, it was claimed that the boundary conditions in \eqref{eq:orbifold} with $ab \neq 0$, together with the conventional Dirichlet and Neumann boundary conditions, give the complete classification of simple boundary conditions of the compact free boson theory at rational radii.

The explicit boundary states can be obtained using the ``hidden'' $\mathrm{SU}(2)$ symmetry, which we will explain more about momentarily from a perspective slightly different from the original work \cite{Callan:1994ub}.
By using this hidden symmetry,  
one finds that the solution to the bootstrap equation  
can be obtained as a projection of the boundary state in the $\hat{\mathfrak{su}}(2)_1$ WZW model.
In the decompactification limit,
the boundary state can be expressed as \cite{Gaberdiel:2001zq}
\begin{equation}\label{eq:Dg}
\ket{B_g} = \sum_{j \in \fr{1}{2}\bb{Z}_+} \sum^j_{m=-j} D^j_{m,-m}(g) \kett{j,m,m}\, ,
\end{equation}
where $\kett{j,m,m}$ is the Virasoro Ishibashi state, and we define the matrix in terms of the $\mathrm{SU}(2)$ representation as
\begin{equation}
D^j_{m,-m}(g) = \braket{j,m|g|j,-m}\, .
\end{equation}
The explicit form of the matrix $D$ is given by
\begin{equation}
D^j_{m,n}(g) =
\sum_{l = \max(0, n - m)}^{\min(j - m, j + n)}
\frac{\left[(j + m)! (j - m)! (j + n)! (j - n)!\right]^{\frac{1}{2}}}
{(j - m - l)! (j + n - l)! l! (m - n + l)!}
a^{j + n - l} (a^*)^{j - m - l} b^l (-b^*)^{m - n + l}\,,
\end{equation}
where $a$ and $b$ are matrix elements of $g \in \mathrm{SU}(2)$ given in \eqref{eq:SU2_quotient}.

For the case of finite radius $R = p/q \in \mathbb{Q}$,
the spectrum includes only states with the momentum number $n$ and winding number $m$ satisfying
\begin{equation}
m - n \in p \mathbb{Z} \quad \wedge \quad m + n \in q \mathbb{Z}\, .
\end{equation}
The linear combinations of possible Ishibashi states such that the interval Hilbert space has the desired properties can be obtained as
\begin{equation}
\ket{B_g} = 2^{-\fr{1}{4}}\sqrt{pq} \sum_{j\in \fr{1}{2}\bb{Z}_+} \sum_{m,n \in \bb{Z}}D^j_{m,n}\pa{P^+_q P^-_p(g)}\kett{j,m,n}\, ,
\end{equation}
where $P^-_p$ is the projector onto the subspace satisfying $m - n \in p \mathbb{Z}$,  
and $P^+_q$ is the projector onto the subspace satisfying $m + n \in q \mathbb{Z}$.  
This can be written in the following explicit form,
\begin{equation}
D^j_{m,n}\pa{P^+_q P^-_p(g)} = \fr{1}{pq}\sum_{l=0}^{p-1}\sum_{k=0}^{q-1}D_{m,n}^j\pa{\Gamma^k_q \Gamma_p^l g \Gamma_p^{-l} \Gamma_q^k}\, ,
\end{equation}
where
\begin{equation}
\Gamma_p =
  \left(
    \begin{array}{cc}
      e^{\fr{\pi i}{p}} & 0  \\
      0 & e^{\fr{-\pi i}{p}}  \\
    \end{array}
  \right)
  \in \mathrm{SU}(2)\, ,
\end{equation}
and similarly for $\Gamma_q$.

Finally, at irrational radii, besides the usual Dirichlet and Neumann boundary conditions, there exists a one-parameter family of the so-called Friedan boundary conditions \cite{Janik:2001hb}, which are non-compact (i.e., have continuous spectrum in the interval Hilbert space).
We will not study them in this paper, but it will be interesting to analyze the higher Berry phase of non-compact boundary conditions as well.

\paragraph{``Non-chiral'' deformations}

In the following, we focus on the case where the radius 
\begin{equation}
    R = N \in \mathbb{N}
\end{equation}
is a positive integer, and discuss the higher Berry connection and curvature on the boundary conformal manifold $M = \mathrm{SU}(2)/\mathbb{Z}_N$.\footnote{When $R=2$, this quotient is different from $\mathrm{SO}(3) = \mathrm{SU}(2)/\mathbb{Z}_2$. See \eqref{eq:SU2_quotient}.}
The boundary exactly marginal deformations in this case were studied in \cite{Recknagel:1998ih,Callan:1993mw,Callan:1994ub}.
Starting from a Neumann boundary condition, where the dual boson is set to zero, $\widehat{X}| = 0$, the boundary exactly marginal operators are given by\footnote{In our convention, the boundary vertex operator $e^{inX}$ on a Neumann boundary condition has the conformal dimension $\Delta = \frac{n^2}{R^2}$.}
\begin{equation} \label{eq:nonchiral_psi}
\psi_1(x) = \sqrt{2} \cos \pa{NX(x)}\, , \quad
\psi_2(x) = \sqrt{2} \sin \pa{NX(x)}\, , \quad
J(x) = \sqrt{2}i N \partial X (x) \,.
\end{equation}
The deformation by $\psi_1$ and $\psi_2$ preserves the Virasoro symmetry,  
but generally breaks the $\mathfrak{u}(1)$ symmetry at the boundary $z=\bar{z}=x$ as follows 
\cite{Callan:1994ub,Recknagel:1998ih},
\begin{equation} \label{eq:nonchiral_gluing}
\begin{aligned}
J(z) &= \sin (\sqrt{2}\xi) \psi_2(x)+\cos(\sqrt{2} \xi) \bar{J}(\bar{z})\, , \ \ \ \ \ \text{for $\psi_1$ deformation with the coupling $\xi$} \,,\\
J(z) &= -\sin (\sqrt{2}\xi) \psi_1(x)+\cos(\sqrt{2} \xi) \bar{J}(\bar{z})\, , \ \ \ \text{for $\psi_2$ deformation with the coupling $\xi$} \,.
\end{aligned}
\end{equation}

In general, on a Neumann boundary condition, the chiral and anti-chiral $\hat{\mathfrak{u}}(1)$ currents satisfy the gluing condition $J(z) = \bar{J}(\bar{z})$, whereas on a Dirichlet boundary condition, they satisfy $J(z) = -\bar{J}(\bar{z})$.
From this perspective, one sees that the non-chiral deformation \eqref{eq:nonchiral_gluing} can rotate the Neumann point $\xi = 0$ into a Dirichlet-like point $\sqrt{2}\xi = \pi n$ for an odd integer $n$.
However, this Dirichlet-like boundary differs from the usual, simple Dirichlet boundary. 
It corresponds to a superposition of $R =N \in \mathbb{N}$ copies of Dirichlet boundary conditions, as we mentioned earlier, which corresponds to the singular locus in the moduli space \eqref{eq:orbifold} with $b=0$.
This can also be understood from the fact that exactly marginal deformations do not change the $g$-function or the boundary entropy.
At an arbitrary radius, the $g$-function of a Neumann boundary condition is $R$ times that of a Dirichlet boundary condition.

\paragraph{Hidden $\mathrm{SU}(2)$ symmetry}

The boundary exactly marginal operators \eqref{eq:nonchiral_psi} generate an $\hat{\mathfrak{su}}(2)_1$ current algebra on the Neumann boundary condition \cite{Callan:1994ub}, even though the bulk of the CFT does not have such a current algebra for $R \neq 1$.
Let us explain the bulk origin of this boundary $\hat{\mathfrak{su}}(2)_1$ current algebra, or the ``hidden symmetry.''
We will see that, even though the marginal deformations $\psi_1$ and $\psi_2$ in \eqref{eq:nonchiral_psi} were referred to as being non-chiral in \cite{Recknagel:1998ih}, they still originate from bulk chiral current operators, which however are ``hidden'' in the sense that they live at the endpoints of topological defect lines.
We focus on the left-moving current algebra, but one can similarly show that there exists a hidden right-moving current algebra. 
They are identified at the boundary.

First, recall that the conformal dimension of the bulk vertex operator (normal-ordering is implicit)
\begin{equation}
    V_{n,w} (z, \bar{z}) =  e^{inX + iw \widehat{X}}
\end{equation}
is given by
\begin{equation}
    (h, \bar{h}) = \left( 
        \frac{1}{4}\left(\frac{n}{R} + wR\right)^2 , \frac{1}{4}\left(\frac{n}{R} - wR\right)^2 
    \right) \,.
\end{equation}
If both $n$ and $w$ are integers, then $V_{n,w}(z,\bar{z})$ is a genuine local operator.
If either of them is not an integer, then $V_{n,w} (z, \bar{z})$ is a non-local defect operator living at the end of a topological line.

For $R=N\in\mathbb{N}$, the two vertex operators with $n = \pm N$, $w= \pm 1/N$ become holomorphic currents.
That is,
\begin{equation}
    V_{\pm N, \pm \frac{1}{N}} (z) =  e^{\pm (iNX + i\frac{\widehat{X}}{N}) } ~~~~\text{has} ~~~~ (h, \bar{h}) = (1,0) \,.
\end{equation}
(We can similarly find anti-holomorphic current operators $V_{\pm N, \mp \frac{1}{N}} (\bar{z})$.)
Combined with $\partial X (z)$, they generate the $\mathfrak{su}(2)_1$ current algebra,
\begin{equation}
    J^3 (z) = J(z) = \sqrt{2}i N \partial X (z) \,, \quad J^{\pm} (z) = V_{\pm N, \pm \frac{1}{N}} (z) \,,
\end{equation}
with OPE
\begin{align} \label{eq:nonchiral_su2}
\begin{split}
    J^3 (z) J^3 (w) &\sim \frac{1}{(z-w)^2} \,,\\
    J^+ (z) J^- (w) &\sim \frac{1}{(z-w)^2} + \frac{\sqrt{2} J^3 (w)}{(z-w)} \,,\\
    J^3 (z) J^\pm (w) &\sim \frac{\pm \sqrt{2}J^\pm (w)}{(z-w)}  
    \,.
\end{split}
\end{align}

\begin{figure}[t!]
\centering
\includegraphics[width=.8\textwidth]{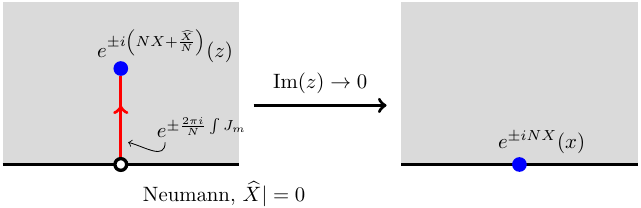}
\caption{Bulk holomorphic operators $e^{\pm i \left( NX + \frac{\widehat{X}}{N} \right)}(z)$ attached to $\mathbb{Z}_N$ momentum symmetry lines are brought to the Neumann boundary condition and become the boundary exactly marginal operators $e^{\pm iNX} (x)$. The topological lines generating the momentum symmetry (shown in red) can topologically end on the Neumann boundary condition.
}\label{fig:nonchiral}
\end{figure}

However, the current algebra \eqref{eq:nonchiral_su2} is not a symmetry of the bulk CFT, unless $R=N = 1$ (which is the self-dual radius).
This is because the vertex operators $J^\pm (z) = V_{\pm N, \pm \frac{1}{N}} (z)$, having fractional winding numbers $w = \pm 1/N$, are not genuine local operators in the bulk.
Instead, they are attached to topological lines corresponding to elements of the $\mathbb{Z}_N$ subgroup of the momentum symmetry.
Schematically,
\begin{equation} \label{eq:branch_cut}
    V_{\pm N, \pm \frac{1}{N}} (z) \times \exp \left( 
        \pm \frac{2\pi i}{N} \int_z  J_m
    \right)
\end{equation}
are well-defined objects inside the bulk of the CFT, where $J_m =  \frac{-iR^2}{2\pi} \star dX$ is the Noether current of the momentum symmetry.
Put differently, the topological line is the branch cut due to the fractional winding number.

Now consider bringing the bulk (non-local) current operators \eqref{eq:branch_cut} to a Neumann boundary condition $\ket{B_0}$.
First of all, since the Neumann boundary condition preserves the momentum symmetry, the topological line $\exp \left[ \pm \frac{2\pi i}{N} \int_z  J_m \right]$ is absorbed by the boundary and disappears.
Next, on the Neumann boundary condition, the dual boson $\widehat{X}$ is set to zero in $V_{\pm N, \pm \frac{1}{N}} (z) =  e^{\pm (iNX + i\frac{\widehat{X}}{N}) }$.
Finally, since the operators \eqref{eq:branch_cut} are holomorphic, the bulk-to-boundary OPE is regular.
Putting all things together, we have
\begin{equation}
    \lim_{\mathrm{Im}(z) \rightarrow 0^+} \left[ 
        V_{\pm N, \pm \frac{1}{N}} (z) \times \exp \left( 
        \pm \frac{2\pi i}{N} \int_z  J_m
    \right)
    \right] \ket{B_0} = e^{\pm iN X (x)} \ket{B_0} \,,
\end{equation}
where $x = \mathrm{Re}(z)$.
In other words, the bulk non-local defect operators \eqref{eq:branch_cut} become the exactly marginal operators $e^{\pm iN X (x)}$ when brought to the Neumann boundary condition.
See Figure \ref{fig:nonchiral}.
The vertex operators $V_{\pm N, \pm \frac{1}{N}} (z)$, having momentum quantum numbers $\pm N$, are neutral under the $\mathbb{Z}_N$ subgroup of the momentum symmetry.
Therefore, the corresponding boundary operators $e^{\pm iNX}$ (together with $\partial X$) are self-local \cite{Recknagel:1998ih}.

Then, on the boundary, the three operators $\partial X (x)$, $e^{\pm i N X (x)}$ (or equivalently $\partial X (x)$, $\psi_{1,2} (x)$) obey the same OPE \eqref{eq:nonchiral_su2} as their bulk avatars, since the bulk-to-boundary OPE is regular.
Namely, the exactly marginal boundary operators \eqref{eq:nonchiral_psi} generate an $\hat{\mathfrak{su}}(2)_1$ current algebra on the boundary.

\paragraph{Fractional higher Berry curvature}

Following Section \ref{sec:WZW},
to compute the higher Berry connection and curvature,
we are interested in the disk partition function
\begin{align}
\langle 0| \exp[ i \oint d\theta 
(\omega^a +
\delta \omega^a(\theta) )
J^a (\theta)] | B_0\rangle \,,
\end{align}
where $J^a$'s are the exactly marginal operators \eqref{eq:nonchiral_psi} which are, as we explained, $\mathrm{SU}(2)$ current operators.
The disk partition function can be studied perturbatively in $\delta w^a$,
much the same way as in Section \ref{sec:WZW}. 
We then find that the higher Berry curvature 
is given in terms of
the $\mathrm{SU}(2)$ WZ term
at level 1.
While this may look innocuous, 
it would actually mean the fractional quantization of the integral of the higher Berry curvature
over the moduli space.
That is,
\begin{equation} \label{eq: frac}
    \frac{1}{2\pi} \int_M \mathcal{H}_{\mathrm{WZ}} = \frac{1}{N} \,,
\end{equation}
where $M = \mathrm{SU}(2)/\mathbb{Z}_N$.
Although we do not yet have a complete understanding of this phenomenon, the apparent lack of integral quantization may be related to the presence of orbifold singularities in the parameter space, along which the boundary conditions are non-simple.

It is known that gerbes on an orbifold $X/G$ are classified by the third orbifold cohomology $H^3_{\rm orb}(X/G;\mathbb{Z})$\cite{AMBP_2004__11_2_155_0,Lupercio:2002kn,AdemLeidaRuanOnline}.
In particular, $H^3_{\rm orb}(\mathrm{SU}(2)/\mathbb{Z}_N;\mathbb{Z})$ is essentially computed by Kawasaki in \cite{Kawasaki1973} as a cohomology of Lens complexes and it is isomorphic to $\mathbb{Z}$.\footnote{The Lens complex $L(b_2,b')$ for $(b_0,b_1,b_2)=(1,N,N)$ corresponds to $\mathrm{SU}(2)/\mathbb{Z}_N$, and Thm.2 in \cite{Kawasaki1973} implies $H^3_{\rm orb}(\mathrm{SU}(2)/\mathbb{Z}_N;\mathbb{Z})\simeq\mathbb{Z}$.} In general, an orbifold that is realized as a quatient of a manifold $X$ by a finite group $G$ is called a global quotient orbifold, and its orbifold cohomology is isomorphic to the Borel equivariant cohomology $H^n_{G}(X;\mathbb{Z})=H^n(X//G;\mathbb{Z})$ where $X//G=EG\times_{G}X/G$. 
When $X=\mathrm{SU}(2)=S^3$ and $G=\mathbb{Z}_N$, 
the Borel fibration $S^3 \to S^3//\mathbb{Z}_N \to B\mathbb{Z}_N$ induces an exact sequence 
\begin{eqnarray}
    0\to 
    H_{3}(S^3;\mathbb{Z}) 
    \simeq\mathbb{Z}
    \xrightarrow{\times N} H_{3}(S^3//\mathbb{Z}_N;\mathbb{Z}) 
    \simeq\mathbb{Z}
    \xrightarrow{/ N} H_{3}
    (B\mathbb{Z}_N;\mathbb{Z})
    \simeq\mathbb{Z}_N
    \to 0\,.
\end{eqnarray}
Thus, the generator of $H_3(\mathrm{SU(2)}/\mathbb{Z}_N;\mathbb{Z})$ can be regarded as ``$\frac{1}{N}S^3$'' in the homology theory.
Eq.~(\ref{eq: frac}) is normalized so that its integral over $S^3$ is quantized to an integer, 
and it implies that the gerbe constructed from the non-chiral deformation represents a generator of 
\begin{eqnarray}
    H^{3}_{\rm orb}(\mathrm{SU}(2)/\mathbb{Z}_N;\mathbb{Z})\simeq\mathbb{Z}\,.
\end{eqnarray}

\subsection{(1+1)d SPTs and symmetric boundaries} \label{sec:SPT}

In the context of tensor networks and gapped ground states of condensed matter systems, even when the parameter space is trivial, the triple inner product of MPS can still extract interesting topological invariants if we impose extra global symmetries \cite{liu2024multiwavefunctionoverlapmulti}.
In this section, we will discuss an analogue of this in BCFT.
That is, here we take a slight digression and consider a single, simple conformal boundary condition which preserves a global symmetry, instead of a family of boundary conditions which has been the main focus so far.

We begin by reviewing how in tensor networks the group cohomology invariant of a (1+1)d SPT state is computed using the triple inner product.
Consider (1+1)d SPT phases protected by an on-site unitary and finite symmetry group $G$. 
These phases are characterized by an SPT invariant valued in $H^2(G; \mathrm{U}(1))$ \cite{Chen_2013}.\footnote{We use the same notation for the group cohomology and the ordinary (e.g., singular) cohomology, but the distinction should be clear from the context.}
As discussed in \cite{liu2024multiwavefunctionoverlapmulti},
this invariant can be extracted from a triple inner product of three MPS, each representing the same physical state but written in different MPS gauges. These different MPS representations can be generated from a reference MPS by acting with a symmetry operation $g \in G$ on the physical degrees of freedom, which in turn induces a corresponding transformation in the MPS gauge.

Let us briefly recall the calculation of the higher Berry phase and triple inner product for generic MPS.  
We consider three MPS
$\Psi_{\alpha,\beta,\gamma}$
with MPS tensors 
$A_{\alpha,\beta,\gamma}$.
They are related by MPS gauge transformations,
$A_{\alpha} = g^{\ }_{\alpha\beta} A_{\beta} g^{\dag}_{\alpha\beta}$.
The triple inner product
of the three MPS 
can be calculated as
\begin{align}
&
\int \Psi_{\alpha}*\Psi_{\beta}*\Psi_{\gamma}
=
\mathrm{tr}\,(
\Lambda^L_{\beta\alpha}
\Lambda^R_{\beta\gamma}
\Lambda^R_{\gamma\alpha}
)
=
\mathrm{tr}\, 
(\Lambda^L_{\alpha} g_{\alpha\beta} g_{\beta\gamma} g_{\gamma\alpha})\, ,
\end{align}
where $\Lambda^{L,R}_{\alpha\beta}$ are the left and right fixed points 
of the mixed transfer matrices.
The formula can be applied when the MPS gauge transformations 
are related to the physical 
symmetry ($\equiv G$) of MPS (e.g., when MPS represent SPT phases). 
By taking two group elements $g,h\in G$ and also the identity, 
we can take ``$\alpha=g$'', ``$\beta=1$'', ``$\gamma=h$'',
then,
\begin{align}
\int \Psi_{g}*\Psi_{1}*\Psi_{h}
=
\mathrm{tr}\,(
\Lambda^L_{gh}
\Lambda^R_{g}
\Lambda^R_{h}
)
=
\mathrm{tr}\,(
\Lambda^L V_{gh}^{\dagger}
V_{g}
V_{h}
)
= e^{i\phi_{g,h}}.
\end{align}
Here, 
$V_g$ is the (projective) symmetry action of $G$ on the auxiliary Hilbert space of MPS, and 
$\phi_{g,h}$ is the group cocycle phase,
$V_{g}\, V_{h} = 
e^{i \phi_{g,h}}
V_{gh}$.

\begin{figure}[t!]
\centering
\includegraphics[width=.55\textwidth]{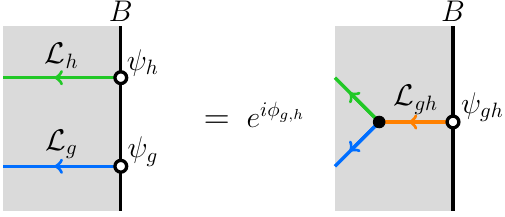}
\caption{For each group element $g \in G$, there is  a topological junction operator $\psi_g$ where the symmetry line operator $\mathcal{L}_g$ topologically ends on the conformal boundary $B$. The product of two such junction operators is determined by the group cocycle phase $e^{i\phi_{g,h}}$.
}\label{fig:boundary_cocycle}
\end{figure}

Let us discuss an analogous computation on the BCFT side.
Consider a (1+1)d CFT with a finite group global symmetry $G$, and a $G$-symmetric conformal boundary condition $B$.
Let $\{ \mathcal{L}_g : g \in G \}$ be the set of topological line operators generating the $G$ symmetry.
The boundary state $\ket{B}$ is invariant under the action of $G$ symmetry,
\begin{equation} \label{eq:symmetric_B}
    \mathcal{L}_g \ket{B} = \ket{B} ~~~~\text{for all $g \in G$} \,.
\end{equation}
Equivalently, for each $g \in G$, there is a topological junction operator $\psi_g$ which allows the line operator $\mathcal{L}_g$ to topologically end on the boundary condition $B$ \cite{Choi:2023xjw}.
The topological junction operator $\psi_g$ is unique for each $g \in G$ up to multiplication by nonzero complex numbers.
The product of two such junction operators, $\psi_g$ and $\psi_h$, is governed by the group cocycle phase $e^{i\phi_{g,h}} \in U(1)$.
See Figure \ref{fig:boundary_cocycle}.

\begin{figure}[t]
       \begin{minipage}{0.6\textwidth}
           \centering
           \begin{equation*}
              \adjincludegraphics[scale=1.0,trim={10pt 10pt 10pt 10pt},valign = c]{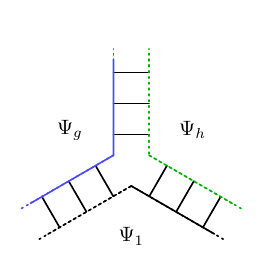}
              \;\;=\;\;
              \adjincludegraphics[scale=1.0,trim={10pt 10pt 10pt 10pt},valign = c]{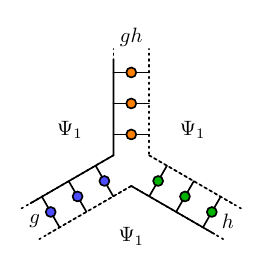}
           \end{equation*}
           \caption*{(a)}    
       \end{minipage}
       \hspace{10pt}
       \begin{minipage}{0.3\textwidth}
           \centering
           \begin{equation*}
              \adjincludegraphics[scale=1.1,trim={10pt 10pt 10pt 10pt},valign = c]{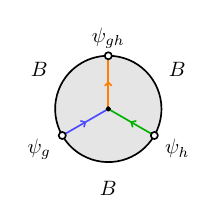}
           \end{equation*}
           \caption*{(b)}
       \end{minipage}
       \caption{
       Calculation of the SPT invariant. (a) In the tensor network picture, the SPT invariant is obtained by computing the triple inner product where the symmetry operators are inserted on the physical legs. (b) As the corresponding quantity, the SPT invariant can be obtained by computing the disk partition function with symmetry defects.
       } 
       \label{fig: SPT}
\end{figure}

The analogue on the BCFT side of the triple inner product of the MPS for a $G$-SPT phase, in three different MPS gauges, corresponds to a disk amplitude with symmetry line defects inserted, as illustrated in Figure \ref{fig: SPT}.
The disk amplitude is easily computed, by absorbing the topological line defects into the boundary using various recombination rules.
In particular, we first use the multiplication rule of junction operators in Figure \ref{fig:boundary_cocycle}, and then shrink the bubbles of topological line defects.
We refer to \cite[Section 2.2]{Choi:2024tri} and references therein, which summarize the full set of recombination rules and consistency conditions.
Following the conventions of \cite{Choi:2024tri},
we obtain 
\begin{align} \label{eq:SPT_bcft}
\begin{split}
    \adjincludegraphics[scale=1.0,trim={10pt 10pt 10pt 10pt},valign = c]{SPT_bcft.pdf}
    &= e^{i \phi_{g,h}} \times
    \adjincludegraphics[scale=1.0,trim={10pt 10pt 10pt 10pt},valign = c]{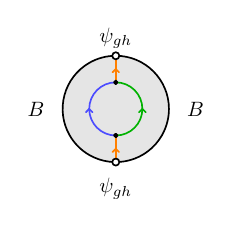}
    = e^{i \phi_{g,h}} \times
    \adjincludegraphics[scale=1.0,trim={10pt 10pt 10pt 10pt},valign = c]{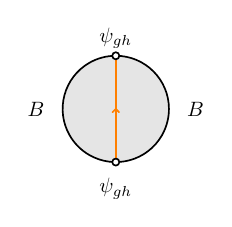}\\
    &= \exp \left[ i \phi_{g,h} \right] \times \frac{g_B}{|G|^{1/4}} \,,
\end{split}
\end{align}
where, $g_B = \langle 0 | B \rangle$ is the $g$-function of the boundary condition $B$, and $|G|$ is the order of the group $G$. 
The overall normalization $\frac{g_B}{|G|^{1/4}}$ is conventional.
Consistency conditions of the recombination rules of topological lines with the boundary condition require the phase factor $\exp \left[ i \phi_{g,h} \right]$ to be a group 2-cocycle \cite{Ostrik:2001xnt,Choi:2023xjw}.
Under the phase redefinitions of the junction operators, $\exp \left[ i \phi_{g,h} \right]$ changes by a coboundary, and its cohomology class in $H^2(G;\mathrm{U}(1))$ is well-defined.
Therefore, we see that the BCFT computation, up to an overall normalization (which is conventional), gives the same group cohomology invariant as in the tensor network formulation.

The above parallel between SPTs and symmetric boundaries generalizes to the case involving finite non-invertible symmetries described by a fusion category $\mathcal{C}$ \cite{Bhardwaj_2018,Chang_2019,Thorngren:2019iar}, which we explain briefly.
Let us first discuss the tensor network side.
In this case, the mixed transfer matrix twisted by $x\in\mathcal{C}$ has degenerate fixed points, even if the MPS is injective.
Let us denote the label of degeneracy by $i$ and let $\Lambda_{x}^{(i) L/R}$ be the fixed point tensors.
Then, one can evaluate the triple inner porduct of $\Psi_x,\Psi_y$ and $\Psi_1$ as 
\begin{align} \label{eq:nonabelian_cocycle}
    \tr{\, \big(\Lambda_{x\otimes y}^{(k) L}\, \Lambda_{x}^{(i) R}\, \Lambda_{y}^{(j) R}\big)}\, .
\end{align}
In particular, one can regard it as a non-abelian cocycle over $\mathcal{C}$ or equivalently the $L$-symbol \cite{inamura202411dsptphasesfusion} defining a fiber functor of $\mathcal{C}$, and this is the topological invariant of $\mathcal{C}$-symmetric SPT phases.

On the BCFT side,
we take a fusion category $\mathcal{C}$ of topological line defects in the bulk (1+1)d CFT, and a ``strongly symmetric'' boundary condition $B$ \cite{Choi:2023xjw}, meaning that for each topological line operator $\mathcal{L}_x$ associated to an object $x$ in the fusion category $\mathcal{C}$, the boundary state $\mathcal{C}$ is an eigenstate, similarly to \eqref{eq:symmetric_B} (the eigenvalues are given by the quantum dimension of the line operator).
For each $\mathcal{L}_x$, now in general there are multiple junction operators $\psi_{x}^i$ where $i$ labels the degeneracy.
This is analogous to the degeneracy label for the fixed points of the mixed transfer matrix on the tensor network side.
The disk amplitude in \eqref{eq:SPT_bcft}, which now includes additional labels for the junction operators, can again be shown to be given by $L$-symbols defining a fiber functor of $\mathcal{C}$, similar to the tensor network case.\footnote{The $L$-symbols were referred to as $\widetilde{F}$-symbols in \cite{Choi:2023xjw,Choi:2024tri}. They are matrix elements, in a chosen basis, of the associator of a module category over $\mathcal{C}$ \cite{Ostrik:2001xnt} (see also \cite[Appendix A]{Choi:2023xjw}).}

We comment that the calculation similar to \eqref{eq:SPT_bcft} can also be performed with a ``weakly symmetric'' boundary condition \cite{Choi:2023xjw}, which allows every topological line in $\mathcal{C}$ to topologically end on the boundary.
In that case, the analogous computation gives us particular components of the $L$-symbols (or $\widetilde{F}$-symbols) for the module category that the boundary condition belongs to.
However, unless the boundary condition is strongly symmetric, this does not correspond to the non-abelian cocycle \eqref{eq:nonabelian_cocycle} defining a fiber functor of $\mathcal{C}$.

\section{Loop space connections on boundary conformal manifolds}
\label{Loop space connections on boundary conformal manifolds}

We have employed modulated boundary states $|B(\xi)\rangle$ to compute the correlation function of bcc operators, focusing on a specific modulation profile—namely, a piecewise constant configuration with step-function singularities.
However, one might ask whether we can take advantage of more general modulations rather than restricting ourselves to such special cases. This naturally motivates the consideration of generic variations of the modulated boundary states,
\begin{align}
\begin{split}
|B(\xi+\delta \xi)\rangle &= |B(\xi)\rangle
 + \oint d\theta \sum_a \delta \xi^a(\theta) \frac{\delta}{\delta \xi^a(\theta)}
|B(\xi)\rangle \\
&\quad  +
\frac{1}{2!}
\oint d\theta\oint d\theta' \sum_{a,b} \delta \xi^a(\theta)\delta \xi^b(\theta')
\frac{\delta^2}{\delta \xi^a(\theta)\delta \xi^b(\theta')}
|B(\xi)\rangle
 +
\cdots \,.
\end{split}
\end{align}
This line of thought naturally leads to a (Berry) connection in the loop space
\cite{Mickelsson:1985ty,
Stone:1989jc,Iso:1990xk}, as we elaborate below. 
We recall that, in our examples, 
$|B(\xi)\rangle$
is constructed
starting from some reference $|B_0\rangle$,
$|B(\xi)\rangle  = U(\xi) | B_0\rangle$.
Here, as before,
$U(\xi)$ is given in terms of
a set of boundary marginal operators as 
$U(\xi) = \exp \big[i \oint d\theta \xi^a(\theta) \psi_a(\theta)\big]$,
and we 
assume $\psi_a(\theta)$ satisfies
an affine Lie 
(current) algebra,
\begin{align}
[\psi_a(\theta), \psi_b(\theta')]
= i f^{abc} \psi_c(\theta)\delta(\theta-\theta')
+
\frac{ik}{2\pi}\delta^{ab}
\partial_{\theta} \delta(\theta-\theta')\, .
\end{align}
This setup covers the examples we studied in the previous section.
It is worth emphasizing  
that the boundary marginal operators $\{\psi_a\}$ are not necessarily related to the bulk current algebra (e.g., Section \ref{sec:non-chiral}).
We also note, in principle, other choices—such as using Virasoro generators—are possible (although they come from dimension-two operator).
With this setup, 
we can formally think of 
$|B(\xi)\rangle$ as a section of a line bundle 
over the loop space $LM$.

The situation is reminiscent of coherent states created from the vacuum
(highest weight state) of the chiral current algebra
\cite{Stone:1989jc,Iso:1990xk}.
In the current case, however, 
our theory and marginal deformations are not chiral.
Moreover,
$|B_0\rangle$ and $|B(\xi)\rangle$ are  
defined on a non-chiral Hilbert space, and require proper normalization. 

Instead of viewing the state
$|B(\xi)\rangle$ as a section of a line bundle, 
one can adopt an alternative view, at least for some of the examples we discussed above. 
The state overlap
of the type
\eqref{eq:WZW_modulated_unfolded}
or
\eqref{eq: fermion}
can be thought of as, after doubling, the partition function of a chiral theory
on $S^2$ in the presence of a background gauge field. 
The phase of the chiral fermion path integral
has a phase ambiguity, 
and hence it defines a line bundle over the loop space, often referred to as 
``determinant line bundle'' or ``Pfaffian bundle.'' 

Let us now have a closer look 
at the line bundle defined by $|B(\xi)\rangle$.
As a warm-up, let us first consider a family of states created from 
the CFT vacuum $|0\rangle$
\cite{Stone:1989jc,Iso:1990xk},\footnote{
In \cite{Stone:1989jc,Iso:1990xk},
chial theories 
and coherent states created from the chiral vacuum were considered. 
Here in our case, both chiral and anti-chiral sectors are included and $|0\rangle$ is the vacuum of the full CFT Hilbert space.
Nevertheless, the calculations are almost identical.}
\begin{align}
| g(\xi) \rangle =U(\xi) |0\rangle\, .
\end{align}
Here, we assume the vacuum  
is normalized, 
$\langle 0|0\rangle =1$. 
We can then introduce the functional Berry connection
or the Berry connection in the loop space 
$LM$ from the first variation of 
$|g(\xi)\rangle$ as 
\begin{align}
{\cal A}(\xi)&=
\oint d\theta \sum_a {\cal A}^a(\xi) \delta \xi^{a}(\theta) = \langle g(\xi)|g(\xi+\delta \xi)\rangle-1
\, ,
\end{align}
much the same way as the regular Berry connection.
(On the right-hand side of the above equation, we only retain the first variation.)
Explicitly, it can be calculated as
\begin{align}
\label{eq: loop space conn 1}
{\cal A}(\xi) 
&=
\langle 0| (W(\xi, \delta \xi)-1)|0\rangle
\nonumber \\
&=
\oint d\theta
K\left(g^{-1} \delta g, \langle 0|\psi(\theta)|0\rangle\right) 
+
\frac{i k}{2 \pi} \oint d\theta
\int_0^1 ds\,
K
\left(\tilde{g}^{-1} \delta \tilde{g}, \partial_{\theta}\left(\tilde{g}^{-1} \partial_s \tilde{g}\right) \right)\, . 
\end{align}
Here, we parameterize the group manifold $M$ as
$g(\theta) = \exp [i \xi^a(\theta) T^a]$
using the generators 
$T^a$ satisfying $[T^a,T^b]=i f^{abc} T^c$,
and correspondingly define
$\psi(\theta)= \psi_a(\theta)T^a$.
(See Footnote \ref{footnote killing} for our convention of the generators $T^a$ and the Killing form $K$.)
We also 
introduced an extension of $g(\theta)$
using the parameter $s$, 
$\tilde{g}(\theta, s) = 
\exp[ i s \xi^a(\theta)T^a]$.
The operator $W(\xi,\delta \xi)$ 
in the first line is given by the following formula
(valid to first order in $\delta \xi$)
\begin{align}
\label{formula}
&
U(\xi+\delta \xi) = U(\xi)W(\xi,\delta\xi)\, , 
\nonumber \\
W(\xi,\delta\xi)-1
&=\oint d\theta
K\left(g^{-1} \delta g, \psi(\theta)\right)  
+\frac{i k}{2 \pi} \oint d\theta
\int_0^1 ds\,
K\left(\tilde{g}^{-1} 
\delta \tilde{g}, \partial_{\theta}\left(\tilde{g}^{-1} \partial_s \tilde{g}\right) \right)\, . 
\end{align}
The second term in the last line of 
\eqref{formula}
is closely related to the WZ term and the WZW action on the group manifold $M$.
In fact, for a one-parameter family of deformations $\xi(\theta,t)$,
we can integrate the loop space connection over $t$,
and see that the time-integral of the second term 
in \eqref{formula} is precisely the WZW action at the critical coupling. 
On the other hand, the first term,
which was called the finite-size term in Ref.\ \cite{Iso:1990xk}, 
can be non-zero depending on the choice (representation) of the vacuum $|0\rangle$,
but
vanishes in the thermodynamic limit as $\sim 1/L$ where $L$ is the circumference of the spatial circle. 
This term plays a role in properly reducing the phase space of the WZW model, 
and also explains why the 
representations of the WZW model at  fixed level $k$
truncate, e.g., 
for $\mathrm{SU}(2)_k$, 
the allowed representations
satisfy $0\le 2J\le k$.

For our case, we are interested in the family of boundary states, 
$|B(\xi)\rangle = U(\xi)|B_0\rangle$.
As we did for $|g(\xi)\rangle$,
we may want to introduce the loop space connection for $|B(\xi)\rangle$,
by considering
the overlap,
$
\langle B(\xi)|B(\xi+\delta \xi)\rangle
$.
The problem is that boundary states need to be properly normalized/regularized.
There appear to be various ways to do this. 
For example, we can first regularize the reference state
$|B_0\rangle$,
$|B_0\rangle \to 
|B_{0}^{\epsilon}\rangle \equiv e^{-\epsilon H_c} |B_0\rangle$,
and then apply $U(\xi)$.
Here, 
$H_c= L_0 + \bar{L}_0 -c/12$
is the time-evolution operator in the closed channel and $\epsilon>0$ 
is the regularization parameter.
We then consider 
$\langle B^{\epsilon}(\xi)| B^{\epsilon}(\xi+\delta \xi)\rangle
:=
\langle B_0|
e^{-\epsilon H_c} \,  U^{-1}(\xi) U(\xi+\delta \xi)\,
e^{-\epsilon H_c} |B_0\rangle
$.
The ratio
$\langle B^{\epsilon}(\xi)| B^{\epsilon}(\xi+\delta \xi)\rangle/
\langle B^{\epsilon}_0| B^{\epsilon}_0 \rangle
\sim
e^{-\fr{(2\pi)^2}{4\epsilon}\fr{c}{24}}
\langle B^{\epsilon}(\xi)| B^{\epsilon}(\xi+\delta \xi)\rangle
$
may remain finite
in the limit $\epsilon \to 0$
and can be used as a regularized overlap.
The regularized overlap can be calculated, 
using \eqref{formula},
as
\begin{align}
\langle B^{\epsilon}(\xi)| B^{\epsilon}(\xi+\delta \xi)\rangle
&=
\langle B_0|
e^{-\epsilon H_c} \,  
W(\xi,\xi+\delta \xi)
\,
e^{-\epsilon H_c} |B_0\rangle\, .
\end{align}
As in \eqref{eq: loop space conn 1},
the loop space connection in this case
consists of two contributions, 
coming from the 
first and second terms in 
\eqref{formula}.
The second contribution
is identical to the  
second term in
\eqref{eq: loop space conn 1},
up to an overall normalization of 
the (regularized) boundary states,
and hence related to the WZ term and WZW action.
On the other hand, the first term
is given in terms of the cylinder one-point function
of $\psi_a(\theta)$,
$
\langle B^{\epsilon}_0|  \psi_a(\theta) |B^{\epsilon}_0\rangle
=
\langle B_0| e^{-\epsilon H_c} \psi_a(\theta) e^{- \epsilon H_c}|B_0\rangle
$:
\begin{align}
\oint d\theta
K\left(g^{-1} \delta g, \langle B^{\epsilon}_0|\psi(\theta)|B_0^{\epsilon}\rangle\right) 
=
\frac{1}{2 h^{\vee}}
\oint d\theta\, 
(g^{-1}\delta g)^a(\theta) 
\langle B^{\epsilon}_0|\psi_a(\theta)|B^{\epsilon}_0\rangle
\, .
\end{align}
By translation symmetry of the regularized reference state $|B^{\epsilon}_0\rangle$, the one-point function depends only on the zero mode,
$
\langle B^{\epsilon}_0|  \psi_{a,0} |B^{\epsilon}_0\rangle
$.
Namely, 
this contribution 
comes  
solely from the uniform deformation, 
and can be interpreted as the regular Berry connection.

Alternatively, we could consider the regularized overlap
$
\langle B(\xi)| e^{-\epsilon H_c} |B(\xi+\delta \xi)\rangle
$
and take 
the opposite limit, $\epsilon\to \infty$.
In the context of boundary critical phenomena, 
this limit corresponds to the thermodynamic limit, 
$\epsilon \sim L/\beta \to \infty$,
where we have quantum field theory on a finite spatial interval of length $L$ and the compact dimension is regarded as a thermal cycle with inverse temperature $\beta$. 
In this limit, the free energy is extensive
and proportional to $L/\beta$.
In our current context, however, the roles of space and (Euclidean) time directions are interchanged, 
and it is not entirely clear if this is the relevant limit. Taken literally, the overlap behaves as $\sim 
e^{{\it const.}/L}$ 
and does not scale extensively. 
Nevertheless, this could simply reflect the fact that we are focusing on  
gapped ground states (that are smeared by
the CFT Euclidean time evolution $e^{- \epsilon H_c}$).
Accepting the overlap as a formal definition of the inner product,  
this limit is given in terms of the disk partition functions,
\begin{align}
\lim_{\epsilon \to \infty}
\langle B(\xi)| e^{-\epsilon H_c} |B(\xi+\delta \xi)\rangle
=
\langle B(\xi)|0\rangle \langle 0|B(\xi+\delta \xi)\rangle
e^{\epsilon \frac{c}{12}}\, .
\end{align}
Using again the formula \eqref{formula},
\begin{align}
\langle B(\xi+\delta \xi)| 0\rangle
&=
\langle B_0| U(\xi) W(\xi,\delta \xi)|0\rangle
\sim
e^{i \Phi(\xi,\delta \xi)}
\langle B(\xi) |0\rangle\, ,
\end{align}
where we  assumed 
$\psi_a(\theta)|0\rangle =0$
and 
\begin{align}
e^{i \Phi(\xi,\delta\xi)}
&\sim 1
+\frac{i k}{2 \pi} \oint d\theta
\int_0^1 ds\,
K\left(
\tilde{g}^{-1} \delta \tilde{g},
\partial_{\theta}
\left(\tilde{g}^{-1} 
\partial_s \tilde{g}\right) \right)\, .
\label{eq: Michelsonn}
\end{align}
Hence, we have obtained the loop space Berry connection for a family of boundary states, much the same way as the family of states $|g(\xi)\rangle$.

From a geometric point of view, $\ket{B(\xi)}$ forms a line bundle over $LM$.
The nontrivial topology of this line bundle may also be regarded as an anomaly in the space of \emph{boundary} coupling constants, following \cite{Cordova_2020a,Cordova_2020b}.
In particular, when $M$ is a simply connected compact Lie group $G$, 
\eqref{eq: Michelsonn} implies that the holonomy of the line bundle is given by the WZW action $\exp{(iS_{\rm WZW})}$ up to the finite-size correction term.
Therefore, this bundle is isomorphic to  Michelson's line bundle \cite{Mickelsson:1985ty,Waldorf_2016,KohnoCFTTopology,Brylinski} with the Chern class $[k]\in H^
2(LG;\mathbb{Z})\simeq \mathbb{Z}$.
This is the simplest example of a non-trivial anomaly in the space of boundary coupling constants.

In order to make a connection 
between the loop space connection 
(the functional derivative version of the Berry connection)
and the higher Berry connection,
we note transgression and regression maps
\cite{Brylinski}.
They map the Dixmier-Douady class of $M$
to a first Chern class 
in the loop space of $M$, $LM$, and vice versa.
Specifically, the transgression is the following relation:
\begin{align}
\oint d\theta\, {\cal A}_{a}[ \xi(\theta)] V^{a}[ \xi(\theta)]
=
\oint d\theta\, {\cal B}_{ab}(\xi) V^{a}[\xi(\theta)] \dot{\xi}^{b}(\theta)
\end{align}
for arbitrary tangent vectors in the loop space 
$LM$, 
$V^{a}[\xi(\theta)] \in T {L}M$.
In particular, a closed loop $\gamma$ in $LM$ defines a two-dimensional surface $\Sigma$ in $M$, and the holonomy along $\gamma$ gives the higher Berry phase over the surface $\Sigma$.

As a concrete example, let us consider the free boson field with $B$-field 
(Section \ref{sec:bosons}).
The first-order variation can be evaluated by using the above formula,
\begin{align}
U(\xi)^{-1}U(\xi+\delta \xi)
=
1
-i \oint d\theta\, P_{a}(\theta) \delta\xi^a(\theta)
+
\frac{1}{2} 
\oint d\theta 
\partial_{\theta}
\xi^a(\theta) \delta \xi^b(\theta)
\frac{i B_{ab}}{\pi}\, .
\end{align}
Relating, by transgression, the last term in the loop space connection
to the 2-form connection,
\begin{align}
\oint d\theta {\cal A}_{a}[\xi(\theta)] V^{a}[\xi(\theta)]
=
\oint d\theta 
\frac{i B_{ab}}{2\pi}
\dot{\xi}^b(\theta)
 V^{a}[\xi(\theta)]\, ,
\end{align}
we conclude the corresponding 2-form connection is $B_{ab}/2\pi$.

Several comments are in order. 
First, while we consider the second variation when computing the higher Berry connection, the loop space connection is derived from the first variation.
Second, in the calculation of the higher Berry connection, we set the reference point $\xi$ to be a constant configuration.
In contrast, in the loop space connection, 
we considered a generic inhomogeneous 
configuration $\xi$ as a reference.
Third, similar to the regular and higher Berry connections,
there is a gauge ambiguity 
for the loop space connection.
For example, inhomogeneous Dirichlet boundary states
in the free boson theory with $B$-field
can be constructed in the coherent state form as in Eq.\ (21) in \cite{Callan:1988wz},
without using $U(\xi)$.
This construction does not use the momentum operator $P_a(\theta)$, but only the coordinate operators
$X^a(\theta)$. 
Consequently, the boundary state does not depend on $B$ at all. Hence, it appears that the loop space connection is zero.

\section{Summary and discussion} \label{sec:summary}

Let us summarize our work.
We start with a field theory perspective.
In this work, we uncovered higher structures on boundary conformal manifolds in (1+1)d BCFT. 
In particular, we introduced and studied the 2-form higher Berry connection and the associated 3-form curvature.
They are defined in terms of 3-point correlation functions of bcc operators, which correspond to discontinuous jumps between different conformal boundary conditions.
This is in contrast to the more standard paradigm where various geometric structrues on conformal manifolds are accessed through the correlation functions of (exactly marginal) local operators on a single member of the family, and it suggests that there might be new structures to be uncovered even for the usual conformal manifolds of bulk CFTs or those of higher-dimensional defects.
In various examples, such as Narain CFTs and WZW models, we find that our higher Berry connection can be identified with the NS-NS $B$-field in string theory, when the boundary conformal manifold corresponds to a position moduli space of a D-brane.
This gives an independent check, beyond the intuition from tensor networks, that our definition of the higher Berry connection and curvature in BCFT is physically sensible.

From the perspective of quantum many-body physics in general, we translated the lattice formulation of the higher Berry phase to the continuum, quantum field theory language. 
To achieve this continuum description, we must appropriately smear the discrete MPS representation. 
Intuitively, this involves “filling in” the empty regions 
in Figure \ref{fig: concept}(a) or (b) with a continuum field theory. Concretely, we replace the entanglement Hamiltonian $\Lambda^L_x$
of the MPS for a half-space with an appropriate BCFT Hamiltonian.
Notably, for a certain class of integrable lattice models, the spectrum of $\Lambda^L_x$ coincides with that of the corresponding BCFT. 
Our discussion may shed light on why the entanglement spectrum
of gapped integrable models is given by the spectrum of BCFT. 
(See \cite{Cho_2017} for related discussion.)
This continuum formulation enables the analytical treatment of various examples.
In contrast to  
the previous (field-theoretical) formulation of the higher Berry phase, e.g., 
\cite{KS20-1, Hsin_2020,
2024CMaPh.405..191A}, our BCFT formulation provides a formula for the 2-form connection, in addition to the 3-form curvature.  

We note that our work is connected to a wide range of existing literature.
These include, for example, noncommutative geometry in BCFT, string and string field theories
\cite{Seiberg:1999vs,Schomerus:2002dc},
as well as works on Wess-Zumino terms and their gerbe structures \cite{gawkedzki1988topological,
Arcioni_2004,
runkel2008gerbeholonomysurfacesdefectnetworks}.
We also highlight that the tools from BCFT serve as powerful building blocks for
exploring quantum many-body systems, quantum field theories, and aspects of quantum spacetime
\cite{hung2024buildingquantumspacetimesbcft,
Brehm_2022,
brehm2024latticemodelscftsurfaces,
sopenko2023chiraltopologicallyorderedstates,
cheng2025precisionreconstructionrationalcft}.
It would be interesting to see how our formalism 
fits in these contexts
and explore
higher structures.

Finally, we conclude by listing open questions and future directions.

\paragraph{Spelling out full higher structures on boundary conformal manifolds}

First, there remain several foundational questions about our formalism. While our BCFT formulation provides explicit expressions for the 2-form connection and 3-form curvature, the full higher structure (the gerbe structure) has not been fully spelled out. In this work, we primarily introduced and utilized the triple inner product on boundary conformal manifolds in a setting involving three physically distinct states. However, as in Ref.\ \cite{ohyama2023higher}, one may also consider the triple inner product of three physically identical states taken from different coordinate patches of the boundary conformal manifold. In the corresponding MPS formulation, such states are related by different MPS gauges. It remains unclear whether an analog of these MPS gauge degrees of freedom exists in the BCFT framework. Clarifying these issues is left for future work.

\paragraph{Non-perturbative method}
In this work, we explicitly computed the 3-point functions of bcc operators in various examples.  
As exemplified by \cite{Runkel_1999}, one standard approach is to determine the OPE coefficients of bcc operators as solutions to the bootstrap equations.  
The OPE coefficients of bcc operators in minimal models and $\hat{\mathfrak{su}}(2)_k$ WZW models (for the case of rational boundaries) are given by the fusion matrices of conformal blocks,\footnote{Since conformal blocks satisfy the pentagon identity, they are closely related to the $F$- and $L$-symbols \cite{Moore:1988qv,Fuchs_2002}. 
} which can be understood from the perspective of solving the bootstrap equations.
In principle, the OPE coefficients of the bcc operators we are interested in should be determined in a similar way.  
In our case, however, the representations are twisted by line operators that induce exactly marginal deformations,  
which makes solving the bootstrap equations more difficult.  
Nevertheless, if this could be achieved, it would allow for an abstract and non-perturbative treatment of the present analysis, leading to a significant advancement.%}

\paragraph{Connection with tensor network methods}
While our BCFT formulation of the higher Berry phase is inspired by the MPS formulation, it is desirable to make the parallel more precice. 
The BCFT framework may also offer insights that feed back into the lattice-based approaches to 
many-body quantum systems. For example,  
it is known that boundary states can be realized as 
MPS using the tensor renormalization group method \cite{Iino_2019,Iino_2020,Iino_2021}. 
See also 
\cite{ueda2023fixedpointtensorfourpointfunction,
cheng2025precisionreconstructionrationalcft}
for discussions on how fixed-point tensor networks
relate to data in (B)CFT.
Since the fixed point of the transfer matrix of the matrix product boundary state constructed in this way is expected to give the bcc operator, it is anticipated that the correlation functions of the bcc operator can be computed numerically.

\paragraph{Open string field theory}

One of the hints that there must exist a 2-form connection on boundary conformal manifolds, in the form that we have defined in Section \ref{sec:berry_BCFT}, came from the mathematical similarity between the tensor network formulation of the higher Berry phase \cite{ohyama2023higher} and Witten's cubic open string field theory \cite{Witten:1985cc}.
On the tensor network side, there are clear analogues of the ``star product'' and ``integration'' which appear in the open string field theory, as was observed in \cite{ohyama2023higher} and briefly reviewed in Section \ref{sec:triple_MPS}.
Under the analogy, the triple inner product of three MPS corresponds to the cubic interaction vertex (of the lightest modes) in open string field theory.
In terms of the worldsheet BCFT, the cubic vertex is defined by a map
\begin{equation} \label{eq:cubic_star}
    \int (\cdot) * (\cdot) * (\cdot) : \mathcal{H} \otimes \mathcal{H} \otimes \mathcal{H} \rightarrow \mathbb{C} \,,
\end{equation}
where $\mathcal{H}$ is the open string Hilbert space with some specified boundary conditions, including the $bc$ ghosts.
Concretely, given three states $\ket{\mathcal{O}_i} \in \mathcal{H}$, $i=1,2,3$,
\eqref{eq:cubic_star} is given by
\begin{equation} \label{eq:cubic_star_uhp}
    \int \mathcal{O}_1 * \mathcal{O}_2 * \mathcal{O}_3 = \langle f_1 \circ \mathcal{O}_1 (0) f_2 \circ \mathcal{O}_2 (0) f_3 \circ \mathcal{O}_3 (0) \rangle_{\mathit{UHP}}\,,
\end{equation}
where $f_i$'s are appropriate conformal transformations.
Explicit forms of the conformal transformations can be found, for instance, in \cite{Rastelli:2000iu,Schnabl:2005gv,Erler:2019vhl}.
If we naively apply \eqref{eq:cubic_star_uhp} to the bcc operators between three conformal boundary conditions $B_{\alpha}$, $B_{\beta}$, and $B_{\gamma}$, we obtain
\begin{equation}
    \int \psi_{\alpha \beta} * \psi_{\beta \gamma } * \psi_{\gamma \alpha} = \left( \frac{2}{3\sqrt{3}} \right)^{\Delta_{\alpha \beta}+\Delta_{\beta \gamma}+ \Delta_{\gamma \alpha}} g_M c(\alpha , \beta , \gamma) \,.
\end{equation}
This contains the information about the higher Berry connection, as it includes the OPE coefficient $c(\alpha , \beta , \gamma)$, but moreover it also depends on the scaling dimensions of the bcc operators.
It would be interesting to further clarify the connection to open string field theory, and investigate, if any, possible applications of our formalism in string field theory and vice versa.
Notably, bcc operators have been used to construct solutions to open string field theory in
\cite{Kiermaier_2011}.

\paragraph{Higher structures on bulk conformal manifolds}
In this work, we investigated the existence of higher structures defined on boundary conformal manifolds in (1+1)d CFTs.
It is natural to expect that similar higher structures also exist on conformal manifolds of bulk marginal deformations.
See \cite{perezlona2025categorifiedstructuresmodulispaces,
Witten:1982hu,
Distler:1992gi,
Periwal:1989mx,
Sharpe:2024rwb,
Baggio:2017aww} for related discussions.
For instance, given an exactly marginal bulk operator, one can construct a conformal interface by turning on the marginal deformation only on half of space, which is a natural analogue of the bcc operator between two conformal boundary conditions \cite{Bachas:2001vj}.
Richer geometric structures are expected to emerge from
correlation functions of such conformal interfaces, involving junctions formed by multiple conformal interfaces.
A closely related situation appears in the context of tensor networks, where the definition of the higher Berry phase in (2+1)d gives rise to a 2-gerbe-like geometric structure\cite{qi2023charting,ohyama2024higher_b}.
Inspired by this analogy, we expect that correlation functions of conformal junction operators may encode a similar 2-gerbe-like structure in the space of (1+1)d CFTs.
We leave a detailed investigation of this possibility for future work.

\paragraph{Interface conformal manifolds}
Our definition of higher Berry connections and curvatures can be extended to conformal interfaces.  
Since boundaries and interfaces are related by the folding trick, this extension to conformal interfaces may be straightforward.
However, this does not imply that the interface conformal manifold is trivial,
since the BCFT obtained via folding may acquire new continuous symmetries \cite{Antinucci:2025uvj}.
An interesting direction would be to focus on how interface-specific quantities behave on the interface conformal manifold.  
One characteristic quantity of conformal boundaries is the boundary entropy \cite{Affleck:1991tk}, and the corresponding quantity for interfaces is the interface entropy.  
However, the interface entropy remains constant on the interface conformal manifold, hence, it is not particularly interesting in this context.  
In contrast, conformal interfaces possess other characteristic quantities,  
such as the transmission coefficient \cite{Quella:2006de} and the effective central charge \cite{Sakai:2008tt},
which are unique to interfaces.  
It would be interesting to explore how these quantities are related to the higher structure of the interface conformal manifold. 

\paragraph{Higher dimenions}
Another interesting direction is to develop a similar formalism in higher dimensions. 
We note that in the lattice many-body physics
context, the higher Berry phase in higher dimensions 
has been discussed
\cite{KS20-1,Hsin_2023,ohyama2024higher_a, sommer2024higher_b,2024CMaPh.405..191A}.
In higher dimensions, it is also possible to consider defects with codimension greater than one.
Even for general $p$-dimensional defects, one can define the defect conformal manifold,  
and it should be possible to study its higher structure using higher Berry connections and curvatures.
However, the interpretation from the perspective of tensor networks remains unclear.

\paragraph{Extended QFT and multi-wave function overlap}
The definition of the multi-wavefunction overlap is enabled by the locality of the theory and is expected to encode information about its extended structure.
Here, the extended structure refers to the following type of structure: 
In ordinary functorial quantum field theory, it is formulated as a monoidal functor from the bordism category of spacetimes to the category of vector spaces.
However, this formulation does not fully capture the locality structure of the theory.
To address this, the framework of extended quantum field theory (extended QFT) has been proposed, which seeks to capture the locality of the theory by replacing the bordism category and the category of vector spaces with higher categories.
In particular, extended structures in Chern–Simons theory have been extensively studied and it has been proposed that the object assigned to a point in Chern–Simons theory corresponds to the fusion category of boundary conditions in the WZW model \cite{henriques2017chernsimonstheoryassignspoint}.
The triple inner product may reflect the fusion structure of this category and could offer insight into how field-theoretic data can be extracted practically from point-assigned data.
A more general field-theoretic formulation of the multi-wavefunction overlap may thus provide a pathway toward a deeper understanding of extended quantum field theory.

\section*{Acknowledgments}
The authors thank Hiroshi Isono, Ho Tat Lam, Juan Maldacena, Kantaro Ohmori, Brandon C. Rayhaun, Nathan Seiberg, Sahand Seifnashri, Ophelia Sommer, Nikita Sopenko, and Yifan Wang
for discussion and useful comments.
Special thanks go to
Xueda Wen who lets us know of their
related works
\cite{Wen}, and also kindly agreed to coordinate submissions of our papers to arXiv. 
We thank the Yukawa
Institute for Theoretical Physics at Kyoto University, where this work was initiated during the YITP-T-22-02
on ``Novel Quantum States in Condensed Matter 2022.'' 
Y.C. gratefully acknowledges funding provided by the Roger Dashen Member Fund and the Fund for Memberships in Natural Sciences at the Institute for Advanced Study.
H.H. is supported by NSF QLCI grant OMA-2120757.
Y.K. is supported by the INAMORI Frontier Program at Kyushu University and JSPS KAKENHI Grant Number 23K20046.
S.O. is supported by the European Union's Horizon 2020 research and innovation programme through grant no. 863476 (ERC-CoG SEQUAM). 
S.O. is also supported by JSPS KAKENHI Grant Number 24K00522.
S.R. is supported by a Simons Investigator Grant from the Simons Foundation
(Award No. 566116).

\appendix

\section{Thouless charge pump in BCFT} \label{app:thouless}

In the main text, our focus was mainly on parameterized families of conformal boundary conditions without any additional global symmetries imposed (except in Section \ref{sec:SPT}, where we assumed a finite $G$-symmetry, but without any nontrivial conformal manifold).
In this Appendix, we discuss an example of a family of boundary conditions, where each member of the family preserves an internal global symmetry.
In particular, we study the $S^1$ family of Dirichlet boundary conditions in the $c=1$ compact boson CFT, where each member of the $S^1$ family preserves the $U(1)$ winding symmetry of the theory.\footnote{One can equivalently study the $T$-dual version, that is, the $S^1$ family of Neumann boundary conditions preserving the $U(1)$ momentum symmetry of the compact boson CFT.}
This gives a BCFT realization of the ordinary Thouless charge pump \cite{PhysRevB.27.6083}, as we explain below.\footnote{The discussion is also analogous to the ``particle on a circle'' quantum mechanics in \cite{Gaiotto:2017yup}.}

\subsection{Coupling to background gauge field}

Consider the $c=1$ compact boson CFT defined on the upper half-plane with the boundary sitting at $y = \mathrm{Im} z = 0$.
We have the $S^1$-parametrized family of Dirichlet boundary conditions, where the value of the boson field $X$ is fixed to be a constant $\xi$ at the boundary,
\begin{equation} \label{eq:Dirichlet}
    X|_{y=0} = \xi \,, \quad \xi \sim \xi + 2\pi \,. 
\end{equation}
The Dirichlet boundary condition can alternatively be described by introducing a Lagrange multiplier field $\widehat{X} \sim \widehat{X} + 2\pi$ living on the boundary $y =0$,
\begin{equation} \label{eq:Dirichlet_Lag}
    \frac{R^2}{4\pi} \int_{y>0} dX \wedge \star dX  + \frac{i}{2\pi} \int_{y=0} X d\widehat{X}  - \frac{i\xi}{2\pi} \int_{y=0} d \widehat{X} \,. 
\end{equation}
In this presentation, the parameter $\xi$ appears as the boundary theta angle for the Lagrange multiplier field $\widehat{X}$ on the boundary.
Integrating out $\widehat{X}$, one recovers the Dirichlet boundary condition \eqref{eq:Dirichlet}.
The Lagrange multiplier field $\widehat{X}$ can also be thought of as the restriction of the dual boson field to the boundary, which is freely fluctuating on the Dirichlet boundary (and hence the notation $\widehat{X}$).

Let us briefly recall the action of global symmetries on the Dirichlet boundary conditions.
At a generic radius $R$, the compact boson CFT has the symmetry
\begin{equation}
    \left(\mathrm{U}(1)_m \times \mathrm{U}(1)_w \right) \rtimes \mathbb{Z}_2^C \,.
\end{equation}
The momentum symmetry $\mathrm{U}(1)_m$ is generated by the current $J_m = \frac{-iR^2}{2\pi} \star dX$ and the winding symmetry $\mathrm{U}(1)_w$ is generated by the current $J_w = \frac{1}{2\pi} dX$.
The conservation equations are given by $dJ_m = dJ_w = 0$.
The charge conjugation symmetry $\mathbb{Z}_2^C$ acts as $X \rightarrow -X$ and similarly on the dual boson.
The momentum and winding symmetry operators are respectively given by
\begin{align} \label{eq:symmetry_ops}
\begin{split}
    U_\alpha (\gamma) &= e^{i\alpha \oint_\gamma \frac{-iR^2}{2\pi} \star dX} \,, \quad \alpha \in [0,2\pi] \,,\\
    V_\alpha (\gamma) &= e^{i\alpha \oint_\gamma \frac{1}{2\pi} dX} \,, \quad \alpha \in [0,2\pi] \,,\\
\end{split}
\end{align}
where $\alpha$ is a $\mathrm{U}(1)$ group parameter, and $\gamma$ is an arbitrary closed loop in spacetime.
The Nother currents are normalized such that the $\mathrm{U}(1)$ charges take integer values.

The family of Dirichlet boundary conditions \eqref{eq:Dirichlet_Lag} preserves the $\mathrm{U}(1)_w$ winding symmetry, but breaks the $\mathrm{U}(1)_m$ momentum symmetry.
The charge conjugation symmetry maps $\xi \rightarrow -\xi$, and is preserved for $\xi = 0 , \pi$.
By acting the winding symmetry operator $V_\alpha$ in \eqref{eq:symmetry_ops} to the Dirichlet boundary condition \eqref{eq:Dirichlet_Lag}, we see that under the winding symmetry, the boundary field $\widehat{X}$ gets shifted,
\begin{equation} \label{eq:winding_boundary}
    V_\alpha : ~~~~\widehat{X} \rightarrow \widehat{X} + \alpha \,.
\end{equation}
If we do not have any local operator insertion on the boundary, then the bulk-boundary Lagrangian \eqref{eq:Dirichlet_Lag} is invariant under the shift \eqref{eq:winding_boundary}, and therefore Dirichlet boundary condition is symmetric under the action of $V_\alpha$.
On the other hand, if we have an insertion of, say, a boundary local operator $e^{i w\widehat{X}}$ with $w\in \mathbb{Z}$, then it transforms under the action of $V_\alpha$ as $e^{iw\widehat{X}} \rightarrow e^{iw\alpha} e^{iw\widehat{X}}$, i.e. the boundary operator $e^{i w\widehat{X}}$ has charge $w$ under the winding symmetry (see, for instance, \cite[Appendix C]{Oshikawa_2006}).

Now, we couple the theory to a background gauge field $A$ for the $\mathrm{U}(1)_w$ winding symmetry.
In the bulk, the coupling is achieved by adding to the Lagrangian the minimal coupling term $A \wedge J_w = \frac{1}{2\pi} A \wedge dX$ (modulo classical counterterms).
On the boundary, since the winding symmetry shifts the boundary field $\widehat{X}$ as in \eqref{eq:winding_boundary}, we have to promote $d\widehat{X} \rightarrow d\widehat{X} - A$ to maintain gauge invariance.
Therefore, the combined bulk-boundary action in the presence of the background gauge field $A$ becomes
\begin{align}
\begin{split}
    &\frac{R^2}{4\pi} \int_{y>0}  dX \wedge \star dX  + \frac{i}{2\pi} \int_{y>0} A \wedge dX \\ & ~~~ + \frac{i}{2\pi} \int_{y=0} \phi (d \widehat{X} - A )  - \frac{i\xi}{2\pi} \int_{y=0} (d \widehat{X} - A) + iQ \int_{y=0} A \,.
\end{split}
\end{align}
We have added a decoupled probe charge $Q \in \mathbb{Z}$ on the boundary, which gives the last term.

As we vary the parameter $\xi$ by $2\pi$, we see that one unit of charge is pumped on the boundary,
\begin{equation}
    Q \rightarrow Q+1  ~~~~~\text{as}~~~~~ \xi \rightarrow \xi + 2\pi \,.
\end{equation}
This is the Thouless charge pump, realized in BCFT.
One may also view it as a mixed 't Hooft anomaly between the $S^1$ space of boundary coupling constant $\xi$ and the $\mathrm{U}(1)_w$ global symmetry \cite{Cordova_2020a,Cordova_2020b}.
Physically, there occurs a level crossing in the interval Hilbert space with two Diriclet boundary conditions, as we tune the parameter for one of the boundaries.
We explain this below.

\subsection{Spectral flow in the interval Hilbert space}

Let us analyze the above example with a slightly different approach.
The boundary state for the Diriclet boundary condition \eqref{eq:Dirichlet} is given by
\begin{equation}
    \ket{D(\xi)} = \frac{1}{2^{1/4} \sqrt{R}} \sum_{n \in \mathbb{Z}} e^{in \xi} | (n,0) \rangle \!\rangle \,,
\end{equation}
where $| (n,0) \rangle \rangle$ is the $\hat{\mathfrak{u}}(1)$ current algebra Ishibashi state associated to the vertex operator $e^{inX}$.
The open string channel partition function between two Dirichlet boundary conditions $\xi_1$ and $\xi_2$ is
\begin{equation}
    \langle D(\xi_1) | \tilde{q}^{\frac{1}{2}(L_0 + \bar{L}_0 - \frac{c}{12})} | D(\xi_2) \rangle = \frac{1}{\eta(q)}\sum_{w \in \mathbb{Z}} q^{R^2 \left(w+ \frac{\xi_2 - \xi_1}{2\pi} \right)^2 } \,.
\end{equation}
We see that the (current algebra) primary states in the interval Hilbert space are given by
\begin{equation}
    \text{Primary states:}~~~~~\ket{w} \in \mathcal{H}_{\xi_1 \xi_2} \,.
\end{equation}
The integer $w \in \mathbb{Z}$ labels the winding symmetry charge of the state $\ket{w}$ (and its descendants).
Under the state-operator correspondence, $\ket{w}$ is mapped to the boundary vertex operator $e^{iw \widehat{X}}$ in the previous subsection.
Its conformal dimension is given by
\begin{equation}
    \Delta_w = R^2 \left(w+ \frac{\xi_2 - \xi_1}{2\pi} \right)^2 \,.
\end{equation}

Now, as we tune the parameter $\xi_2$ while keeping $\xi_1$ to be fixed, there is a spectral flow in the interval Hilbert space,
\begin{equation}
    \ket{w} \rightarrow \ket{w+1}  ~~~~~\text{as}~~~~~ \xi_2 \rightarrow \xi_2 + 2\pi \,.
\end{equation}
This is an alternative way to detect the Thouless pump.
At $\xi_2 = \xi_1 \pm \pi$, there is a level crossing and the spectrum is doubly-degenerate,
\begin{equation}
    \Delta_{w} = \Delta_{-w \mp 1} ~~~ \text{when}~~~\xi_2 = \xi_1 \pm \pi \,.
\end{equation}

\section{Ground state energy flow under chiral deformation}\label{app:flow}

Here, we investigate how the ground state energy of the interval Hilbert space changes when the right boundary of the interval is deformed by a chiral deformation, using a bootstrap approach.  
We consider a diagonal rational CFT with the chiral algebra $\hat{\mathfrak{g}}_k \otimes \hat{\mathfrak{g}}_k$ generated by the currents $J^a$.

The spectrum of the interval Hilbert space can be extracted from the modular $S$ transformation of the following amplitude,
\begin{equation}
\bra{B_a} \ex{2\pi i \tau \pa{L_0 -\fr{c}{24}} } \ket{B_a^{\xi}}  =
\bra{B_a} \ex{2\pi i \tau \pa{L_0 -\fr{c}{24}} +2\pi i \xi \cdot J_0} \ket{B_a} = \sum_i B^i_{a^+}B^i_a \chi_i(\tau,\xi)\, ,
\end{equation}
where the unspecialized character is defined by \cite{Kac:1990gs}
\begin{equation}
\chi_i(\tau,\xi) \equiv
\mathrm{Tr}_i\, \ex{2\pi i \tau \pa{L_0 -\fr{c}{24}} +2\pi i \xi \cdot J_0}\, .
\end{equation}
The boundary state $\ket{B_a}$ is a solution to the bootstrap equation with the trivial gluing condition,  
and its coefficients are given by the following expression,
\begin{equation}
B_a^i=\fr{S_{ai}}{\sqrt{S_{0i}}}\, ,
\end{equation}
where $S_{ij}$ is the modular S-matrix.
We denote the charge conjugate of $a$ by $a^+$,  
and we have $B^i_{a^+} = (B^i_a)^*$.
The boundary state $\ket{B_a^{\xi}}$ is the boundary state deformed by the chiral deformation,  
and it can be simply written using $U_\xi = \ex{2\pi i \xi \cdot J_0}$ as
\begin{equation}
\ket{B_a^{\xi}} = U_\xi \ket{B_a}\, ,
\end{equation}
as explained around (\ref{eq:BC}).
The resulting open string partition function is
\begin{equation}
\ex{-2\pi i \fr{k}{2\tau} K(\xi\cdot J, \xi\cdot J)}\sum_i N_{a a^+}^i \chi_i\pa{-\fr{1}{\tau},\fr{\xi}{\tau}}
=
\sum_i N_{a a^+}^i \mathrm{Tr}_i\, \ex{-\fr{2\pi i}{\tau} \pa{L_0 +\xi \cdot J_0 + \fr{k}{2} K(\xi\cdot J, \xi\cdot J) -\fr{c}{24} }}\, ,
\end{equation}
where we recall that $K$ is the Killing form of $\mathfrak{g}$ defined by $K(A,B)\equiv \fr{1}{2 h^{\vee}}\tr\pa{\mathrm{ad}(A)\mathrm{ad}(B)}$ with the dual coxeter number $h^{\vee}$, and $N_{ij}^k$ is the fusion coefficient satisfying the Verlinde formula,
\begin{equation}
N_{ij}^k = \sum_l \fr{S_{il} S_{jl} \overline{S}_{kl}}{S_{0l}}\, .
\end{equation}
From this open string partition function,  
one finds that the conformal dimension of the bcc operator  
between the boundary conditions before and after the deformation  
is given by the following expression,
\begin{equation}\label{eq:bcc_flow}
\Delta_{bcc} =  \fr{k}{2} K(\xi\cdot J, \xi\cdot J)\, .
\end{equation}

\bibliographystyle{JHEP}
\bibliography{ref}

\end{document}